\documentclass[reprint,10pt,superscriptaddress,twocolumn,aps,prl]{revtex4-2}
\usepackage{amsmath}
\usepackage{amssymb}
\usepackage{graphicx}
\usepackage[colorlinks=true,citecolor=blue,linktocpage=true,linkcolor=blue,filecolor=blue,urlcolor=magenta]{hyperref}
\usepackage{physics}
\usepackage{bbm}
\usepackage{color}
\usepackage{dcolumn}
\usepackage{color}

\newcommand{\Am}[0]{\hat{a}}
\newcommand{\Ap}[0]{\hat{a}^\dagger}

\newcommand{\beginsupplement}{%
            \setcounter{table}{0}
            \renewcommand{\thetable}{S\arabic{table}}%
            \setcounter{equation}{0}
            \renewcommand{\theequation}{S\arabic{equation}}%
            \setcounter{figure}{0}
            \renewcommand{\thefigure}{S\arabic{figure}}%
}

\makeatletter
\def\maketitle{
\@author@finish
\title@column\titleblock@produce
\suppressfloats[t]}
\makeatother

\begin{document}
\title{
All-microwave manipulation of superconducting qubits with a fixed-frequency transmon coupler
}
\author{Shotaro Shirai}
\email{shirai-shotaro@g.ecc.u-tokyo.ac.jp}
\affiliation{Komaba Institute for Science (KIS), The University of Tokyo, Meguro-ku, Tokyo, 153-8902, Japan}
\author{Yuta Okubo}
\affiliation{Komaba Institute for Science (KIS), The University of Tokyo, Meguro-ku, Tokyo, 153-8902, Japan}
\author{Kohei Matsuura}
\affiliation{Department of Applied Physics, Graduate School of Engineering, The University of Tokyo, Bunkyo-ku, Tokyo 113-8656, Japan}
\author{\\Alto Osada}
\affiliation{Komaba Institute for Science (KIS), The University of Tokyo, Meguro-ku, Tokyo, 153-8902, Japan}
\affiliation{PRESTO, Japan Science and Technology Agency, Kawaguchi-shi, Saitama}
\author{Yasunobu Nakamura}
\affiliation{Department of Applied Physics, Graduate School of Engineering, The University of Tokyo, Bunkyo-ku, Tokyo 113-8656, Japan}
\affiliation{RIKEN Center for Quantum Computing (RQC), Wako, Saitama 351--0198, Japan}
\author{Atsushi Noguchi}
\email{u-atsushi@g.ecc.u-tokyo.ac.jp}
\affiliation{Komaba Institute for Science (KIS), The University of Tokyo, Meguro-ku, Tokyo, 153-8902, Japan}
\affiliation{RIKEN Center for Quantum Computing (RQC), Wako, Saitama 351--0198, Japan}
\affiliation{Inamori Research Institute for Science (InaRIS), Kyoto-shi, Kyoto 600-8411, Japan}

\date{\today}

\begin{abstract}
All-microwave control of fixed-frequency superconducting quantum computing circuits is advantageous for minimizing the noise channels and wiring costs. Here we introduce a swap interaction between two data transmons assisted by the third-order nonlinearity of a coupler transmon under a microwave drive. We model the interaction analytically and numerically and use it to implement an all-microwave controlled-Z gate. The gate based on the coupler-assisted swap transition maintains high drive efficiency and small residual interaction over a wide range of detuning between the data transmons.
\end{abstract}

\maketitle

Quantum information science has evolved with the discovery and proposal of promising applications and is now entering the phase of testing them using actual quantum hardware. However, currently-available quantum hardware is still vulnerable to environmental noise and energy loss. Hence, implementing quantum error correction~\cite{KITAEV20032, bravyi1998quantum} in a scalable approach is essential to demonstrate their potential and is thus being pursued in many physical systems~\cite{Ladd2010, Kjaergaard_review, Brown2016, Saffman_2016, kloeffel2013prospects, doi:10.1063/1.5115814}.

Superconducting circuits are one of the leading platforms toward realization of fault-tolerant quantum computing~\cite{doi:10.1063/1.5089550, Chen2021}. Among various types of qubits, fixed-frequency transmon~\cite{Koch2007} is a promising building block thanks to its long coherence time and small wiring overhead. For the architecture using fixed-frequency transmons, various all-microwave two-qubit gates have been proposed \cite{CR_2010Rigetti, bSWAP_2012Poletto, MAP_2013Chow, RIP_2016Paik, NogFastCT, FogiCZ_2020krinner, siZZle_2021Mitchell, PRXQuantum.2.040336}, and the cross-resonance (CR) gate is the most commonly-used entangling gate~\cite{CR_2010Rigetti, CR_1, CR_procedure, zz_cl_IBM_reso}. In those schemes, however, weak anharmonicity of transmons results in a residual static ZZ interaction, which causes coherent errors and reduces the fidelity of operations. Therefore, it is of importance to suppress the residual interaction while maintaining the gate operation speed. A widely-adopted method for the purpose is to set the detuning between neighboring transmons to be in the so-called straddling regime, i.e., within the limited anharmonicity~\cite{Koch2007}, though there remain some unwanted higher-order transitions to be avoided. The so-called frequency-crowding problem hinders the straightforward design of the circuits~\cite{FirstPrinciples, opt_f_crowding}. Recently, this problem has been addressed partially via frequency tuning using post-fabrication techniques such as laser annealing~\cite{granata2008trimming, Hertzberg2021, doi:10.1126/sciadv.abi6690, doi:10.1063/5.0102092}, but further tolerance in design parameters is still desirable.

In this study, we propose and experimentally demonstrate a drive-efficient single-excitation exchange interaction between two transmons that allows all-microwave controlled-Z~(CZ) gate over a wide range of detuning between data transmons. In this scheme, the interaction is activated by applying a microwave drive to a coupler transmon whose third-order nonlinearity plays a central role. The process can be understood as four-wave mixing involving three qubits and a drive microwave photon. We have therefore named it Coupler-Assisted Swap (CAS) interaction or transition. Note that a similar mechanism is used to exchange a single photon between two cavities~\cite{PhysRevX.10.021038}.

Remarkably, the CAS transition relies neither on the less-coherent higher energy levels outside the qubit subspace nor on the direct transverse coupling between the data transmons. At the same time, the latter can in turn be utilized for the suppression of unwanted ZZ coupling~\cite{zz_cl_IBM_reso, zz_cl_Princeton}. This also widens the choice of the qubit detuning in the device design.

The circuit under consideration [Figs.\,\ref{fig:sample_jqsv24}(a) and (b)] consists of three fixed-frequency transmons, with the total Hamiltonian being modeled as coupled Duffing oscillators under the rotating-wave approximation,
\begin{figure}
    \centering
    \includegraphics[width=8.6cm]{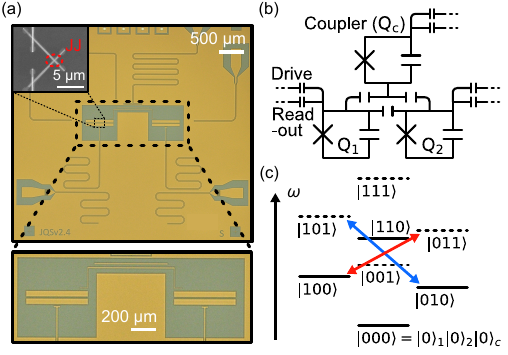}
    \caption{(a) Optical images of a fabricated superconducting circuit (top) and three coupled transmons (bottom). Most of the structures are made from TiN electrodes (yellow) on a Si substrate (gray). Inset:\,Scanning electron micrograph of an Al/AlO$_x$/Al Josephson junction fabricated with the in-situ bandage technique~\cite{insitu_bandage}. (b) Equivalent circuit diagram of the coupled transmon system, where readout resonators, Purcell filters, and drive lines are omitted. Only the coupling capacitors connected to them are depicted. Q$_1$, Q$_2$, and Q$_c$ represent the two data qubits and one coupler qubit, respectively. (c) Energy-level diagram of the system eigenstates $\ket{ijk}=\ket{i}_{1}\ket{j}_{2}\ket{k}_c$ ($i,j,k \in \{0, 1\}$) truncated to the first excited state of each transmon. The blue and red arrows are the CAS transitions activated by microwave drives. The dashed energy levels involve the single excitation of the coupler.}
    \label{fig:sample_jqsv24}
\end{figure}
\begin{align}
    \hat{H}/\hbar &= \sum_i \left(\omega_i\Ap_i\Am_i + \frac{\alpha_i}{2}\Ap_i\Ap_i\Am_i\Am_i\right) + \sum_{i\neq c} g_{ic}(\Ap_i \Am_c + \Am_i\Ap_c),
    \label{eq:H}
\end{align}
where $\hbar$ is the reduced Planck constant, $\omega_i$ and $\alpha_i$~($i \in \{1, 2, c\})$ are the fundamental frequency and anharmonicity of each transmon, $\Am_i$ and $\Ap_i$ are the annihilation and creation operators, and $g_{ic}$ is the transverse coupling strength between the data transmon Q$_i$ and the coupler transmon Q$_c$. Here we assume the dispersive regime $|g_{ic}/\Delta_{ic}| \ll 1$ , where $ \Delta_{ic}=\omega_i-\omega_c$. For the moment, we omit the direct coupling between the data qubits, $g_{12}$, and consider up to the third excited state of each transmon.

To induce the interaction between the data qubits, we apply a microwave drive
\begin{align}
    \hat{H}_d/\hbar = \Omega_d \cos{\omega_d t} \left(\Ap_c + \Am_c \right)
    \label{eq:Hd} 
\end{align}
to the coupler qubit, where $\omega_d$ and $\Omega_d$ are the drive frequency and amplitude, respectively. To find an analytical expression of the induced CAS interaction strength, we expand the drive term to the second order of $g_{ic}$ using the Schrieffer-Wolff transformation [See Supplemental Material]. Then, we obtain the effective drive term in the Hamiltonian,
\begin{align}
    \hat{H}'_d &\approx \hat{H}_d + [\hat{S},\hat{H}_d] + \frac{1}{2}[\hat{S}_1,[\hat{S}_1,\hat{H}_d]], \label{eq:Hd_eff_2D}
\end{align}
where the anti-Hermitian operator $\hat{S} = \hat{S}_1 + \hat{S}_2$ fulfills the conditions
\begin{align}
    [\hat{H}_0,\hat{S}_1] &+ \hat{O}_1 = 0, \\
    [\hat{H}_0,\hat{S}_2] &+ \hat{O}_2 = 0.
\end{align}
Here, $\hat{O}_1$ is the off-diagonal part of Eq.\,\eqref{eq:H}, corresponding to the coupling term, and $\hat{H}_0$ is the rest. $\hat{O}_2$ is the off-diagonal part of $\frac{1}{2}[\hat{O}_1, \hat{S}_1]$. The effective drive term, Eq.\,\eqref{eq:Hd_eff_2D}, due to the third-order nonlinearity of the coupler, contains many transition matrix elements between eigenstates in the Hilbert space spanned by the three transmons. Among them, we focus on the CAS transitions between data qubits assisted by the single-photon excitation of the nonlinear coupler, such as $\ket{010} \leftrightarrow \ket{101}$ and $\ket{100} \leftrightarrow \ket{011}$, respectively illustrated by the blue and red arrows in Fig.\,\ref{fig:sample_jqsv24}(c), where $\ket{ijk}=\ket{i}_{1}\ket{j}_{2}\ket{k}_c$ ($i,j,k \in \{0, 1\}$). Here, we refer to them as the blue and red CAS transitions at the frequencies of $\omega_b$ and $\omega_r$, respectively. We also assume $\omega_1 > \omega_2$ without loss of generality. From Eq.\,\eqref{eq:Hd_eff_2D}, analytical expressions for the drive-induced oscillation frequencies are calculated under the rotating-wave approximation as
\begin{align}
    \Omega_b &\approx 2\bra{010}\hat{H}'_d\ket{101}/\hbar \notag\\
    &= \frac{2g_{1c} g_{2c} \alpha_c \Omega_d}{\Delta_{12}(\omega_c-\omega_1+\alpha_c)(\omega_c-\omega_2)}, \label{eq:gb}\\
    \Omega_r &\approx 2\bra{100}\hat{H}'_d\ket{011}/\hbar \notag\\
    &= \frac{-2g_{1c} g_{2c} \alpha_c \Omega_d}{\Delta_{12}(\omega_c-\omega_2+\alpha_c)(\omega_c-\omega_1)}, \label{eq:gr}
\end{align}
respectively for the blue and red CAS transitions, where $\Delta_{12} = \omega_1 - \omega_2$. The CAS-based CZ gate can be realized by applying a resonant 2$\pi$-pulse of the blue (red) CAS transition, where the state $\ket{010}$\,($\ket{100}$) acquires the geometric phase of $\pi$ after a round trip~\cite{Erik_GP}. 

\begin{figure}
    \centering
    \includegraphics[width=8.6cm]{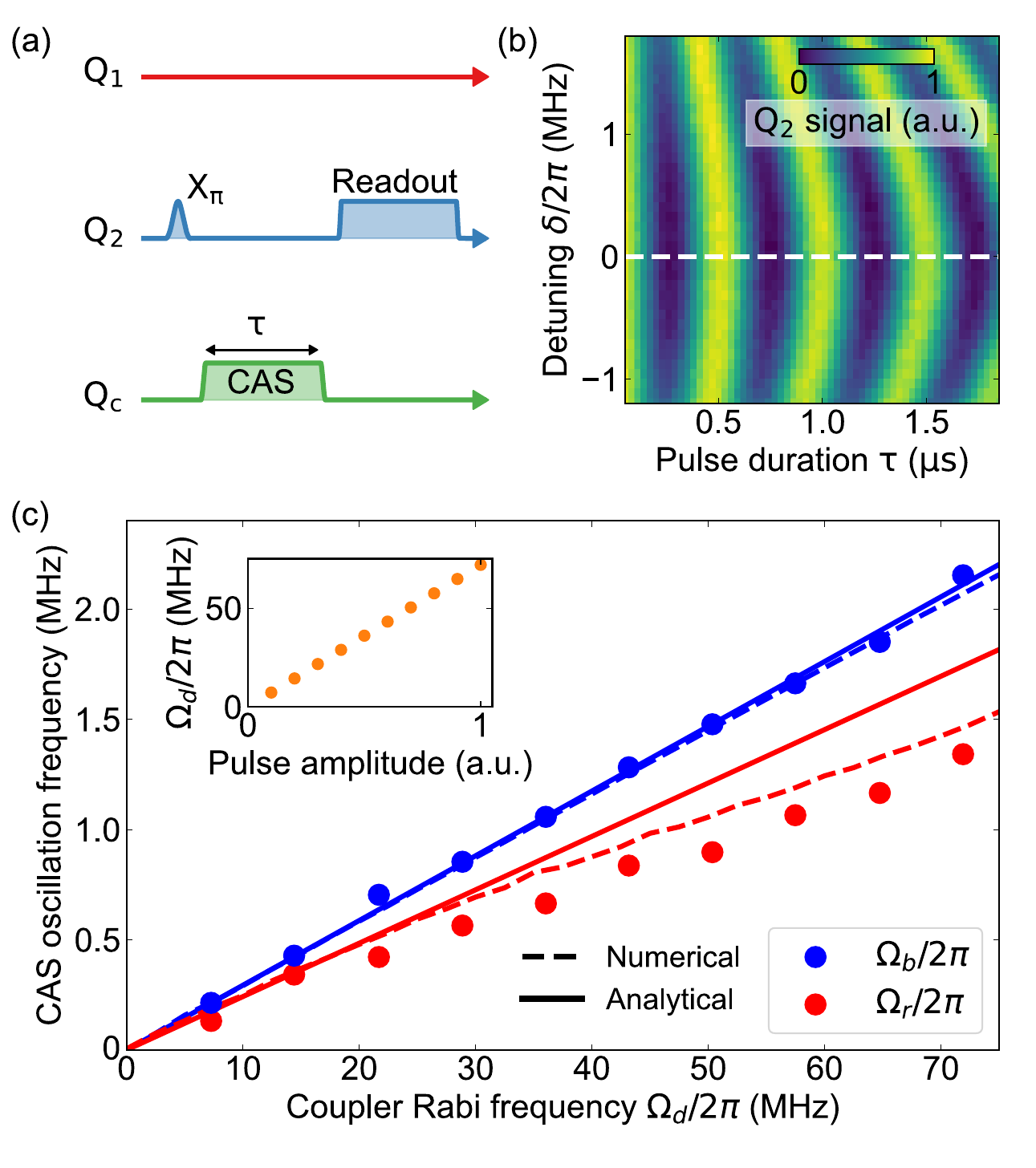}
    \caption{(a) Pulse sequence for the data transmons, Q$_1$ and Q$_2$, and the coupler transmon $\mathrm{Q_c}$ to measure the CAS oscillation frequency between the states indicated by the blue arrow in Fig.\,\ref{fig:sample_jqsv24}(c). To activate the transition, we prepare Q$_2$ in the first excited state with a $\pi$ pulse, and then apply a drive pulse to the coupler. (b) Chevron pattern of the blue CAS transition as a function of the detuning $\delta = \omega_d - \omega_b$ and the pulse duration $\tau$. The white dashed line, $\delta = 0$, shows the resonance condition for the blue CAS transition at $\omega_b/2\pi \simeq 6.4207$~GHz. The data is obtained for the coupler drive amplitude $\Omega_d/2\pi = 72$~MHz. Note that the blue CAS transition frequency $\omega_b$ depends on $\Omega_d$ through the ac Stark shift and the associated correlated oscillations of the excited-state populations of the three transmons are separately observed [See Supplemental Material]. (c) Blue and red CAS oscillation frequencies obtained from the fitting. The blue and red solid lines are analytical evaluations respectively using Eqs.\,\eqref{eq:gb} and \eqref{eq:gr} with experimentally-determined parameters. The dashed lines are the numerical simulations based on Eqs.\,\eqref{eq:H} and \eqref{eq:Hd} using QuTiP~\cite{JOHANSSON20121760, JOHANSSON20131234}. Inset: $\Omega_d$ calibration result by driving the fundamental mode of the coupler qubit as a function of the pulse amplitude.}
    \label{fig:calib_CAS}
\end{figure}

In the experiment, we use a circuit consisting of three capacitively coupled fixed-frequency transmons~\cite{Koch2007}, two $\lambda/4$ coplanar-waveguide~(CPW) readout resonators and Purcell filters~\cite{reed2010fast} for data transmons, and one $\mathrm{\lambda}/2$-CPW readout resonator for the coupler transmon\,[Fig.\,\ref{fig:sample_jqsv24}(a)]. The device parameters are the following: The fundamental frequencies of the data transmons and the coupler transmon are $\omega_{1}/2\pi\simeq5.641\,$GHz, $\omega_{2}/2\pi\simeq5.507\,$GHz and $\omega_{c}/2\pi\simeq6.317 \,$GHz, respectively. The third-order nonlinearities of the transmons are  $\alpha_{1}/2\pi\simeq-300\,$MHz, $\alpha_{2}/2\pi\simeq-303\,$MHz and $\alpha_{c}/2\pi\simeq-381\,$MHz, and the transverse coupling strengths between the data transmons and the coupler transmon are $g_{1c}/2\pi\simeq40\,$MHz and $g_{2c}/2\pi\simeq31\,$MHz. The direct transverse coupling between data transmons is estimated to be $g_{12}/2\pi\simeq1.9\,$MHz by fitting the measurement result of the ZZ interaction, and the static ZZ interaction strength between data transmons is estimated as $\xi_0/2\pi\simeq-1.5\,$kHz [See Supplemental Material].
The transmons, Q$_1$, Q$_2$ and Q$_c$, have energy relaxation times $T_1$ of $95\,\mathrm{\mu s}$, $108\,\mathrm{\mu s}$ and $15\,\mathrm{\mu s}$, Ramsey dephasing times $T_2^*$ of $76\,\mathrm{\mu s}$, $81\,\mathrm{\mu s}$ and $15\,\mathrm{\mu s}$, and echo dephasing times $T_2^e$ of $88\,\mathrm{\mu s}$, $166\,\mathrm{\mu s}$ and $18\,\mathrm{\mu s}$, respectively. Part of the reason for the lower coherence of the coupler transmon is presumably due to its narrower electrodes and concentrated electric field~\cite{PR_IBM_Analytical}. This can be improved by design modifications.

For single-qubit gates, we use a Gaussian pulse with its FWHM $\sigma=7.5\,$ns, total gate length $4\sigma$, and with derivative removal by adiabatic modulation (DRAG)~\cite{DRAG_leak}. For the CAS transitions, we apply to the coupler a flat-top drive pulse with Gaussian-shaped edges of $\sigma = 10$~ns and a total edge length of $4\sigma$.

We first measure the CAS oscillation frequencies as a function of the drive amplitude $\Omega_d$. As shown in Fig.\,\ref{fig:calib_CAS}(a), for the blue CAS transition, we prepare the system in $\ket{010}$ and then apply a coupler drive with a given $\Omega_d$ and with various drive frequencies and pulse lengths. By fitting the resulting oscillations in the excited state population of Q$_2$~[Fig.\,\ref{fig:calib_CAS}(b)], we obtain the oscillation frequency $\Omega_b$, which is plotted with blue dots in Fig.\,\ref{fig:calib_CAS}(c) as a function of $\Omega_d$. Similarly, $\Omega_r$ for the red CAS transition is obtained. 

\begin{figure}
    \centering
    \includegraphics[width=8.6cm]{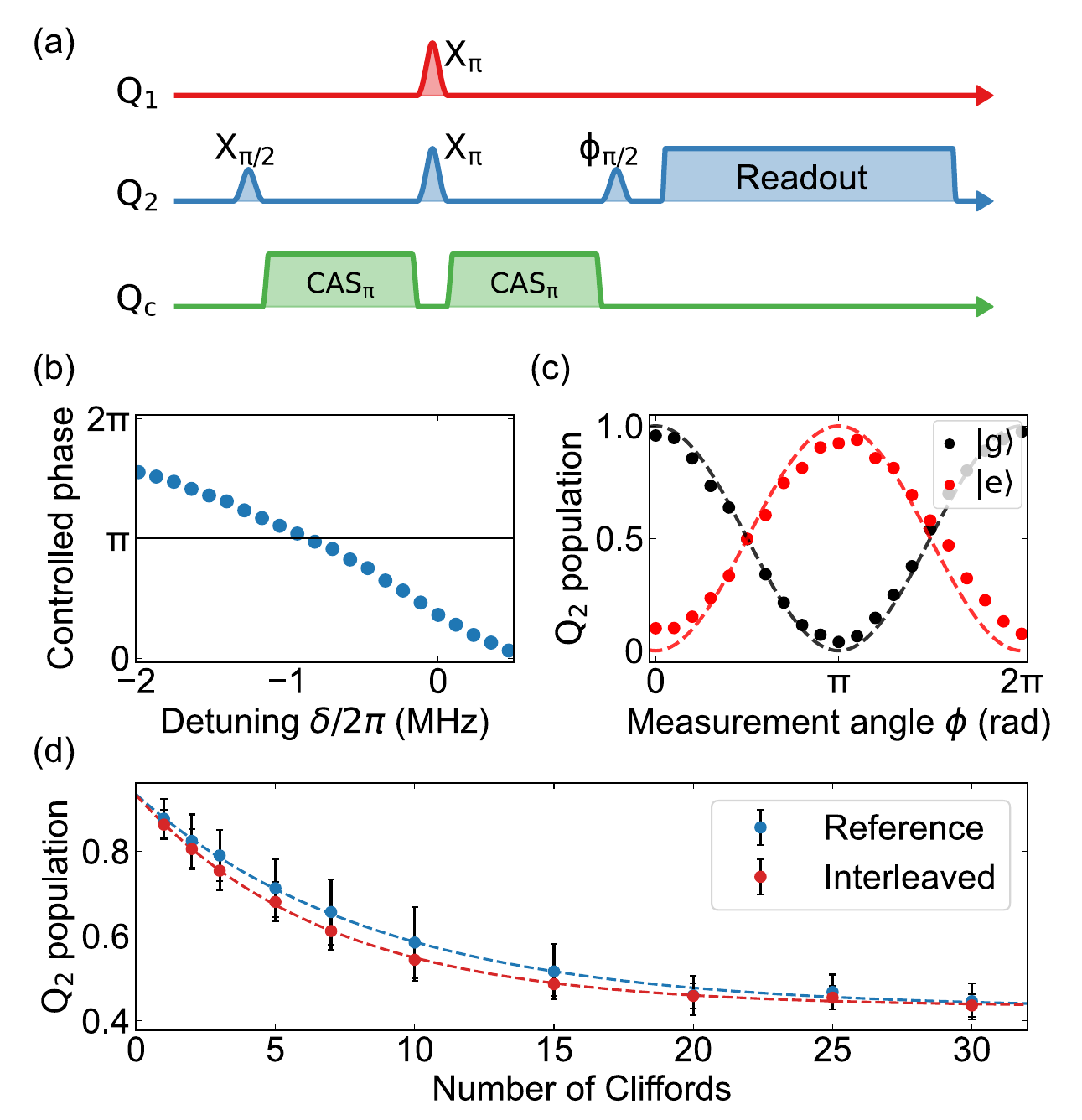}
    \caption{(a) Pulse sequence for measuring the control phase using the Joint Amplification of ZZ (JAZZ) protocol~\cite{GARBOW1982504, SuppressZZ_TQ_CSFQ}. The measurement angle $\phi$ is swept to find an optimal CAS drive frequency for the CZ gate. (b) Controlled phase measured as a function of $\delta = \omega_d - \omega_b$, where $\omega_b/2\pi \simeq 6.4157$~GHz for the drive amplitude of $\Omega_d/2\pi = 75$~MHz. For each drive frequency, we adjust the pulse length so that the coupler returns to the ground state. (c)~Ramsey fringes measured with the calibrated detuning of the blue CAS drive. A $\pi$ phase shift is observed depending on the states of the control transmon $\mathrm{Q_1}$. The vertical axis is the signal of $\mathrm{Q_2}$ normalized to the responses of the ground and excited states of $\mathrm{Q_2}$. The black and red dashed curves represent the functions of the ideal CZ gate. (d) Interleaved randomized benchmarking\,(IRB). Blue and red dots are the averaged experimental results of the reference RB and IRB, respectively. The number of randomly-generated RB sequences used is 30, and the error bars represent 95\% confidence. Dashed lines are fitting curves to the decay model. The horizontal axis is the number of Clifford gates applied. All single-qubit Clifford gates consist of two $X_{\pi/2}$ gates and three virtual-Z gates, and the length of the CZ gate is $504\,$ns. Thus, the average duration of the two-qubit Clifford gate is $945\,$ns, where each spacing between two successive pulses is set to $6\,$ns \cite{Barends2014}.}
    \label{fig:calib_cphase}
\end{figure}

To check the validity of our theoretical model, we also plot in Fig.\,\ref{fig:calib_CAS}(c) the analytically obtained values from Eqs.\,\eqref{eq:gb} and~\eqref{eq:gr} and the numerical ones from Eqs.\,\eqref{eq:H} and~\eqref{eq:Hd}. For the blue CAS transition, our model is in good agreement with the experimental result. For the red CAS transition, the numerical calculation is also in good agreement with the experimental result, but the analytical model shows a deviation in the strong drive regime. This could be an off-resonant effect of a single-photon transition ($\ket{110}\leftrightarrow\ket{201}$) and two-photon transitions ($\ket{000}\leftrightarrow\ket{002}$, $\ket{010}\leftrightarrow\ket{102}$) near $\omega_r$. The blue CAS oscillation frequency fits better as there are no near disturbing transitions on the higher frequency side of $\omega_c$ because of the negative anharmonicity of the transmon. This can be an advantage for relaxing the frequency crowding problem.

We next implement the CZ gate using the blue CAS transition. We first determine the relation between the CAS drive detuning and the pulse duration by fitting the chevron pattern with the drive amplitude $\Omega_d/2\pi = 75$~MHz, which is slightly larger than the experiments presented in Fig.\,\ref{fig:calib_CAS} and the resulting blue CAS oscillation frequency is about 2.2 MHz.
We then calibrate the amount of controlled phase shift using the Joint Amplification of ZZ~(JAZZ) sequence~\cite{GARBOW1982504, SuppressZZ_TQ_CSFQ} shown in Fig.\,\ref{fig:calib_cphase}(a). In this sequence, $\mathrm{Q_2}$ is detected in the excited state when the amount of the controlled phase shift is $\pi$ and the final measurement angle $\phi$ is 0. By sweeping $\phi$ and fitting the result with a cosine function, the amount of the control phase is obtained from the phase shift of the cosine function. Figure\,\ref{fig:calib_cphase}(b) shows the obtained phase shift as a function of the CAS drive detuning. The optimal drive frequency and flat-top duration are obtained by interpolating the result. The associated local phase shift induced by the CAS drive on each qubit is evaluated and canceled with a virtual-Z gate \cite{mckay2017efficient} to implement the CAS-based CZ gate. Through the interleaved randomized benchmarking (IRB)~\cite{magesan2012efficient} of the calibrated CAS-based CZ gate, a fidelity of 97.8(6)\% is obtained [Fig.\,\ref{fig:calib_cphase}(d)]. The master-equation simulation with our device parameters yields 97.8\% fidelity for the CZ gate, which is mainly limited by the short coherence time of the coupler qubit. This implies that the CAS-based CZ gate can be improved further by optimizing the design parameter and coherence time of the coupler.

\begin{figure}
    \centering
    \includegraphics[width=8.6cm]{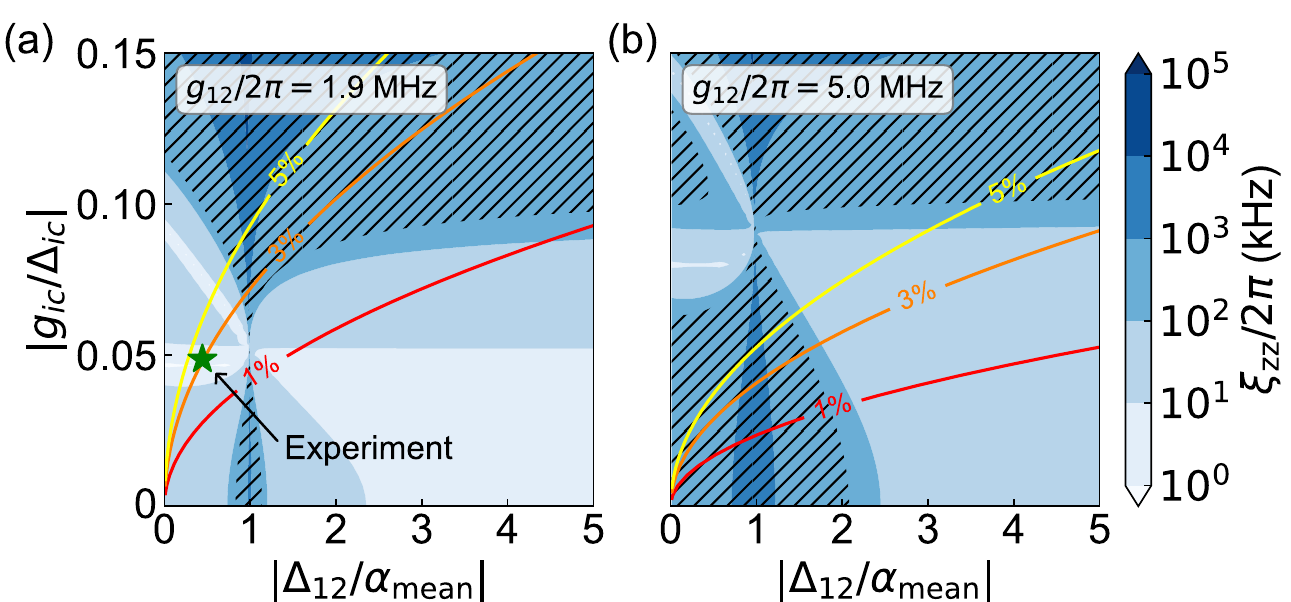}
    \caption{Residual ZZ interaction strength $\xi_\mathrm{ZZ}$ and the drive efficiency $\eta_b$ of the blue CAS transition as a function of the detuning $\Delta_{12}$ and transverse coupling strength $g_{ic}$ normalized by the mean anharmonicity $\alpha_\mathrm{mean} = (\alpha_1 + \alpha_2)/2$ and detuning $\Delta_{ic},\,(i\in \{1,2\})$, respectively. Here, $\xi_\mathrm{ZZ}$ is calculated through numerical diagonalization of Eq.\,\eqref{eq:H}~(filled contour plot) using (a)~the current and (b)~prospective design parameters with the direct transverse coupling $g_{12}$. The drive efficiency is defined as $\eta_b = \Omega_b/\Omega_d$ from Eq.\,\eqref{eq:gb} (contour line plot). As the prospective design parameters, we set $(\omega_c - \omega_1)/2\pi = 0.6$~GHz, $\omega_2/2\pi = 5.0$~GHz, and $\alpha_i/2\pi = (-0.20,\,-0.20,\,-0.45)$~GHz for $i = (1,2,c)$. The sweep parameters are $\omega_1$ and $g_{1c}/\Delta_{1c}=g_{2c}/\Delta_{2c}$, and the shaded areas indicate the residual ZZ interaction strength larger than 150~kHz. The green star in~(a) indicates the condition in the current experiment.}
    \label{fig:eta_sim}
\end{figure}

Finally, using Figs.\,\ref{fig:eta_sim}(a) and (b), we discuss dependencies of the residual ZZ interaction strength between the data qubits, $\xi_\mathrm{ZZ}$, and the drive efficiency of the blue CAS-based CZ gate rate, $\eta_b=\Omega_b/\Omega_d$, on the current and prospective design parameters. Here, we numerically diagonalize Eq.\,\eqref{eq:H} to calculate the residual ZZ-interaction strength when the coupler is in the ground state. Note that in these calculations, the term $g_\mathrm{12}(\hat{a}^\dagger_1\hat{a}_2 + \hat{a}_1\hat{a}^\dagger_2)$, which has been ignored so far, is added to Eq.\,\eqref{eq:H} to see the effect of direct coupling. As reported in previous studies~\cite{zz_cl_IBM_reso, zz_cl_Princeton}, the direct coupling $g_{12}$ can suppress the residual ZZ interaction by canceling the one mediated by the coupler. As shown in Fig.\,\ref{fig:eta_sim}(a), the straddling regime ($|\Delta_{12}/\alpha_\mathrm{mean}| < 1$) gives high drive efficiency and low residual ZZ interaction for the parameter set. On the other hand, we can also achieve practical performance far outside the straddling regime by selecting appropriate values of parameters, especially of $g_{12}$. In Fig.\,\ref{fig:eta_sim}(b), we set $g_{12}/2\pi=5\,$MHz as an example. This parameter set enables implementation of the blue CAS-based CZ gates of 100--200 ns for $\Omega_d /2\pi = 200$~MHz in a wide range of the detuning ($2 \lesssim \Delta_{12}/\alpha_\mathrm{mean} \lesssim 4$) between the data transmons while keeping the residual ZZ coupling $<$100~kHz. The coherent error due to the residual ZZ interaction during the non-commuting single-qubit gates can be mitigated with an optimal-control pulse~\cite{KHANEJA2005296, PRXQuantum.2.040324} or a composite pulse robust to frequency shift \cite{PhysRevA.67.042308, Cummins_2000}. We can also apply an active residual ZZ interaction cancellation using an off-resonant microwave drive near the blue CAS transition [See Supplemental Material] or coupler-qubit transition~\cite{ChinaZZcl}.

In conclusion, we have investigated and analytically modeled the four-wave-mixing interaction among three superconducting qubits under a microwave drive. By using the interaction, we demonstrated the coupler-assisted-swap-based control-Z gate between two fixed-frequency transmons mediated by a fixed-frequency transmon coupler. The drive efficiency of the gate has a practical value in a wide parameter range, providing an alternative solution to the frequency crowding problem and a new design paradigm for superconducting quantum processors. Moreover, a physically-efficient parity measurement could be realized by measuring the coupler after a pulse sequence of simultaneous $\pi$-pulses to the blue and red CAS transitions. An alternative pulse sequence of two $\pi$-pulses to the blue\,(red) CAS transition sandwiching $\pi$-pulses to the data qubits would also work.



\begin{acknowledgments}
\emph{Acknowledgement}---The authors acknowledge K. Kusuyama, K. Nittoh, and L. Szikszai for their support in the device fabrication, H. Terai and Y. Hishida  for providing the TiN films, H. Goto, M. Shigefuji, and K. Taniguchi for fruitful discussions, and Y. Sunada, K. Heya, and T. Miyamura for sharing the measurement codes. This work was partly supported by JST ERATO (Grant No.\,JPMJER1601), MEXT Q-LEAP (Grant No.\,JPMXS0118068682), and JSPS KAKENHI (Grant No.\,JP22J15257).

\emph{Data availability}---The experimental datasets and source codes used to generate the results of this manuscript are available in the [Zenodo], at [\url{https://doi.org/10.5281/zenodo.7885632}].
\end{acknowledgments}

\bibliographystyle{apsrev4-2}
\bibliography{uni.bib}

\begin{thebibliography}{54}%
\makeatletter
\providecommand \@ifxundefined [1]{%
 \@ifx{#1\undefined}
}%
\providecommand \@ifnum [1]{%
 \ifnum #1\expandafter \@firstoftwo
 \else \expandafter \@secondoftwo
 \fi
}%
\providecommand \@ifx [1]{%
 \ifx #1\expandafter \@firstoftwo
 \else \expandafter \@secondoftwo
 \fi
}%
\providecommand \natexlab [1]{#1}%
\providecommand \enquote  [1]{``#1''}%
\providecommand \bibnamefont  [1]{#1}%
\providecommand \bibfnamefont [1]{#1}%
\providecommand \citenamefont [1]{#1}%
\providecommand \href@noop [0]{\@secondoftwo}%
\providecommand \href [0]{\begingroup \@sanitize@url \@href}%
\providecommand \@href[1]{\@@startlink{#1}\@@href}%
\providecommand \@@href[1]{\endgroup#1\@@endlink}%
\providecommand \@sanitize@url [0]{\catcode `\\12\catcode `\$12\catcode
  `\&12\catcode `\#12\catcode `\^12\catcode `\_12\catcode `\%12\relax}%
\providecommand \@@startlink[1]{}%
\providecommand \@@endlink[0]{}%
\providecommand \url  [0]{\begingroup\@sanitize@url \@url }%
\providecommand \@url [1]{\endgroup\@href {#1}{\urlprefix }}%
\providecommand \urlprefix  [0]{URL }%
\providecommand \Eprint [0]{\href }%
\providecommand \doibase [0]{https://doi.org/}%
\providecommand \selectlanguage [0]{\@gobble}%
\providecommand \bibinfo  [0]{\@secondoftwo}%
\providecommand \bibfield  [0]{\@secondoftwo}%
\providecommand \translation [1]{[#1]}%
\providecommand \BibitemOpen [0]{}%
\providecommand \bibitemStop [0]{}%
\providecommand \bibitemNoStop [0]{.\EOS\space}%
\providecommand \EOS [0]{\spacefactor3000\relax}%
\providecommand \BibitemShut  [1]{\csname bibitem#1\endcsname}%
\let\auto@bib@innerbib\@empty
\bibitem [{\citenamefont {Kitaev}(2003)}]{KITAEV20032}%
  \BibitemOpen
  \bibfield  {author} {\bibinfo {author} {\bibfnamefont {A.}~\bibnamefont
  {Kitaev}},\ }\href
  {https://doi.org/https://doi.org/10.1016/S0003-4916(02)00018-0} {\bibfield
  {journal} {\bibinfo  {journal} {Annals of Physics}\ }\textbf {\bibinfo
  {volume} {303}},\ \bibinfo {pages} {2} (\bibinfo {year} {2003})}\BibitemShut
  {NoStop}%
\bibitem [{\citenamefont {Bravyi}\ and\ \citenamefont
  {Kitaev}(1998)}]{bravyi1998quantum}%
  \BibitemOpen
  \bibfield  {author} {\bibinfo {author} {\bibfnamefont {S.~B.}\ \bibnamefont
  {Bravyi}}\ and\ \bibinfo {author} {\bibfnamefont {A.~Y.}\ \bibnamefont
  {Kitaev}},\ }\href {https://arxiv.org/abs/quant-ph/9811052} {\bibfield
  {journal} {\bibinfo  {journal} {arXiv preprint quant-ph/9811052}\ } (\bibinfo
  {year} {1998})}\BibitemShut {NoStop}%
\bibitem [{\citenamefont {Ladd}\ \emph {et~al.}(2010)\citenamefont {Ladd},
  \citenamefont {Jelezko}, \citenamefont {Laflamme}, \citenamefont {Nakamura},
  \citenamefont {Monroe},\ and\ \citenamefont {O'Brien}}]{Ladd2010}%
  \BibitemOpen
  \bibfield  {author} {\bibinfo {author} {\bibfnamefont {T.~D.}\ \bibnamefont
  {Ladd}}, \bibinfo {author} {\bibfnamefont {F.}~\bibnamefont {Jelezko}},
  \bibinfo {author} {\bibfnamefont {R.}~\bibnamefont {Laflamme}}, \bibinfo
  {author} {\bibfnamefont {Y.}~\bibnamefont {Nakamura}}, \bibinfo {author}
  {\bibfnamefont {C.}~\bibnamefont {Monroe}},\ and\ \bibinfo {author}
  {\bibfnamefont {J.~L.}\ \bibnamefont {O'Brien}},\ }\href
  {https://doi.org/10.1038/nature08812} {\bibfield  {journal} {\bibinfo
  {journal} {Nature}\ }\textbf {\bibinfo {volume} {464}},\ \bibinfo {pages}
  {45} (\bibinfo {year} {2010})}\BibitemShut {NoStop}%
\bibitem [{\citenamefont {Kjaergaard}\ \emph {et~al.}(2020)\citenamefont
  {Kjaergaard}, \citenamefont {Schwartz}, \citenamefont {Braum\"{u}ller},
  \citenamefont {Krantz}, \citenamefont {Wang}, \citenamefont {Gustavsson},\
  and\ \citenamefont {Oliver}}]{Kjaergaard_review}%
  \BibitemOpen
  \bibfield  {author} {\bibinfo {author} {\bibfnamefont {M.}~\bibnamefont
  {Kjaergaard}}, \bibinfo {author} {\bibfnamefont {M.~E.}\ \bibnamefont
  {Schwartz}}, \bibinfo {author} {\bibfnamefont {J.}~\bibnamefont
  {Braum\"{u}ller}}, \bibinfo {author} {\bibfnamefont {P.}~\bibnamefont
  {Krantz}}, \bibinfo {author} {\bibfnamefont {J.~I.-J.}\ \bibnamefont {Wang}},
  \bibinfo {author} {\bibfnamefont {S.}~\bibnamefont {Gustavsson}},\ and\
  \bibinfo {author} {\bibfnamefont {W.~D.}\ \bibnamefont {Oliver}},\ }\href
  {https://doi.org/10.1146/annurev-conmatphys-031119-050605} {\bibfield
  {journal} {\bibinfo  {journal} {Annual Review of Condensed Matter Physics}\
  }\textbf {\bibinfo {volume} {11}},\ \bibinfo {pages} {369} (\bibinfo {year}
  {2020})}\BibitemShut {NoStop}%
\bibitem [{\citenamefont {Brown}\ \emph {et~al.}(2016)\citenamefont {Brown},
  \citenamefont {Kim},\ and\ \citenamefont {Monroe}}]{Brown2016}%
  \BibitemOpen
  \bibfield  {author} {\bibinfo {author} {\bibfnamefont {K.~R.}\ \bibnamefont
  {Brown}}, \bibinfo {author} {\bibfnamefont {J.}~\bibnamefont {Kim}},\ and\
  \bibinfo {author} {\bibfnamefont {C.}~\bibnamefont {Monroe}},\ }\href
  {https://doi.org/10.1038/npjqi.2016.34} {\bibfield  {journal} {\bibinfo
  {journal} {npj Quantum Information}\ }\textbf {\bibinfo {volume} {2}},\
  \bibinfo {pages} {16034} (\bibinfo {year} {2016})}\BibitemShut {NoStop}%
\bibitem [{\citenamefont {Saffman}(2016)}]{Saffman_2016}%
  \BibitemOpen
  \bibfield  {author} {\bibinfo {author} {\bibfnamefont {M.}~\bibnamefont
  {Saffman}},\ }\href {https://doi.org/10.1088/0953-4075/49/20/202001}
  {\bibfield  {journal} {\bibinfo  {journal} {Journal of Physics B: Atomic,
  Molecular and Optical Physics}\ }\textbf {\bibinfo {volume} {49}},\ \bibinfo
  {pages} {202001} (\bibinfo {year} {2016})}\BibitemShut {NoStop}%
\bibitem [{\citenamefont {Kloeffel}\ and\ \citenamefont
  {Loss}(2013)}]{kloeffel2013prospects}%
  \BibitemOpen
  \bibfield  {author} {\bibinfo {author} {\bibfnamefont {C.}~\bibnamefont
  {Kloeffel}}\ and\ \bibinfo {author} {\bibfnamefont {D.}~\bibnamefont
  {Loss}},\ }\href {https://doi.org/10.1146/annurev-conmatphys-030212-184248}
  {\bibfield  {journal} {\bibinfo  {journal} {Annual Review of Condensed Matter
  Physics}\ }\textbf {\bibinfo {volume} {4}},\ \bibinfo {pages} {51} (\bibinfo
  {year} {2013})}\BibitemShut {NoStop}%
\bibitem [{\citenamefont {Slussarenko}\ and\ \citenamefont
  {Pryde}(2019)}]{doi:10.1063/1.5115814}%
  \BibitemOpen
  \bibfield  {author} {\bibinfo {author} {\bibfnamefont {S.}~\bibnamefont
  {Slussarenko}}\ and\ \bibinfo {author} {\bibfnamefont {G.~J.}\ \bibnamefont
  {Pryde}},\ }\href {https://doi.org/10.1063/1.5115814} {\bibfield  {journal}
  {\bibinfo  {journal} {Applied Physics Reviews}\ }\textbf {\bibinfo {volume}
  {6}},\ \bibinfo {pages} {041303} (\bibinfo {year} {2019})}\BibitemShut
  {NoStop}%
\bibitem [{\citenamefont {Krantz}\ \emph {et~al.}(2019)\citenamefont {Krantz},
  \citenamefont {Kjaergaard}, \citenamefont {Yan}, \citenamefont {Orlando},
  \citenamefont {Gustavsson},\ and\ \citenamefont
  {Oliver}}]{doi:10.1063/1.5089550}%
  \BibitemOpen
  \bibfield  {author} {\bibinfo {author} {\bibfnamefont {P.}~\bibnamefont
  {Krantz}}, \bibinfo {author} {\bibfnamefont {M.}~\bibnamefont {Kjaergaard}},
  \bibinfo {author} {\bibfnamefont {F.}~\bibnamefont {Yan}}, \bibinfo {author}
  {\bibfnamefont {T.~P.}\ \bibnamefont {Orlando}}, \bibinfo {author}
  {\bibfnamefont {S.}~\bibnamefont {Gustavsson}},\ and\ \bibinfo {author}
  {\bibfnamefont {W.~D.}\ \bibnamefont {Oliver}},\ }\href
  {https://doi.org/10.1063/1.5089550} {\bibfield  {journal} {\bibinfo
  {journal} {Applied Physics Reviews}\ }\textbf {\bibinfo {volume} {6}},\
  \bibinfo {pages} {021318} (\bibinfo {year} {2019})}\BibitemShut {NoStop}%
\bibitem [{\citenamefont {Chen}\ \emph {et~al.}(2021)\citenamefont {Chen},
  \citenamefont {Satzinger}, \citenamefont {Atalaya}, \citenamefont {Korotkov},
  \citenamefont {Dunsworth}, \citenamefont {Sank}, \citenamefont {Quintana},
  \citenamefont {McEwen}, \citenamefont {Barends}, \citenamefont {Klimov},
  \citenamefont {Hong}, \citenamefont {Jones}, \citenamefont {Petukhov},
  \citenamefont {Kafri}, \citenamefont {Demura}, \citenamefont {Burkett},
  \citenamefont {Gidney}, \citenamefont {Fowler}, \citenamefont {Paler},
  \citenamefont {Putterman}, \citenamefont {Aleiner}, \citenamefont {Arute},
  \citenamefont {Arya}, \citenamefont {Babbush}, \citenamefont {Bardin},
  \citenamefont {Bengtsson}, \citenamefont {Bourassa}, \citenamefont
  {Broughton}, \citenamefont {Buckley}, \citenamefont {Buell}, \citenamefont
  {Bushnell}, \citenamefont {Chiaro}, \citenamefont {Collins}, \citenamefont
  {Courtney}, \citenamefont {Derk}, \citenamefont {Eppens}, \citenamefont
  {Erickson}, \citenamefont {Farhi}, \citenamefont {Foxen}, \citenamefont
  {Giustina}, \citenamefont {Greene}, \citenamefont {Gross}, \citenamefont
  {Harrigan}, \citenamefont {Harrington}, \citenamefont {Hilton}, \citenamefont
  {Ho}, \citenamefont {Huang}, \citenamefont {Huggins}, \citenamefont {Ioffe},
  \citenamefont {Isakov}, \citenamefont {Jeffrey}, \citenamefont {Jiang},
  \citenamefont {Kechedzhi}, \citenamefont {Kim}, \citenamefont {Kitaev},
  \citenamefont {Kostritsa}, \citenamefont {Landhuis}, \citenamefont {Laptev},
  \citenamefont {Lucero}, \citenamefont {Martin}, \citenamefont {McClean},
  \citenamefont {McCourt}, \citenamefont {Mi}, \citenamefont {Miao},
  \citenamefont {Mohseni}, \citenamefont {Montazeri}, \citenamefont
  {Mruczkiewicz}, \citenamefont {Mutus}, \citenamefont {Naaman}, \citenamefont
  {Neeley}, \citenamefont {Neill}, \citenamefont {Newman}, \citenamefont {Niu},
  \citenamefont {O'Brien}, \citenamefont {Opremcak}, \citenamefont {Ostby},
  \citenamefont {Pat{\'o}}, \citenamefont {Redd}, \citenamefont {Roushan},
  \citenamefont {Rubin}, \citenamefont {Shvarts}, \citenamefont {Strain},
  \citenamefont {Szalay}, \citenamefont {Trevithick}, \citenamefont
  {Villalonga}, \citenamefont {White}, \citenamefont {Yao}, \citenamefont
  {Yeh}, \citenamefont {Yoo}, \citenamefont {Zalcman}, \citenamefont {Neven},
  \citenamefont {Boixo}, \citenamefont {Smelyanskiy}, \citenamefont {Chen},
  \citenamefont {Megrant}, \citenamefont {Kelly},\ and\ \citenamefont
  {AI}}]{Chen2021}%
  \BibitemOpen
  \bibfield  {author} {\bibinfo {author} {\bibfnamefont {Z.}~\bibnamefont
  {Chen}}, \bibinfo {author} {\bibfnamefont {K.~J.}\ \bibnamefont {Satzinger}},
  \bibinfo {author} {\bibfnamefont {J.}~\bibnamefont {Atalaya}}, \bibinfo
  {author} {\bibfnamefont {A.~N.}\ \bibnamefont {Korotkov}}, \bibinfo {author}
  {\bibfnamefont {A.}~\bibnamefont {Dunsworth}}, \bibinfo {author}
  {\bibfnamefont {D.}~\bibnamefont {Sank}}, \bibinfo {author} {\bibfnamefont
  {C.}~\bibnamefont {Quintana}}, \bibinfo {author} {\bibfnamefont
  {M.}~\bibnamefont {McEwen}}, \bibinfo {author} {\bibfnamefont
  {R.}~\bibnamefont {Barends}}, \bibinfo {author} {\bibfnamefont {P.~V.}\
  \bibnamefont {Klimov}}, \bibinfo {author} {\bibfnamefont {S.}~\bibnamefont
  {Hong}}, \bibinfo {author} {\bibfnamefont {C.}~\bibnamefont {Jones}},
  \bibinfo {author} {\bibfnamefont {A.}~\bibnamefont {Petukhov}}, \bibinfo
  {author} {\bibfnamefont {D.}~\bibnamefont {Kafri}}, \bibinfo {author}
  {\bibfnamefont {S.}~\bibnamefont {Demura}}, \bibinfo {author} {\bibfnamefont
  {B.}~\bibnamefont {Burkett}}, \bibinfo {author} {\bibfnamefont
  {C.}~\bibnamefont {Gidney}}, \bibinfo {author} {\bibfnamefont {A.~G.}\
  \bibnamefont {Fowler}}, \bibinfo {author} {\bibfnamefont {A.}~\bibnamefont
  {Paler}}, \bibinfo {author} {\bibfnamefont {H.}~\bibnamefont {Putterman}},
  \bibinfo {author} {\bibfnamefont {I.}~\bibnamefont {Aleiner}}, \bibinfo
  {author} {\bibfnamefont {F.}~\bibnamefont {Arute}}, \bibinfo {author}
  {\bibfnamefont {K.}~\bibnamefont {Arya}}, \bibinfo {author} {\bibfnamefont
  {R.}~\bibnamefont {Babbush}}, \bibinfo {author} {\bibfnamefont {J.~C.}\
  \bibnamefont {Bardin}}, \bibinfo {author} {\bibfnamefont {A.}~\bibnamefont
  {Bengtsson}}, \bibinfo {author} {\bibfnamefont {A.}~\bibnamefont {Bourassa}},
  \bibinfo {author} {\bibfnamefont {M.}~\bibnamefont {Broughton}}, \bibinfo
  {author} {\bibfnamefont {B.~B.}\ \bibnamefont {Buckley}}, \bibinfo {author}
  {\bibfnamefont {D.~A.}\ \bibnamefont {Buell}}, \bibinfo {author}
  {\bibfnamefont {N.}~\bibnamefont {Bushnell}}, \bibinfo {author}
  {\bibfnamefont {B.}~\bibnamefont {Chiaro}}, \bibinfo {author} {\bibfnamefont
  {R.}~\bibnamefont {Collins}}, \bibinfo {author} {\bibfnamefont
  {W.}~\bibnamefont {Courtney}}, \bibinfo {author} {\bibfnamefont {A.~R.}\
  \bibnamefont {Derk}}, \bibinfo {author} {\bibfnamefont {D.}~\bibnamefont
  {Eppens}}, \bibinfo {author} {\bibfnamefont {C.}~\bibnamefont {Erickson}},
  \bibinfo {author} {\bibfnamefont {E.}~\bibnamefont {Farhi}}, \bibinfo
  {author} {\bibfnamefont {B.}~\bibnamefont {Foxen}}, \bibinfo {author}
  {\bibfnamefont {M.}~\bibnamefont {Giustina}}, \bibinfo {author}
  {\bibfnamefont {A.}~\bibnamefont {Greene}}, \bibinfo {author} {\bibfnamefont
  {J.~A.}\ \bibnamefont {Gross}}, \bibinfo {author} {\bibfnamefont {M.~P.}\
  \bibnamefont {Harrigan}}, \bibinfo {author} {\bibfnamefont {S.~D.}\
  \bibnamefont {Harrington}}, \bibinfo {author} {\bibfnamefont
  {J.}~\bibnamefont {Hilton}}, \bibinfo {author} {\bibfnamefont
  {A.}~\bibnamefont {Ho}}, \bibinfo {author} {\bibfnamefont {T.}~\bibnamefont
  {Huang}}, \bibinfo {author} {\bibfnamefont {W.~J.}\ \bibnamefont {Huggins}},
  \bibinfo {author} {\bibfnamefont {L.~B.}\ \bibnamefont {Ioffe}}, \bibinfo
  {author} {\bibfnamefont {S.~V.}\ \bibnamefont {Isakov}}, \bibinfo {author}
  {\bibfnamefont {E.}~\bibnamefont {Jeffrey}}, \bibinfo {author} {\bibfnamefont
  {Z.}~\bibnamefont {Jiang}}, \bibinfo {author} {\bibfnamefont
  {K.}~\bibnamefont {Kechedzhi}}, \bibinfo {author} {\bibfnamefont
  {S.}~\bibnamefont {Kim}}, \bibinfo {author} {\bibfnamefont {A.}~\bibnamefont
  {Kitaev}}, \bibinfo {author} {\bibfnamefont {F.}~\bibnamefont {Kostritsa}},
  \bibinfo {author} {\bibfnamefont {D.}~\bibnamefont {Landhuis}}, \bibinfo
  {author} {\bibfnamefont {P.}~\bibnamefont {Laptev}}, \bibinfo {author}
  {\bibfnamefont {E.}~\bibnamefont {Lucero}}, \bibinfo {author} {\bibfnamefont
  {O.}~\bibnamefont {Martin}}, \bibinfo {author} {\bibfnamefont {J.~R.}\
  \bibnamefont {McClean}}, \bibinfo {author} {\bibfnamefont {T.}~\bibnamefont
  {McCourt}}, \bibinfo {author} {\bibfnamefont {X.}~\bibnamefont {Mi}},
  \bibinfo {author} {\bibfnamefont {K.~C.}\ \bibnamefont {Miao}}, \bibinfo
  {author} {\bibfnamefont {M.}~\bibnamefont {Mohseni}}, \bibinfo {author}
  {\bibfnamefont {S.}~\bibnamefont {Montazeri}}, \bibinfo {author}
  {\bibfnamefont {W.}~\bibnamefont {Mruczkiewicz}}, \bibinfo {author}
  {\bibfnamefont {J.}~\bibnamefont {Mutus}}, \bibinfo {author} {\bibfnamefont
  {O.}~\bibnamefont {Naaman}}, \bibinfo {author} {\bibfnamefont
  {M.}~\bibnamefont {Neeley}}, \bibinfo {author} {\bibfnamefont
  {C.}~\bibnamefont {Neill}}, \bibinfo {author} {\bibfnamefont
  {M.}~\bibnamefont {Newman}}, \bibinfo {author} {\bibfnamefont {M.~Y.}\
  \bibnamefont {Niu}}, \bibinfo {author} {\bibfnamefont {T.~E.}\ \bibnamefont
  {O'Brien}}, \bibinfo {author} {\bibfnamefont {A.}~\bibnamefont {Opremcak}},
  \bibinfo {author} {\bibfnamefont {E.}~\bibnamefont {Ostby}}, \bibinfo
  {author} {\bibfnamefont {B.}~\bibnamefont {Pat{\'o}}}, \bibinfo {author}
  {\bibfnamefont {N.}~\bibnamefont {Redd}}, \bibinfo {author} {\bibfnamefont
  {P.}~\bibnamefont {Roushan}}, \bibinfo {author} {\bibfnamefont {N.~C.}\
  \bibnamefont {Rubin}}, \bibinfo {author} {\bibfnamefont {V.}~\bibnamefont
  {Shvarts}}, \bibinfo {author} {\bibfnamefont {D.}~\bibnamefont {Strain}},
  \bibinfo {author} {\bibfnamefont {M.}~\bibnamefont {Szalay}}, \bibinfo
  {author} {\bibfnamefont {M.~D.}\ \bibnamefont {Trevithick}}, \bibinfo
  {author} {\bibfnamefont {B.}~\bibnamefont {Villalonga}}, \bibinfo {author}
  {\bibfnamefont {T.}~\bibnamefont {White}}, \bibinfo {author} {\bibfnamefont
  {Z.~J.}\ \bibnamefont {Yao}}, \bibinfo {author} {\bibfnamefont
  {P.}~\bibnamefont {Yeh}}, \bibinfo {author} {\bibfnamefont {J.}~\bibnamefont
  {Yoo}}, \bibinfo {author} {\bibfnamefont {A.}~\bibnamefont {Zalcman}},
  \bibinfo {author} {\bibfnamefont {H.}~\bibnamefont {Neven}}, \bibinfo
  {author} {\bibfnamefont {S.}~\bibnamefont {Boixo}}, \bibinfo {author}
  {\bibfnamefont {V.}~\bibnamefont {Smelyanskiy}}, \bibinfo {author}
  {\bibfnamefont {Y.}~\bibnamefont {Chen}}, \bibinfo {author} {\bibfnamefont
  {A.}~\bibnamefont {Megrant}}, \bibinfo {author} {\bibfnamefont
  {J.}~\bibnamefont {Kelly}},\ and\ \bibinfo {author} {\bibfnamefont {G.~Q.}\
  \bibnamefont {AI}},\ }\href {https://doi.org/10.1038/s41586-021-03588-y}
  {\bibfield  {journal} {\bibinfo  {journal} {Nature}\ }\textbf {\bibinfo
  {volume} {595}},\ \bibinfo {pages} {383} (\bibinfo {year}
  {2021})}\BibitemShut {NoStop}%
\bibitem [{\citenamefont {Koch}\ \emph {et~al.}(2007)\citenamefont {Koch},
  \citenamefont {Yu}, \citenamefont {Gambetta}, \citenamefont {Houck},
  \citenamefont {Schuster}, \citenamefont {Majer}, \citenamefont {Blais},
  \citenamefont {Devoret}, \citenamefont {Girvin},\ and\ \citenamefont
  {Schoelkopf}}]{Koch2007}%
  \BibitemOpen
  \bibfield  {author} {\bibinfo {author} {\bibfnamefont {J.}~\bibnamefont
  {Koch}}, \bibinfo {author} {\bibfnamefont {T.~M.}\ \bibnamefont {Yu}},
  \bibinfo {author} {\bibfnamefont {J.}~\bibnamefont {Gambetta}}, \bibinfo
  {author} {\bibfnamefont {A.~A.}\ \bibnamefont {Houck}}, \bibinfo {author}
  {\bibfnamefont {D.~I.}\ \bibnamefont {Schuster}}, \bibinfo {author}
  {\bibfnamefont {J.}~\bibnamefont {Majer}}, \bibinfo {author} {\bibfnamefont
  {A.}~\bibnamefont {Blais}}, \bibinfo {author} {\bibfnamefont {M.~H.}\
  \bibnamefont {Devoret}}, \bibinfo {author} {\bibfnamefont {S.~M.}\
  \bibnamefont {Girvin}},\ and\ \bibinfo {author} {\bibfnamefont {R.~J.}\
  \bibnamefont {Schoelkopf}},\ }\href
  {https://doi.org/10.1103/PhysRevA.76.042319} {\bibfield  {journal} {\bibinfo
  {journal} {Phys. Rev. A}\ }\textbf {\bibinfo {volume} {76}},\ \bibinfo
  {pages} {042319} (\bibinfo {year} {2007})}\BibitemShut {NoStop}%
\bibitem [{\citenamefont {Rigetti}\ and\ \citenamefont
  {Devoret}(2010)}]{CR_2010Rigetti}%
  \BibitemOpen
  \bibfield  {author} {\bibinfo {author} {\bibfnamefont {C.}~\bibnamefont
  {Rigetti}}\ and\ \bibinfo {author} {\bibfnamefont {M.}~\bibnamefont
  {Devoret}},\ }\href {https://doi.org/10.1103/PhysRevB.81.134507} {\bibfield
  {journal} {\bibinfo  {journal} {Phys. Rev. B}\ }\textbf {\bibinfo {volume}
  {81}},\ \bibinfo {pages} {134507} (\bibinfo {year} {2010})}\BibitemShut
  {NoStop}%
\bibitem [{\citenamefont {Poletto}\ \emph {et~al.}(2012)\citenamefont
  {Poletto}, \citenamefont {Gambetta}, \citenamefont {Merkel}, \citenamefont
  {Smolin}, \citenamefont {Chow}, \citenamefont {C\'orcoles}, \citenamefont
  {Keefe}, \citenamefont {Rothwell}, \citenamefont {Rozen}, \citenamefont
  {Abraham}, \citenamefont {Rigetti},\ and\ \citenamefont
  {Steffen}}]{bSWAP_2012Poletto}%
  \BibitemOpen
  \bibfield  {author} {\bibinfo {author} {\bibfnamefont {S.}~\bibnamefont
  {Poletto}}, \bibinfo {author} {\bibfnamefont {J.~M.}\ \bibnamefont
  {Gambetta}}, \bibinfo {author} {\bibfnamefont {S.~T.}\ \bibnamefont
  {Merkel}}, \bibinfo {author} {\bibfnamefont {J.~A.}\ \bibnamefont {Smolin}},
  \bibinfo {author} {\bibfnamefont {J.~M.}\ \bibnamefont {Chow}}, \bibinfo
  {author} {\bibfnamefont {A.~D.}\ \bibnamefont {C\'orcoles}}, \bibinfo
  {author} {\bibfnamefont {G.~A.}\ \bibnamefont {Keefe}}, \bibinfo {author}
  {\bibfnamefont {M.~B.}\ \bibnamefont {Rothwell}}, \bibinfo {author}
  {\bibfnamefont {J.~R.}\ \bibnamefont {Rozen}}, \bibinfo {author}
  {\bibfnamefont {D.~W.}\ \bibnamefont {Abraham}}, \bibinfo {author}
  {\bibfnamefont {C.}~\bibnamefont {Rigetti}},\ and\ \bibinfo {author}
  {\bibfnamefont {M.}~\bibnamefont {Steffen}},\ }\href
  {https://doi.org/10.1103/PhysRevLett.109.240505} {\bibfield  {journal}
  {\bibinfo  {journal} {Phys. Rev. Lett.}\ }\textbf {\bibinfo {volume} {109}},\
  \bibinfo {pages} {240505} (\bibinfo {year} {2012})}\BibitemShut {NoStop}%
\bibitem [{\citenamefont {Chow}\ \emph {et~al.}(2013)\citenamefont {Chow},
  \citenamefont {Gambetta}, \citenamefont {Cross}, \citenamefont {Merkel},
  \citenamefont {Rigetti},\ and\ \citenamefont {Steffen}}]{MAP_2013Chow}%
  \BibitemOpen
  \bibfield  {author} {\bibinfo {author} {\bibfnamefont {J.~M.}\ \bibnamefont
  {Chow}}, \bibinfo {author} {\bibfnamefont {J.~M.}\ \bibnamefont {Gambetta}},
  \bibinfo {author} {\bibfnamefont {A.~W.}\ \bibnamefont {Cross}}, \bibinfo
  {author} {\bibfnamefont {S.~T.}\ \bibnamefont {Merkel}}, \bibinfo {author}
  {\bibfnamefont {C.}~\bibnamefont {Rigetti}},\ and\ \bibinfo {author}
  {\bibfnamefont {M.}~\bibnamefont {Steffen}},\ }\href
  {https://doi.org/10.1088/1367-2630/15/11/115012} {\bibfield  {journal}
  {\bibinfo  {journal} {New Journal of Physics}\ }\textbf {\bibinfo {volume}
  {15}},\ \bibinfo {pages} {115012} (\bibinfo {year} {2013})}\BibitemShut
  {NoStop}%
\bibitem [{\citenamefont {Paik}\ \emph {et~al.}(2016)\citenamefont {Paik},
  \citenamefont {Mezzacapo}, \citenamefont {Sandberg}, \citenamefont {McClure},
  \citenamefont {Abdo}, \citenamefont {C\'orcoles}, \citenamefont {Dial},
  \citenamefont {Bogorin}, \citenamefont {Plourde}, \citenamefont {Steffen},
  \citenamefont {Cross}, \citenamefont {Gambetta},\ and\ \citenamefont
  {Chow}}]{RIP_2016Paik}%
  \BibitemOpen
  \bibfield  {author} {\bibinfo {author} {\bibfnamefont {H.}~\bibnamefont
  {Paik}}, \bibinfo {author} {\bibfnamefont {A.}~\bibnamefont {Mezzacapo}},
  \bibinfo {author} {\bibfnamefont {M.}~\bibnamefont {Sandberg}}, \bibinfo
  {author} {\bibfnamefont {D.~T.}\ \bibnamefont {McClure}}, \bibinfo {author}
  {\bibfnamefont {B.}~\bibnamefont {Abdo}}, \bibinfo {author} {\bibfnamefont
  {A.~D.}\ \bibnamefont {C\'orcoles}}, \bibinfo {author} {\bibfnamefont
  {O.}~\bibnamefont {Dial}}, \bibinfo {author} {\bibfnamefont {D.~F.}\
  \bibnamefont {Bogorin}}, \bibinfo {author} {\bibfnamefont {B.~L.~T.}\
  \bibnamefont {Plourde}}, \bibinfo {author} {\bibfnamefont {M.}~\bibnamefont
  {Steffen}}, \bibinfo {author} {\bibfnamefont {A.~W.}\ \bibnamefont {Cross}},
  \bibinfo {author} {\bibfnamefont {J.~M.}\ \bibnamefont {Gambetta}},\ and\
  \bibinfo {author} {\bibfnamefont {J.~M.}\ \bibnamefont {Chow}},\ }\href
  {https://doi.org/10.1103/PhysRevLett.117.250502} {\bibfield  {journal}
  {\bibinfo  {journal} {Phys. Rev. Lett.}\ }\textbf {\bibinfo {volume} {117}},\
  \bibinfo {pages} {250502} (\bibinfo {year} {2016})}\BibitemShut {NoStop}%
\bibitem [{\citenamefont {Noguchi}\ \emph {et~al.}(2020)\citenamefont
  {Noguchi}, \citenamefont {Osada}, \citenamefont {Masuda}, \citenamefont
  {Kono}, \citenamefont {Heya}, \citenamefont {Wolski}, \citenamefont
  {Takahashi}, \citenamefont {Sugiyama}, \citenamefont {Lachance-Quirion},\
  and\ \citenamefont {Nakamura}}]{NogFastCT}%
  \BibitemOpen
  \bibfield  {author} {\bibinfo {author} {\bibfnamefont {A.}~\bibnamefont
  {Noguchi}}, \bibinfo {author} {\bibfnamefont {A.}~\bibnamefont {Osada}},
  \bibinfo {author} {\bibfnamefont {S.}~\bibnamefont {Masuda}}, \bibinfo
  {author} {\bibfnamefont {S.}~\bibnamefont {Kono}}, \bibinfo {author}
  {\bibfnamefont {K.}~\bibnamefont {Heya}}, \bibinfo {author} {\bibfnamefont
  {S.~P.}\ \bibnamefont {Wolski}}, \bibinfo {author} {\bibfnamefont
  {H.}~\bibnamefont {Takahashi}}, \bibinfo {author} {\bibfnamefont
  {T.}~\bibnamefont {Sugiyama}}, \bibinfo {author} {\bibfnamefont
  {D.}~\bibnamefont {Lachance-Quirion}},\ and\ \bibinfo {author} {\bibfnamefont
  {Y.}~\bibnamefont {Nakamura}},\ }\href
  {https://doi.org/10.1103/PhysRevA.102.062408} {\bibfield  {journal} {\bibinfo
   {journal} {Phys. Rev. A}\ }\textbf {\bibinfo {volume} {102}},\ \bibinfo
  {pages} {062408} (\bibinfo {year} {2020})}\BibitemShut {NoStop}%
\bibitem [{\citenamefont {Krinner}\ \emph {et~al.}(2020)\citenamefont
  {Krinner}, \citenamefont {Kurpiers}, \citenamefont {Royer}, \citenamefont
  {Magnard}, \citenamefont {Tsitsilin}, \citenamefont {Besse}, \citenamefont
  {Remm}, \citenamefont {Blais},\ and\ \citenamefont
  {Wallraff}}]{FogiCZ_2020krinner}%
  \BibitemOpen
  \bibfield  {author} {\bibinfo {author} {\bibfnamefont {S.}~\bibnamefont
  {Krinner}}, \bibinfo {author} {\bibfnamefont {P.}~\bibnamefont {Kurpiers}},
  \bibinfo {author} {\bibfnamefont {B.}~\bibnamefont {Royer}}, \bibinfo
  {author} {\bibfnamefont {P.}~\bibnamefont {Magnard}}, \bibinfo {author}
  {\bibfnamefont {I.}~\bibnamefont {Tsitsilin}}, \bibinfo {author}
  {\bibfnamefont {J.-C.}\ \bibnamefont {Besse}}, \bibinfo {author}
  {\bibfnamefont {A.}~\bibnamefont {Remm}}, \bibinfo {author} {\bibfnamefont
  {A.}~\bibnamefont {Blais}},\ and\ \bibinfo {author} {\bibfnamefont
  {A.}~\bibnamefont {Wallraff}},\ }\href
  {https://doi.org/10.1103/PhysRevApplied.14.044039} {\bibfield  {journal}
  {\bibinfo  {journal} {Phys. Rev. Applied}\ }\textbf {\bibinfo {volume}
  {14}},\ \bibinfo {pages} {044039} (\bibinfo {year} {2020})}\BibitemShut
  {NoStop}%
\bibitem [{\citenamefont {Mitchell}\ \emph {et~al.}(2021)\citenamefont
  {Mitchell}, \citenamefont {Naik}, \citenamefont {Morvan}, \citenamefont
  {Hashim}, \citenamefont {Kreikebaum}, \citenamefont {Marinelli},
  \citenamefont {Lavrijsen}, \citenamefont {Nowrouzi}, \citenamefont
  {Santiago},\ and\ \citenamefont {Siddiqi}}]{siZZle_2021Mitchell}%
  \BibitemOpen
  \bibfield  {author} {\bibinfo {author} {\bibfnamefont {B.~K.}\ \bibnamefont
  {Mitchell}}, \bibinfo {author} {\bibfnamefont {R.~K.}\ \bibnamefont {Naik}},
  \bibinfo {author} {\bibfnamefont {A.}~\bibnamefont {Morvan}}, \bibinfo
  {author} {\bibfnamefont {A.}~\bibnamefont {Hashim}}, \bibinfo {author}
  {\bibfnamefont {J.~M.}\ \bibnamefont {Kreikebaum}}, \bibinfo {author}
  {\bibfnamefont {B.}~\bibnamefont {Marinelli}}, \bibinfo {author}
  {\bibfnamefont {W.}~\bibnamefont {Lavrijsen}}, \bibinfo {author}
  {\bibfnamefont {K.}~\bibnamefont {Nowrouzi}}, \bibinfo {author}
  {\bibfnamefont {D.~I.}\ \bibnamefont {Santiago}},\ and\ \bibinfo {author}
  {\bibfnamefont {I.}~\bibnamefont {Siddiqi}},\ }\href
  {https://doi.org/10.1103/PhysRevLett.127.200502} {\bibfield  {journal}
  {\bibinfo  {journal} {Phys. Rev. Lett.}\ }\textbf {\bibinfo {volume} {127}},\
  \bibinfo {pages} {200502} (\bibinfo {year} {2021})}\BibitemShut {NoStop}%
\bibitem [{\citenamefont {Heya}\ and\ \citenamefont
  {Kanazawa}(2021)}]{PRXQuantum.2.040336}%
  \BibitemOpen
  \bibfield  {author} {\bibinfo {author} {\bibfnamefont {K.}~\bibnamefont
  {Heya}}\ and\ \bibinfo {author} {\bibfnamefont {N.}~\bibnamefont
  {Kanazawa}},\ }\href {https://doi.org/10.1103/PRXQuantum.2.040336} {\bibfield
   {journal} {\bibinfo  {journal} {PRX Quantum}\ }\textbf {\bibinfo {volume}
  {2}},\ \bibinfo {pages} {040336} (\bibinfo {year} {2021})}\BibitemShut
  {NoStop}%
\bibitem [{\citenamefont {Chow}\ \emph {et~al.}(2011)\citenamefont {Chow},
  \citenamefont {C\'orcoles}, \citenamefont {Gambetta}, \citenamefont
  {Rigetti}, \citenamefont {Johnson}, \citenamefont {Smolin}, \citenamefont
  {Rozen}, \citenamefont {Keefe}, \citenamefont {Rothwell}, \citenamefont
  {Ketchen},\ and\ \citenamefont {Steffen}}]{CR_1}%
  \BibitemOpen
  \bibfield  {author} {\bibinfo {author} {\bibfnamefont {J.~M.}\ \bibnamefont
  {Chow}}, \bibinfo {author} {\bibfnamefont {A.~D.}\ \bibnamefont
  {C\'orcoles}}, \bibinfo {author} {\bibfnamefont {J.~M.}\ \bibnamefont
  {Gambetta}}, \bibinfo {author} {\bibfnamefont {C.}~\bibnamefont {Rigetti}},
  \bibinfo {author} {\bibfnamefont {B.~R.}\ \bibnamefont {Johnson}}, \bibinfo
  {author} {\bibfnamefont {J.~A.}\ \bibnamefont {Smolin}}, \bibinfo {author}
  {\bibfnamefont {J.~R.}\ \bibnamefont {Rozen}}, \bibinfo {author}
  {\bibfnamefont {G.~A.}\ \bibnamefont {Keefe}}, \bibinfo {author}
  {\bibfnamefont {M.~B.}\ \bibnamefont {Rothwell}}, \bibinfo {author}
  {\bibfnamefont {M.~B.}\ \bibnamefont {Ketchen}},\ and\ \bibinfo {author}
  {\bibfnamefont {M.}~\bibnamefont {Steffen}},\ }\href
  {https://doi.org/10.1103/PhysRevLett.107.080502} {\bibfield  {journal}
  {\bibinfo  {journal} {Phys. Rev. Lett.}\ }\textbf {\bibinfo {volume} {107}},\
  \bibinfo {pages} {080502} (\bibinfo {year} {2011})}\BibitemShut {NoStop}%
\bibitem [{\citenamefont {Sheldon}\ \emph {et~al.}(2016)\citenamefont
  {Sheldon}, \citenamefont {Magesan}, \citenamefont {Chow},\ and\ \citenamefont
  {Gambetta}}]{CR_procedure}%
  \BibitemOpen
  \bibfield  {author} {\bibinfo {author} {\bibfnamefont {S.}~\bibnamefont
  {Sheldon}}, \bibinfo {author} {\bibfnamefont {E.}~\bibnamefont {Magesan}},
  \bibinfo {author} {\bibfnamefont {J.~M.}\ \bibnamefont {Chow}},\ and\
  \bibinfo {author} {\bibfnamefont {J.~M.}\ \bibnamefont {Gambetta}},\ }\href
  {https://doi.org/10.1103/PhysRevA.93.060302} {\bibfield  {journal} {\bibinfo
  {journal} {Phys. Rev. A}\ }\textbf {\bibinfo {volume} {93}},\ \bibinfo
  {pages} {060302(R)} (\bibinfo {year} {2016})}\BibitemShut {NoStop}%
\bibitem [{\citenamefont {Kandala}\ \emph {et~al.}(2021)\citenamefont
  {Kandala}, \citenamefont {Wei}, \citenamefont {Srinivasan}, \citenamefont
  {Magesan}, \citenamefont {Carnevale}, \citenamefont {Keefe}, \citenamefont
  {Klaus}, \citenamefont {Dial},\ and\ \citenamefont {McKay}}]{zz_cl_IBM_reso}%
  \BibitemOpen
  \bibfield  {author} {\bibinfo {author} {\bibfnamefont {A.}~\bibnamefont
  {Kandala}}, \bibinfo {author} {\bibfnamefont {K.~X.}\ \bibnamefont {Wei}},
  \bibinfo {author} {\bibfnamefont {S.}~\bibnamefont {Srinivasan}}, \bibinfo
  {author} {\bibfnamefont {E.}~\bibnamefont {Magesan}}, \bibinfo {author}
  {\bibfnamefont {S.}~\bibnamefont {Carnevale}}, \bibinfo {author}
  {\bibfnamefont {G.~A.}\ \bibnamefont {Keefe}}, \bibinfo {author}
  {\bibfnamefont {D.}~\bibnamefont {Klaus}}, \bibinfo {author} {\bibfnamefont
  {O.}~\bibnamefont {Dial}},\ and\ \bibinfo {author} {\bibfnamefont {D.~C.}\
  \bibnamefont {McKay}},\ }\href
  {https://doi.org/10.1103/PhysRevLett.127.130501} {\bibfield  {journal}
  {\bibinfo  {journal} {Phys. Rev. Lett.}\ }\textbf {\bibinfo {volume} {127}},\
  \bibinfo {pages} {130501} (\bibinfo {year} {2021})}\BibitemShut {NoStop}%
\bibitem [{\citenamefont {Malekakhlagh}\ \emph {et~al.}(2020)\citenamefont
  {Malekakhlagh}, \citenamefont {Magesan},\ and\ \citenamefont
  {McKay}}]{FirstPrinciples}%
  \BibitemOpen
  \bibfield  {author} {\bibinfo {author} {\bibfnamefont {M.}~\bibnamefont
  {Malekakhlagh}}, \bibinfo {author} {\bibfnamefont {E.}~\bibnamefont
  {Magesan}},\ and\ \bibinfo {author} {\bibfnamefont {D.~C.}\ \bibnamefont
  {McKay}},\ }\href {https://doi.org/10.1103/PhysRevA.102.042605} {\bibfield
  {journal} {\bibinfo  {journal} {Phys. Rev. A}\ }\textbf {\bibinfo {volume}
  {102}},\ \bibinfo {pages} {042605} (\bibinfo {year} {2020})}\BibitemShut
  {NoStop}%
\bibitem [{\citenamefont {Morvan}\ \emph {et~al.}(2022)\citenamefont {Morvan},
  \citenamefont {Chen}, \citenamefont {Larson}, \citenamefont {Santiago},\ and\
  \citenamefont {Siddiqi}}]{opt_f_crowding}%
  \BibitemOpen
  \bibfield  {author} {\bibinfo {author} {\bibfnamefont {A.}~\bibnamefont
  {Morvan}}, \bibinfo {author} {\bibfnamefont {L.}~\bibnamefont {Chen}},
  \bibinfo {author} {\bibfnamefont {J.~M.}\ \bibnamefont {Larson}}, \bibinfo
  {author} {\bibfnamefont {D.~I.}\ \bibnamefont {Santiago}},\ and\ \bibinfo
  {author} {\bibfnamefont {I.}~\bibnamefont {Siddiqi}},\ }\href
  {https://doi.org/10.1103/PhysRevResearch.4.023079} {\bibfield  {journal}
  {\bibinfo  {journal} {Phys. Rev. Research}\ }\textbf {\bibinfo {volume}
  {4}},\ \bibinfo {pages} {023079} (\bibinfo {year} {2022})}\BibitemShut
  {NoStop}%
\bibitem [{\citenamefont {Granata}\ \emph {et~al.}(2008)\citenamefont
  {Granata}, \citenamefont {Vettoliere}, \citenamefont {Petti}, \citenamefont
  {Rippa}, \citenamefont {Ruggiero}, \citenamefont {Mormile},\ and\
  \citenamefont {Russo}}]{granata2008trimming}%
  \BibitemOpen
  \bibfield  {author} {\bibinfo {author} {\bibfnamefont {C.}~\bibnamefont
  {Granata}}, \bibinfo {author} {\bibfnamefont {A.}~\bibnamefont {Vettoliere}},
  \bibinfo {author} {\bibfnamefont {L.}~\bibnamefont {Petti}}, \bibinfo
  {author} {\bibfnamefont {M.}~\bibnamefont {Rippa}}, \bibinfo {author}
  {\bibfnamefont {B.}~\bibnamefont {Ruggiero}}, \bibinfo {author}
  {\bibfnamefont {P.}~\bibnamefont {Mormile}},\ and\ \bibinfo {author}
  {\bibfnamefont {M.}~\bibnamefont {Russo}},\ }in\ \href@noop {} {\emph
  {\bibinfo {booktitle} {Journal of Physics: Conference Series}}},\
  Vol.~\bibinfo {volume} {97}\ (\bibinfo {organization} {IOP Publishing},\
  \bibinfo {year} {2008})\ p.\ \bibinfo {pages} {012110}\BibitemShut {NoStop}%
\bibitem [{\citenamefont {Hertzberg}\ \emph {et~al.}(2021)\citenamefont
  {Hertzberg}, \citenamefont {Zhang}, \citenamefont {Rosenblatt}, \citenamefont
  {Magesan}, \citenamefont {Smolin}, \citenamefont {Yau}, \citenamefont
  {Adiga}, \citenamefont {Sandberg}, \citenamefont {Brink}, \citenamefont
  {Chow},\ and\ \citenamefont {Orcutt}}]{Hertzberg2021}%
  \BibitemOpen
  \bibfield  {author} {\bibinfo {author} {\bibfnamefont {J.~B.}\ \bibnamefont
  {Hertzberg}}, \bibinfo {author} {\bibfnamefont {E.~J.}\ \bibnamefont
  {Zhang}}, \bibinfo {author} {\bibfnamefont {S.}~\bibnamefont {Rosenblatt}},
  \bibinfo {author} {\bibfnamefont {E.}~\bibnamefont {Magesan}}, \bibinfo
  {author} {\bibfnamefont {J.~A.}\ \bibnamefont {Smolin}}, \bibinfo {author}
  {\bibfnamefont {J.-B.}\ \bibnamefont {Yau}}, \bibinfo {author} {\bibfnamefont
  {V.~P.}\ \bibnamefont {Adiga}}, \bibinfo {author} {\bibfnamefont
  {M.}~\bibnamefont {Sandberg}}, \bibinfo {author} {\bibfnamefont
  {M.}~\bibnamefont {Brink}}, \bibinfo {author} {\bibfnamefont {J.~M.}\
  \bibnamefont {Chow}},\ and\ \bibinfo {author} {\bibfnamefont {J.~S.}\
  \bibnamefont {Orcutt}},\ }\href {https://doi.org/10.1038/s41534-021-00464-5}
  {\bibfield  {journal} {\bibinfo  {journal} {npj Quantum Information}\
  }\textbf {\bibinfo {volume} {7}},\ \bibinfo {pages} {129} (\bibinfo {year}
  {2021})}\BibitemShut {NoStop}%
\bibitem [{\citenamefont {Zhang}\ \emph {et~al.}(2022)\citenamefont {Zhang},
  \citenamefont {Srinivasan}, \citenamefont {Sundaresan}, \citenamefont
  {Bogorin}, \citenamefont {Martin}, \citenamefont {Hertzberg}, \citenamefont
  {Timmerwilke}, \citenamefont {Pritchett}, \citenamefont {Yau}, \citenamefont
  {Wang}, \citenamefont {Landers}, \citenamefont {Lewandowski}, \citenamefont
  {Narasgond}, \citenamefont {Rosenblatt}, \citenamefont {Keefe}, \citenamefont
  {Lauer}, \citenamefont {Rothwell}, \citenamefont {McClure}, \citenamefont
  {Dial}, \citenamefont {Orcutt}, \citenamefont {Brink},\ and\ \citenamefont
  {Chow}}]{doi:10.1126/sciadv.abi6690}%
  \BibitemOpen
  \bibfield  {author} {\bibinfo {author} {\bibfnamefont {E.~J.}\ \bibnamefont
  {Zhang}}, \bibinfo {author} {\bibfnamefont {S.}~\bibnamefont {Srinivasan}},
  \bibinfo {author} {\bibfnamefont {N.}~\bibnamefont {Sundaresan}}, \bibinfo
  {author} {\bibfnamefont {D.~F.}\ \bibnamefont {Bogorin}}, \bibinfo {author}
  {\bibfnamefont {Y.}~\bibnamefont {Martin}}, \bibinfo {author} {\bibfnamefont
  {J.~B.}\ \bibnamefont {Hertzberg}}, \bibinfo {author} {\bibfnamefont
  {J.}~\bibnamefont {Timmerwilke}}, \bibinfo {author} {\bibfnamefont {E.~J.}\
  \bibnamefont {Pritchett}}, \bibinfo {author} {\bibfnamefont {J.-B.}\
  \bibnamefont {Yau}}, \bibinfo {author} {\bibfnamefont {C.}~\bibnamefont
  {Wang}}, \bibinfo {author} {\bibfnamefont {W.}~\bibnamefont {Landers}},
  \bibinfo {author} {\bibfnamefont {E.~P.}\ \bibnamefont {Lewandowski}},
  \bibinfo {author} {\bibfnamefont {A.}~\bibnamefont {Narasgond}}, \bibinfo
  {author} {\bibfnamefont {S.}~\bibnamefont {Rosenblatt}}, \bibinfo {author}
  {\bibfnamefont {G.~A.}\ \bibnamefont {Keefe}}, \bibinfo {author}
  {\bibfnamefont {I.}~\bibnamefont {Lauer}}, \bibinfo {author} {\bibfnamefont
  {M.~B.}\ \bibnamefont {Rothwell}}, \bibinfo {author} {\bibfnamefont {D.~T.}\
  \bibnamefont {McClure}}, \bibinfo {author} {\bibfnamefont {O.~E.}\
  \bibnamefont {Dial}}, \bibinfo {author} {\bibfnamefont {J.~S.}\ \bibnamefont
  {Orcutt}}, \bibinfo {author} {\bibfnamefont {M.}~\bibnamefont {Brink}},\ and\
  \bibinfo {author} {\bibfnamefont {J.~M.}\ \bibnamefont {Chow}},\ }\href
  {https://doi.org/10.1126/sciadv.abi6690} {\bibfield  {journal} {\bibinfo
  {journal} {Science Advances}\ }\textbf {\bibinfo {volume} {8}},\ \bibinfo
  {pages} {eabi6690} (\bibinfo {year} {2022})}\BibitemShut {NoStop}%
\bibitem [{\citenamefont {Kim}\ \emph {et~al.}(2022)\citenamefont {Kim},
  \citenamefont {Jünger}, \citenamefont {Morvan}, \citenamefont {Barnard},
  \citenamefont {Livingston}, \citenamefont {Altoé}, \citenamefont {Kim},
  \citenamefont {Song}, \citenamefont {Chen}, \citenamefont {Kreikebaum},
  \citenamefont {Ogletree}, \citenamefont {Santiago},\ and\ \citenamefont
  {Siddiqi}}]{doi:10.1063/5.0102092}%
  \BibitemOpen
  \bibfield  {author} {\bibinfo {author} {\bibfnamefont {H.}~\bibnamefont
  {Kim}}, \bibinfo {author} {\bibfnamefont {C.}~\bibnamefont {Jünger}},
  \bibinfo {author} {\bibfnamefont {A.}~\bibnamefont {Morvan}}, \bibinfo
  {author} {\bibfnamefont {E.~S.}\ \bibnamefont {Barnard}}, \bibinfo {author}
  {\bibfnamefont {W.~P.}\ \bibnamefont {Livingston}}, \bibinfo {author}
  {\bibfnamefont {M.~V.~P.}\ \bibnamefont {Altoé}}, \bibinfo {author}
  {\bibfnamefont {Y.}~\bibnamefont {Kim}}, \bibinfo {author} {\bibfnamefont
  {C.}~\bibnamefont {Song}}, \bibinfo {author} {\bibfnamefont {L.}~\bibnamefont
  {Chen}}, \bibinfo {author} {\bibfnamefont {J.~M.}\ \bibnamefont
  {Kreikebaum}}, \bibinfo {author} {\bibfnamefont {D.~F.}\ \bibnamefont
  {Ogletree}}, \bibinfo {author} {\bibfnamefont {D.~I.}\ \bibnamefont
  {Santiago}},\ and\ \bibinfo {author} {\bibfnamefont {I.}~\bibnamefont
  {Siddiqi}},\ }\href {https://doi.org/10.1063/5.0102092} {\bibfield  {journal}
  {\bibinfo  {journal} {Applied Physics Letters}\ }\textbf {\bibinfo {volume}
  {121}},\ \bibinfo {pages} {142601} (\bibinfo {year} {2022})}\BibitemShut
  {NoStop}%
\bibitem [{\citenamefont {Lescanne}\ \emph {et~al.}(2020)\citenamefont
  {Lescanne}, \citenamefont {Del\'eglise}, \citenamefont {Albertinale},
  \citenamefont {R\'eglade}, \citenamefont {Capelle}, \citenamefont {Ivanov},
  \citenamefont {Jacqmin}, \citenamefont {Leghtas},\ and\ \citenamefont
  {Flurin}}]{PhysRevX.10.021038}%
  \BibitemOpen
  \bibfield  {author} {\bibinfo {author} {\bibfnamefont {R.}~\bibnamefont
  {Lescanne}}, \bibinfo {author} {\bibfnamefont {S.}~\bibnamefont
  {Del\'eglise}}, \bibinfo {author} {\bibfnamefont {E.}~\bibnamefont
  {Albertinale}}, \bibinfo {author} {\bibfnamefont {U.}~\bibnamefont
  {R\'eglade}}, \bibinfo {author} {\bibfnamefont {T.}~\bibnamefont {Capelle}},
  \bibinfo {author} {\bibfnamefont {E.}~\bibnamefont {Ivanov}}, \bibinfo
  {author} {\bibfnamefont {T.}~\bibnamefont {Jacqmin}}, \bibinfo {author}
  {\bibfnamefont {Z.}~\bibnamefont {Leghtas}},\ and\ \bibinfo {author}
  {\bibfnamefont {E.}~\bibnamefont {Flurin}},\ }\href
  {https://doi.org/10.1103/PhysRevX.10.021038} {\bibfield  {journal} {\bibinfo
  {journal} {Phys. Rev. X}\ }\textbf {\bibinfo {volume} {10}},\ \bibinfo
  {pages} {021038} (\bibinfo {year} {2020})}\BibitemShut {NoStop}%
\bibitem [{\citenamefont {Mundada}\ \emph {et~al.}(2019)\citenamefont
  {Mundada}, \citenamefont {Zhang}, \citenamefont {Hazard},\ and\ \citenamefont
  {Houck}}]{zz_cl_Princeton}%
  \BibitemOpen
  \bibfield  {author} {\bibinfo {author} {\bibfnamefont {P.}~\bibnamefont
  {Mundada}}, \bibinfo {author} {\bibfnamefont {G.}~\bibnamefont {Zhang}},
  \bibinfo {author} {\bibfnamefont {T.}~\bibnamefont {Hazard}},\ and\ \bibinfo
  {author} {\bibfnamefont {A.}~\bibnamefont {Houck}},\ }\href
  {https://doi.org/10.1103/PhysRevApplied.12.054023} {\bibfield  {journal}
  {\bibinfo  {journal} {Phys. Rev. Applied}\ }\textbf {\bibinfo {volume}
  {12}},\ \bibinfo {pages} {054023} (\bibinfo {year} {2019})}\BibitemShut
  {NoStop}%
\bibitem [{\citenamefont {Osman}\ \emph {et~al.}(2021)\citenamefont {Osman},
  \citenamefont {Simon}, \citenamefont {Bengtsson}, \citenamefont {Kosen},
  \citenamefont {Krantz}, \citenamefont {P.~Lozano}, \citenamefont
  {Scigliuzzo}, \citenamefont {Delsing}, \citenamefont {Bylander},\ and\
  \citenamefont {Fadavi~Roudsari}}]{insitu_bandage}%
  \BibitemOpen
  \bibfield  {author} {\bibinfo {author} {\bibfnamefont {A.}~\bibnamefont
  {Osman}}, \bibinfo {author} {\bibfnamefont {J.}~\bibnamefont {Simon}},
  \bibinfo {author} {\bibfnamefont {A.}~\bibnamefont {Bengtsson}}, \bibinfo
  {author} {\bibfnamefont {S.}~\bibnamefont {Kosen}}, \bibinfo {author}
  {\bibfnamefont {P.}~\bibnamefont {Krantz}}, \bibinfo {author} {\bibfnamefont
  {D.}~\bibnamefont {P.~Lozano}}, \bibinfo {author} {\bibfnamefont
  {M.}~\bibnamefont {Scigliuzzo}}, \bibinfo {author} {\bibfnamefont
  {P.}~\bibnamefont {Delsing}}, \bibinfo {author} {\bibfnamefont
  {J.}~\bibnamefont {Bylander}},\ and\ \bibinfo {author} {\bibfnamefont
  {A.}~\bibnamefont {Fadavi~Roudsari}},\ }\href
  {https://doi.org/10.1063/5.0037093} {\bibfield  {journal} {\bibinfo
  {journal} {Applied Physics Letters}\ }\textbf {\bibinfo {volume} {118}},\
  \bibinfo {pages} {064002} (\bibinfo {year} {2021})}\BibitemShut {NoStop}%
\bibitem [{\citenamefont {Sjöqvist}(2015)}]{Erik_GP}%
  \BibitemOpen
  \bibfield  {author} {\bibinfo {author} {\bibfnamefont {E.}~\bibnamefont
  {Sjöqvist}},\ }\href {https://doi.org/https://doi.org/10.1002/qua.24941}
  {\bibfield  {journal} {\bibinfo  {journal} {International Journal of Quantum
  Chemistry}\ }\textbf {\bibinfo {volume} {115}},\ \bibinfo {pages} {1311}
  (\bibinfo {year} {2015})}\BibitemShut {NoStop}%
\bibitem [{\citenamefont {Johansson}\ \emph {et~al.}(2012)\citenamefont
  {Johansson}, \citenamefont {Nation},\ and\ \citenamefont
  {Nori}}]{JOHANSSON20121760}%
  \BibitemOpen
  \bibfield  {author} {\bibinfo {author} {\bibfnamefont {J.}~\bibnamefont
  {Johansson}}, \bibinfo {author} {\bibfnamefont {P.}~\bibnamefont {Nation}},\
  and\ \bibinfo {author} {\bibfnamefont {F.}~\bibnamefont {Nori}},\ }\href
  {https://doi.org/https://doi.org/10.1016/j.cpc.2012.02.021} {\bibfield
  {journal} {\bibinfo  {journal} {Computer Physics Communications}\ }\textbf
  {\bibinfo {volume} {183}},\ \bibinfo {pages} {1760} (\bibinfo {year}
  {2012})}\BibitemShut {NoStop}%
\bibitem [{\citenamefont {Johansson}\ \emph {et~al.}(2013)\citenamefont
  {Johansson}, \citenamefont {Nation},\ and\ \citenamefont
  {Nori}}]{JOHANSSON20131234}%
  \BibitemOpen
  \bibfield  {author} {\bibinfo {author} {\bibfnamefont {J.}~\bibnamefont
  {Johansson}}, \bibinfo {author} {\bibfnamefont {P.}~\bibnamefont {Nation}},\
  and\ \bibinfo {author} {\bibfnamefont {F.}~\bibnamefont {Nori}},\ }\href
  {https://doi.org/https://doi.org/10.1016/j.cpc.2012.11.019} {\bibfield
  {journal} {\bibinfo  {journal} {Computer Physics Communications}\ }\textbf
  {\bibinfo {volume} {184}},\ \bibinfo {pages} {1234} (\bibinfo {year}
  {2013})}\BibitemShut {NoStop}%
\bibitem [{\citenamefont {Reed}\ \emph {et~al.}(2010)\citenamefont {Reed},
  \citenamefont {Johnson}, \citenamefont {Houck}, \citenamefont {DiCarlo},
  \citenamefont {Chow}, \citenamefont {Schuster}, \citenamefont {Frunzio},\
  and\ \citenamefont {Schoelkopf}}]{reed2010fast}%
  \BibitemOpen
  \bibfield  {author} {\bibinfo {author} {\bibfnamefont {M.~D.}\ \bibnamefont
  {Reed}}, \bibinfo {author} {\bibfnamefont {B.~R.}\ \bibnamefont {Johnson}},
  \bibinfo {author} {\bibfnamefont {A.~A.}\ \bibnamefont {Houck}}, \bibinfo
  {author} {\bibfnamefont {L.}~\bibnamefont {DiCarlo}}, \bibinfo {author}
  {\bibfnamefont {J.~M.}\ \bibnamefont {Chow}}, \bibinfo {author}
  {\bibfnamefont {D.~I.}\ \bibnamefont {Schuster}}, \bibinfo {author}
  {\bibfnamefont {L.}~\bibnamefont {Frunzio}},\ and\ \bibinfo {author}
  {\bibfnamefont {R.~J.}\ \bibnamefont {Schoelkopf}},\ }\href
  {https://doi.org/10.1063/1.3435463} {\bibfield  {journal} {\bibinfo
  {journal} {Applied Physics Letters}\ }\textbf {\bibinfo {volume} {96}},\
  \bibinfo {pages} {203110} (\bibinfo {year} {2010})}\BibitemShut {NoStop}%
\bibitem [{\citenamefont {Murray}\ \emph {et~al.}(2018)\citenamefont {Murray},
  \citenamefont {Gambetta}, \citenamefont {McClure},\ and\ \citenamefont
  {Steffen}}]{PR_IBM_Analytical}%
  \BibitemOpen
  \bibfield  {author} {\bibinfo {author} {\bibfnamefont {C.~E.}\ \bibnamefont
  {Murray}}, \bibinfo {author} {\bibfnamefont {J.~M.}\ \bibnamefont
  {Gambetta}}, \bibinfo {author} {\bibfnamefont {D.~T.}\ \bibnamefont
  {McClure}},\ and\ \bibinfo {author} {\bibfnamefont {M.}~\bibnamefont
  {Steffen}},\ }\href {https://doi.org/10.1109/TMTT.2018.2841829} {\bibfield
  {journal} {\bibinfo  {journal} {IEEE Transactions on Microwave Theory and
  Techniques}\ }\textbf {\bibinfo {volume} {66}},\ \bibinfo {pages} {3724}
  (\bibinfo {year} {2018})}\BibitemShut {NoStop}%
\bibitem [{\citenamefont {Motzoi}\ \emph {et~al.}(2009)\citenamefont {Motzoi},
  \citenamefont {Gambetta}, \citenamefont {Rebentrost},\ and\ \citenamefont
  {Wilhelm}}]{DRAG_leak}%
  \BibitemOpen
  \bibfield  {author} {\bibinfo {author} {\bibfnamefont {F.}~\bibnamefont
  {Motzoi}}, \bibinfo {author} {\bibfnamefont {J.~M.}\ \bibnamefont
  {Gambetta}}, \bibinfo {author} {\bibfnamefont {P.}~\bibnamefont
  {Rebentrost}},\ and\ \bibinfo {author} {\bibfnamefont {F.~K.}\ \bibnamefont
  {Wilhelm}},\ }\href {https://doi.org/10.1103/PhysRevLett.103.110501}
  {\bibfield  {journal} {\bibinfo  {journal} {Phys. Rev. Lett.}\ }\textbf
  {\bibinfo {volume} {103}},\ \bibinfo {pages} {110501} (\bibinfo {year}
  {2009})}\BibitemShut {NoStop}%
\bibitem [{\citenamefont {Garbow}\ \emph {et~al.}(1982)\citenamefont {Garbow},
  \citenamefont {Weitekamp},\ and\ \citenamefont {Pines}}]{GARBOW1982504}%
  \BibitemOpen
  \bibfield  {author} {\bibinfo {author} {\bibfnamefont {J.}~\bibnamefont
  {Garbow}}, \bibinfo {author} {\bibfnamefont {D.}~\bibnamefont {Weitekamp}},\
  and\ \bibinfo {author} {\bibfnamefont {A.}~\bibnamefont {Pines}},\ }\href
  {https://doi.org/https://doi.org/10.1016/0009-2614(82)83229-6} {\bibfield
  {journal} {\bibinfo  {journal} {Chemical Physics Letters}\ }\textbf {\bibinfo
  {volume} {93}},\ \bibinfo {pages} {504} (\bibinfo {year} {1982})}\BibitemShut
  {NoStop}%
\bibitem [{\citenamefont {Ku}\ \emph {et~al.}(2020)\citenamefont {Ku},
  \citenamefont {Xu}, \citenamefont {Brink}, \citenamefont {McKay},
  \citenamefont {Hertzberg}, \citenamefont {Ansari},\ and\ \citenamefont
  {Plourde}}]{SuppressZZ_TQ_CSFQ}%
  \BibitemOpen
  \bibfield  {author} {\bibinfo {author} {\bibfnamefont {J.}~\bibnamefont
  {Ku}}, \bibinfo {author} {\bibfnamefont {X.}~\bibnamefont {Xu}}, \bibinfo
  {author} {\bibfnamefont {M.}~\bibnamefont {Brink}}, \bibinfo {author}
  {\bibfnamefont {D.~C.}\ \bibnamefont {McKay}}, \bibinfo {author}
  {\bibfnamefont {J.~B.}\ \bibnamefont {Hertzberg}}, \bibinfo {author}
  {\bibfnamefont {M.~H.}\ \bibnamefont {Ansari}},\ and\ \bibinfo {author}
  {\bibfnamefont {B.~L.~T.}\ \bibnamefont {Plourde}},\ }\href
  {https://doi.org/10.1103/PhysRevLett.125.200504} {\bibfield  {journal}
  {\bibinfo  {journal} {Phys. Rev. Lett.}\ }\textbf {\bibinfo {volume} {125}},\
  \bibinfo {pages} {200504} (\bibinfo {year} {2020})}\BibitemShut {NoStop}%
\bibitem [{\citenamefont {Barends}\ \emph {et~al.}(2014)\citenamefont
  {Barends}, \citenamefont {Kelly}, \citenamefont {Megrant}, \citenamefont
  {Veitia}, \citenamefont {Sank}, \citenamefont {Jeffrey}, \citenamefont
  {White}, \citenamefont {Mutus}, \citenamefont {Fowler}, \citenamefont
  {Campbell}, \citenamefont {Chen}, \citenamefont {Chen}, \citenamefont
  {Chiaro}, \citenamefont {Dunsworth}, \citenamefont {Neill}, \citenamefont
  {O'Malley}, \citenamefont {Roushan}, \citenamefont {Vainsencher},
  \citenamefont {Wenner}, \citenamefont {Korotkov}, \citenamefont {Cleland},\
  and\ \citenamefont {Martinis}}]{Barends2014}%
  \BibitemOpen
  \bibfield  {author} {\bibinfo {author} {\bibfnamefont {R.}~\bibnamefont
  {Barends}}, \bibinfo {author} {\bibfnamefont {J.}~\bibnamefont {Kelly}},
  \bibinfo {author} {\bibfnamefont {A.}~\bibnamefont {Megrant}}, \bibinfo
  {author} {\bibfnamefont {A.}~\bibnamefont {Veitia}}, \bibinfo {author}
  {\bibfnamefont {D.}~\bibnamefont {Sank}}, \bibinfo {author} {\bibfnamefont
  {E.}~\bibnamefont {Jeffrey}}, \bibinfo {author} {\bibfnamefont {T.~C.}\
  \bibnamefont {White}}, \bibinfo {author} {\bibfnamefont {J.}~\bibnamefont
  {Mutus}}, \bibinfo {author} {\bibfnamefont {A.~G.}\ \bibnamefont {Fowler}},
  \bibinfo {author} {\bibfnamefont {B.}~\bibnamefont {Campbell}}, \bibinfo
  {author} {\bibfnamefont {Y.}~\bibnamefont {Chen}}, \bibinfo {author}
  {\bibfnamefont {Z.}~\bibnamefont {Chen}}, \bibinfo {author} {\bibfnamefont
  {B.}~\bibnamefont {Chiaro}}, \bibinfo {author} {\bibfnamefont
  {A.}~\bibnamefont {Dunsworth}}, \bibinfo {author} {\bibfnamefont
  {C.}~\bibnamefont {Neill}}, \bibinfo {author} {\bibfnamefont
  {P.}~\bibnamefont {O'Malley}}, \bibinfo {author} {\bibfnamefont
  {P.}~\bibnamefont {Roushan}}, \bibinfo {author} {\bibfnamefont
  {A.}~\bibnamefont {Vainsencher}}, \bibinfo {author} {\bibfnamefont
  {J.}~\bibnamefont {Wenner}}, \bibinfo {author} {\bibfnamefont {A.~N.}\
  \bibnamefont {Korotkov}}, \bibinfo {author} {\bibfnamefont {A.~N.}\
  \bibnamefont {Cleland}},\ and\ \bibinfo {author} {\bibfnamefont {J.~M.}\
  \bibnamefont {Martinis}},\ }\href {https://doi.org/10.1038/nature13171}
  {\bibfield  {journal} {\bibinfo  {journal} {Nature}\ }\textbf {\bibinfo
  {volume} {508}},\ \bibinfo {pages} {500} (\bibinfo {year}
  {2014})}\BibitemShut {NoStop}%
\bibitem [{\citenamefont {McKay}\ \emph {et~al.}(2017)\citenamefont {McKay},
  \citenamefont {Wood}, \citenamefont {Sheldon}, \citenamefont {Chow},\ and\
  \citenamefont {Gambetta}}]{mckay2017efficient}%
  \BibitemOpen
  \bibfield  {author} {\bibinfo {author} {\bibfnamefont {D.~C.}\ \bibnamefont
  {McKay}}, \bibinfo {author} {\bibfnamefont {C.~J.}\ \bibnamefont {Wood}},
  \bibinfo {author} {\bibfnamefont {S.}~\bibnamefont {Sheldon}}, \bibinfo
  {author} {\bibfnamefont {J.~M.}\ \bibnamefont {Chow}},\ and\ \bibinfo
  {author} {\bibfnamefont {J.~M.}\ \bibnamefont {Gambetta}},\ }\href
  {https://doi.org/10.1103/PhysRevA.96.022330} {\bibfield  {journal} {\bibinfo
  {journal} {Phys. Rev. A}\ }\textbf {\bibinfo {volume} {96}},\ \bibinfo
  {pages} {022330} (\bibinfo {year} {2017})}\BibitemShut {NoStop}%
\bibitem [{\citenamefont {Magesan}\ \emph {et~al.}(2012)\citenamefont
  {Magesan}, \citenamefont {Gambetta}, \citenamefont {Johnson}, \citenamefont
  {Ryan}, \citenamefont {Chow}, \citenamefont {Merkel}, \citenamefont
  {da~Silva}, \citenamefont {Keefe}, \citenamefont {Rothwell}, \citenamefont
  {Ohki}, \citenamefont {Ketchen},\ and\ \citenamefont
  {Steffen}}]{magesan2012efficient}%
  \BibitemOpen
  \bibfield  {author} {\bibinfo {author} {\bibfnamefont {E.}~\bibnamefont
  {Magesan}}, \bibinfo {author} {\bibfnamefont {J.~M.}\ \bibnamefont
  {Gambetta}}, \bibinfo {author} {\bibfnamefont {B.~R.}\ \bibnamefont
  {Johnson}}, \bibinfo {author} {\bibfnamefont {C.~A.}\ \bibnamefont {Ryan}},
  \bibinfo {author} {\bibfnamefont {J.~M.}\ \bibnamefont {Chow}}, \bibinfo
  {author} {\bibfnamefont {S.~T.}\ \bibnamefont {Merkel}}, \bibinfo {author}
  {\bibfnamefont {M.~P.}\ \bibnamefont {da~Silva}}, \bibinfo {author}
  {\bibfnamefont {G.~A.}\ \bibnamefont {Keefe}}, \bibinfo {author}
  {\bibfnamefont {M.~B.}\ \bibnamefont {Rothwell}}, \bibinfo {author}
  {\bibfnamefont {T.~A.}\ \bibnamefont {Ohki}}, \bibinfo {author}
  {\bibfnamefont {M.~B.}\ \bibnamefont {Ketchen}},\ and\ \bibinfo {author}
  {\bibfnamefont {M.}~\bibnamefont {Steffen}},\ }\href
  {https://doi.org/10.1103/PhysRevLett.109.080505} {\bibfield  {journal}
  {\bibinfo  {journal} {Phys. Rev. Lett.}\ }\textbf {\bibinfo {volume} {109}},\
  \bibinfo {pages} {080505} (\bibinfo {year} {2012})}\BibitemShut {NoStop}%
\bibitem [{\citenamefont {Khaneja}\ \emph {et~al.}(2005)\citenamefont
  {Khaneja}, \citenamefont {Reiss}, \citenamefont {Kehlet}, \citenamefont
  {Schulte-Herbrüggen},\ and\ \citenamefont {Glaser}}]{KHANEJA2005296}%
  \BibitemOpen
  \bibfield  {author} {\bibinfo {author} {\bibfnamefont {N.}~\bibnamefont
  {Khaneja}}, \bibinfo {author} {\bibfnamefont {T.}~\bibnamefont {Reiss}},
  \bibinfo {author} {\bibfnamefont {C.}~\bibnamefont {Kehlet}}, \bibinfo
  {author} {\bibfnamefont {T.}~\bibnamefont {Schulte-Herbrüggen}},\ and\
  \bibinfo {author} {\bibfnamefont {S.~J.}\ \bibnamefont {Glaser}},\ }\href
  {https://doi.org/https://doi.org/10.1016/j.jmr.2004.11.004} {\bibfield
  {journal} {\bibinfo  {journal} {Journal of Magnetic Resonance}\ }\textbf
  {\bibinfo {volume} {172}},\ \bibinfo {pages} {296} (\bibinfo {year}
  {2005})}\BibitemShut {NoStop}%
\bibitem [{\citenamefont {Baum}\ \emph {et~al.}(2021)\citenamefont {Baum},
  \citenamefont {Amico}, \citenamefont {Howell}, \citenamefont {Hush},
  \citenamefont {Liuzzi}, \citenamefont {Mundada}, \citenamefont {Merkh},
  \citenamefont {Carvalho},\ and\ \citenamefont
  {Biercuk}}]{PRXQuantum.2.040324}%
  \BibitemOpen
  \bibfield  {author} {\bibinfo {author} {\bibfnamefont {Y.}~\bibnamefont
  {Baum}}, \bibinfo {author} {\bibfnamefont {M.}~\bibnamefont {Amico}},
  \bibinfo {author} {\bibfnamefont {S.}~\bibnamefont {Howell}}, \bibinfo
  {author} {\bibfnamefont {M.}~\bibnamefont {Hush}}, \bibinfo {author}
  {\bibfnamefont {M.}~\bibnamefont {Liuzzi}}, \bibinfo {author} {\bibfnamefont
  {P.}~\bibnamefont {Mundada}}, \bibinfo {author} {\bibfnamefont
  {T.}~\bibnamefont {Merkh}}, \bibinfo {author} {\bibfnamefont {A.~R.~R.}\
  \bibnamefont {Carvalho}},\ and\ \bibinfo {author} {\bibfnamefont {M.~J.}\
  \bibnamefont {Biercuk}},\ }\href
  {https://doi.org/10.1103/PRXQuantum.2.040324} {\bibfield  {journal} {\bibinfo
   {journal} {PRX Quantum}\ }\textbf {\bibinfo {volume} {2}},\ \bibinfo {pages}
  {040324} (\bibinfo {year} {2021})}\BibitemShut {NoStop}%
\bibitem [{\citenamefont {Cummins}\ \emph {et~al.}(2003)\citenamefont
  {Cummins}, \citenamefont {Llewellyn},\ and\ \citenamefont
  {Jones}}]{PhysRevA.67.042308}%
  \BibitemOpen
  \bibfield  {author} {\bibinfo {author} {\bibfnamefont {H.~K.}\ \bibnamefont
  {Cummins}}, \bibinfo {author} {\bibfnamefont {G.}~\bibnamefont {Llewellyn}},\
  and\ \bibinfo {author} {\bibfnamefont {J.~A.}\ \bibnamefont {Jones}},\ }\href
  {https://doi.org/10.1103/PhysRevA.67.042308} {\bibfield  {journal} {\bibinfo
  {journal} {Phys. Rev. A}\ }\textbf {\bibinfo {volume} {67}},\ \bibinfo
  {pages} {042308} (\bibinfo {year} {2003})}\BibitemShut {NoStop}%
\bibitem [{\citenamefont {Cummins}\ and\ \citenamefont
  {Jones}(2000)}]{Cummins_2000}%
  \BibitemOpen
  \bibfield  {author} {\bibinfo {author} {\bibfnamefont {H.~K.}\ \bibnamefont
  {Cummins}}\ and\ \bibinfo {author} {\bibfnamefont {J.~A.}\ \bibnamefont
  {Jones}},\ }\href {https://doi.org/10.1088/1367-2630/2/1/006} {\bibfield
  {journal} {\bibinfo  {journal} {New Journal of Physics}\ }\textbf {\bibinfo
  {volume} {2}},\ \bibinfo {pages} {6} (\bibinfo {year} {2000})}\BibitemShut
  {NoStop}%
\bibitem [{\citenamefont {Ni}\ \emph {et~al.}(2022)\citenamefont {Ni},
  \citenamefont {Li}, \citenamefont {Zhang}, \citenamefont {Chu}, \citenamefont
  {Niu}, \citenamefont {Yan}, \citenamefont {Deng}, \citenamefont {Hu},
  \citenamefont {Li}, \citenamefont {Zhong}, \citenamefont {Liu}, \citenamefont
  {Yan}, \citenamefont {Xu},\ and\ \citenamefont {Yu}}]{ChinaZZcl}%
  \BibitemOpen
  \bibfield  {author} {\bibinfo {author} {\bibfnamefont {Z.}~\bibnamefont
  {Ni}}, \bibinfo {author} {\bibfnamefont {S.}~\bibnamefont {Li}}, \bibinfo
  {author} {\bibfnamefont {L.}~\bibnamefont {Zhang}}, \bibinfo {author}
  {\bibfnamefont {J.}~\bibnamefont {Chu}}, \bibinfo {author} {\bibfnamefont
  {J.}~\bibnamefont {Niu}}, \bibinfo {author} {\bibfnamefont {T.}~\bibnamefont
  {Yan}}, \bibinfo {author} {\bibfnamefont {X.}~\bibnamefont {Deng}}, \bibinfo
  {author} {\bibfnamefont {L.}~\bibnamefont {Hu}}, \bibinfo {author}
  {\bibfnamefont {J.}~\bibnamefont {Li}}, \bibinfo {author} {\bibfnamefont
  {Y.}~\bibnamefont {Zhong}}, \bibinfo {author} {\bibfnamefont
  {S.}~\bibnamefont {Liu}}, \bibinfo {author} {\bibfnamefont {F.}~\bibnamefont
  {Yan}}, \bibinfo {author} {\bibfnamefont {Y.}~\bibnamefont {Xu}},\ and\
  \bibinfo {author} {\bibfnamefont {D.}~\bibnamefont {Yu}},\ }\href
  {https://doi.org/10.1103/PhysRevLett.129.040502} {\bibfield  {journal}
  {\bibinfo  {journal} {Phys. Rev. Lett.}\ }\textbf {\bibinfo {volume} {129}},\
  \bibinfo {pages} {040502} (\bibinfo {year} {2022})}\BibitemShut {NoStop}%
\bibitem [{\citenamefont {Winkler}(2003)}]{winkler2003spin}%
  \BibitemOpen
  \bibfield  {author} {\bibinfo {author} {\bibfnamefont {R.}~\bibnamefont
  {Winkler}},\ }\href {https://doi.org/10.1007/b13586} {\emph {\bibinfo {title}
  {{Spin-Orbit Coupling Effects in Two-Dimensional Electron and Hole
  Systems}}}},\ Tracts in Modern Physics\ (\bibinfo  {publisher} {Springer},\
  \bibinfo {address} {Berlin},\ \bibinfo {year} {2003})\BibitemShut {NoStop}%
\bibitem [{\citenamefont {Leskes}\ \emph {et~al.}(2010)\citenamefont {Leskes},
  \citenamefont {Madhu},\ and\ \citenamefont {Vega}}]{Floquet_Leskes}%
  \BibitemOpen
  \bibfield  {author} {\bibinfo {author} {\bibfnamefont {M.}~\bibnamefont
  {Leskes}}, \bibinfo {author} {\bibfnamefont {P.}~\bibnamefont {Madhu}},\ and\
  \bibinfo {author} {\bibfnamefont {S.}~\bibnamefont {Vega}},\ }\href
  {https://doi.org/https://doi.org/10.1016/j.pnmrs.2010.06.002} {\bibfield
  {journal} {\bibinfo  {journal} {Progress in Nuclear Magnetic Resonance
  Spectroscopy}\ }\textbf {\bibinfo {volume} {57}},\ \bibinfo {pages} {345}
  (\bibinfo {year} {2010})}\BibitemShut {NoStop}%
\bibitem [{\citenamefont {Wood}\ and\ \citenamefont
  {Gambetta}(2018)}]{AGF_leak}%
  \BibitemOpen
  \bibfield  {author} {\bibinfo {author} {\bibfnamefont {C.~J.}\ \bibnamefont
  {Wood}}\ and\ \bibinfo {author} {\bibfnamefont {J.~M.}\ \bibnamefont
  {Gambetta}},\ }\href {https://doi.org/10.1103/PhysRevA.97.032306} {\bibfield
  {journal} {\bibinfo  {journal} {Phys. Rev. A}\ }\textbf {\bibinfo {volume}
  {97}},\ \bibinfo {pages} {032306} (\bibinfo {year} {2018})}\BibitemShut
  {NoStop}%
\bibitem [{\citenamefont {Schuster}\ \emph {et~al.}(2005)\citenamefont
  {Schuster}, \citenamefont {Wallraff}, \citenamefont {Blais}, \citenamefont
  {Frunzio}, \citenamefont {Huang}, \citenamefont {Majer}, \citenamefont
  {Girvin},\ and\ \citenamefont {Schoelkopf}}]{PhysRevLett.94.123602}%
  \BibitemOpen
  \bibfield  {author} {\bibinfo {author} {\bibfnamefont {D.~I.}\ \bibnamefont
  {Schuster}}, \bibinfo {author} {\bibfnamefont {A.}~\bibnamefont {Wallraff}},
  \bibinfo {author} {\bibfnamefont {A.}~\bibnamefont {Blais}}, \bibinfo
  {author} {\bibfnamefont {L.}~\bibnamefont {Frunzio}}, \bibinfo {author}
  {\bibfnamefont {R.-S.}\ \bibnamefont {Huang}}, \bibinfo {author}
  {\bibfnamefont {J.}~\bibnamefont {Majer}}, \bibinfo {author} {\bibfnamefont
  {S.~M.}\ \bibnamefont {Girvin}},\ and\ \bibinfo {author} {\bibfnamefont
  {R.~J.}\ \bibnamefont {Schoelkopf}},\ }\href
  {https://doi.org/10.1103/PhysRevLett.94.123602} {\bibfield  {journal}
  {\bibinfo  {journal} {Phys. Rev. Lett.}\ }\textbf {\bibinfo {volume} {94}},\
  \bibinfo {pages} {123602} (\bibinfo {year} {2005})}\BibitemShut {NoStop}%
\bibitem [{\citenamefont {Schneider}\ \emph {et~al.}(2018)\citenamefont
  {Schneider}, \citenamefont {Braum\"uller}, \citenamefont {Guo}, \citenamefont
  {Stehle}, \citenamefont {Rotzinger}, \citenamefont {Marthaler}, \citenamefont
  {Ustinov},\ and\ \citenamefont {Weides}}]{PhysRevA.97.062334}%
  \BibitemOpen
  \bibfield  {author} {\bibinfo {author} {\bibfnamefont {A.}~\bibnamefont
  {Schneider}}, \bibinfo {author} {\bibfnamefont {J.}~\bibnamefont
  {Braum\"uller}}, \bibinfo {author} {\bibfnamefont {L.}~\bibnamefont {Guo}},
  \bibinfo {author} {\bibfnamefont {P.}~\bibnamefont {Stehle}}, \bibinfo
  {author} {\bibfnamefont {H.}~\bibnamefont {Rotzinger}}, \bibinfo {author}
  {\bibfnamefont {M.}~\bibnamefont {Marthaler}}, \bibinfo {author}
  {\bibfnamefont {A.~V.}\ \bibnamefont {Ustinov}},\ and\ \bibinfo {author}
  {\bibfnamefont {M.}~\bibnamefont {Weides}},\ }\href
  {https://doi.org/10.1103/PhysRevA.97.062334} {\bibfield  {journal} {\bibinfo
  {journal} {Phys. Rev. A}\ }\textbf {\bibinfo {volume} {97}},\ \bibinfo
  {pages} {062334} (\bibinfo {year} {2018})}\BibitemShut {NoStop}%
\bibitem [{\citenamefont {Zeytino\ifmmode~\breve{g}\else \u{g}\fi{}lu}\ \emph
  {et~al.}(2015)\citenamefont {Zeytino\ifmmode~\breve{g}\else \u{g}\fi{}lu},
  \citenamefont {Pechal}, \citenamefont {Berger}, \citenamefont {Abdumalikov},
  \citenamefont {Wallraff},\ and\ \citenamefont {Filipp}}]{PhysRevA.91.043846}%
  \BibitemOpen
  \bibfield  {author} {\bibinfo {author} {\bibfnamefont {S.}~\bibnamefont
  {Zeytino\ifmmode~\breve{g}\else \u{g}\fi{}lu}}, \bibinfo {author}
  {\bibfnamefont {M.}~\bibnamefont {Pechal}}, \bibinfo {author} {\bibfnamefont
  {S.}~\bibnamefont {Berger}}, \bibinfo {author} {\bibfnamefont {A.~A.}\
  \bibnamefont {Abdumalikov}}, \bibinfo {author} {\bibfnamefont
  {A.}~\bibnamefont {Wallraff}},\ and\ \bibinfo {author} {\bibfnamefont
  {S.}~\bibnamefont {Filipp}},\ }\href
  {https://doi.org/10.1103/PhysRevA.91.043846} {\bibfield  {journal} {\bibinfo
  {journal} {Phys. Rev. A}\ }\textbf {\bibinfo {volume} {91}},\ \bibinfo
  {pages} {043846} (\bibinfo {year} {2015})}\BibitemShut {NoStop}%
\bibitem [{\citenamefont {Magesan}\ and\ \citenamefont
  {Gambetta}(2020)}]{PhysRevA.101.052308}%
  \BibitemOpen
  \bibfield  {author} {\bibinfo {author} {\bibfnamefont {E.}~\bibnamefont
  {Magesan}}\ and\ \bibinfo {author} {\bibfnamefont {J.~M.}\ \bibnamefont
  {Gambetta}},\ }\href {https://doi.org/10.1103/PhysRevA.101.052308} {\bibfield
   {journal} {\bibinfo  {journal} {Phys. Rev. A}\ }\textbf {\bibinfo {volume}
  {101}},\ \bibinfo {pages} {052308} (\bibinfo {year} {2020})}\BibitemShut
  {NoStop}%
\end{thebibliography}%
@Article{Ladd2010,
author={Ladd, T. D.
and Jelezko, F.
and Laflamme, R.
and Nakamura, Y.
and Monroe, C.
and O'Brien, J. L.},
title={Quantum computers},
journal={Nature},
year={2010},
month={Mar},
day={01},
volume={464},
number={7285},
pages={45-53},
abstract={With basic information processing units (qubits) governed by the exotic phenomena of quantum mechanics, quantum computers have the potential to be far better at performing certain calculations than today's computers using conventional 'bits'. That said, it's far from clear what technology practical quantum computers --- if and when they arrive --- will use. In an extensive review, six researchers from major labs in the field describe the latest work on the hardware for quantum information systems. Current materials are compared --- including the nuclear spins of donor atoms in doped silicon, electron spins in gallium arsenide and nitrogen-vacancy centres in diamond --- and the materials that are yet to come are speculated upon.},
issn={1476-4687},
doi={10.1038/nature08812},
url={https://doi.org/10.1038/nature08812}
}

@article{KITAEV20032,
title = {Fault-tolerant quantum computation by anyons},
journal = {Annals of Physics},
volume = {303},
number = {1},
pages = {2-30},
year = {2003},
issn = {0003-4916},
doi = {https://doi.org/10.1016/S0003-4916(02)00018-0},
url = {https://www.sciencedirect.com/science/article/pii/S0003491602000180},
author = {A.Yu. Kitaev},
abstract = {A two-dimensional quantum system with anyonic excitations can be considered as a quantum computer. Unitary transformations can be performed by moving the excitations around each other. Measurements can be performed by joining excitations in pairs and observing the result of fusion. Such computation is fault-tolerant by its physical nature.}
}

@article{bravyi1998quantum,
  title={Quantum codes on a lattice with boundary},
  author={Bravyi, Sergey B and Kitaev, A Yu},
  journal={arXiv preprint quant-ph/9811052},
  year={1998},
  url = {https://arxiv.org/abs/quant-ph/9811052}
}

@article{doi:10.1063/1.5115814,
author = {Slussarenko,Sergei  and Pryde,Geoff J. },
title = {Photonic quantum information processing: A concise review},
journal = {Applied Physics Reviews},
volume = {6},
number = {4},
pages = {041303},
year = {2019},
doi = {10.1063/1.5115814},
URL = {https://doi.org/10.1063/1.5115814}
}

@article{PhysRevX.10.021038,
  title = {Irreversible Qubit-Photon Coupling for the Detection of Itinerant Microwave Photons},
  author = {Lescanne, Rapha\"el and Del\'eglise, Samuel and Albertinale, Emanuele and R\'eglade, Ulysse and Capelle, Thibault and Ivanov, Edouard and Jacqmin, Thibaut and Leghtas, Zaki and Flurin, Emmanuel},
  journal = {Phys. Rev. X},
  volume = {10},
  issue = {2},
  pages = {021038},
  numpages = {17},
  year = {2020},
  month = {May},
  publisher = {American Physical Society},
  doi = {10.1103/PhysRevX.10.021038},
  url = {https://link.aps.org/doi/10.1103/PhysRevX.10.021038}
}

@article{Saffman_2016,
	doi = {10.1088/0953-4075/49/20/202001},
	url = {https://doi.org/10.1088/0953-4075/49/20/202001},
	year = 2016,
	month = {oct},
	publisher = {{IOP} Publishing},
	volume = {49},
	number = {20},
	pages = {202001},
	author = {M Saffman},
	title = {Quantum computing with atomic qubits and Rydberg interactions: progress and challenges},
	journal = {Journal of Physics B: Atomic, Molecular and Optical Physics},
	abstract = {We present a review of quantum computation with neutral atom qubits. After an overview of architectural options and approaches to preparing large qubit arrays we examine Rydberg mediated gate protocols and fidelity for two- and multi-qubit interactions. Quantum simulation and Rydberg dressing are alternatives to circuit based quantum computing for exploring many body quantum dynamics. We review the properties of the dressing interaction and provide a quantitative figure of merit for the complexity of the coherent dynamics that can be accessed with dressing. We conclude with a summary of the current status and an outlook for future progress.}
}

@article{kloeffel2013prospects,
author = {Kloeffel, Christoph and Loss, Daniel},
title = {Prospects for Spin-Based Quantum Computing in Quantum Dots},
journal = {Annual Review of Condensed Matter Physics},
volume = {4},
number = {1},
pages = {51-81},
year = {2013},
doi = {10.1146/annurev-conmatphys-030212-184248},

URL = { 
        https://doi.org/10.1146/annurev-conmatphys-030212-184248
    
},

    abstract = { Experimental and theoretical progress toward quantum computation with spins in quantum dots (QDs) is reviewed, with particular focus on QDs formed in GaAs heterostructures, on nanowire-based QDs, and on self-assembled QDs. We report on a remarkable evolution of the field, where decoherence—one of the main challenges for realizing quantum computers—no longer seems to be the stumbling block it had originally been considered. General concepts, relevant quantities, and basic requirements for spin-based quantum computing are explained; opportunities and challenges of spin-orbit interaction and nuclear spins are reviewed. We discuss recent achievements, present current theoretical proposals, and make several suggestions for further experiments. }
}

@article{Floquet_Leskes,
title = {Floquet theory in solid-state nuclear magnetic resonance},
journal = {Progress in Nuclear Magnetic Resonance Spectroscopy},
volume = {57},
number = {4},
pages = {345-380},
year = {2010},
issn = {0079-6565},
doi = {https://doi.org/10.1016/j.pnmrs.2010.06.002},
url = {https://www.sciencedirect.com/science/article/pii/S0079656510000798},
author = {Michal Leskes and P.K. Madhu and Shimon Vega},
keywords = {Solid-state NMR, Floquet theory, van Vleck transformation, Average Hamiltonian theory}
}

@Article{Hertzberg2021,
author={Hertzberg, Jared B.
and Zhang, Eric J.
and Rosenblatt, Sami
and Magesan, Easwar
and Smolin, John A.
and Yau, Jeng-Bang
and Adiga, Vivekananda P.
and Sandberg, Martin
and Brink, Markus
and Chow, Jerry M.
and Orcutt, Jason S.},
title={Laser-annealing Josephson junctions for yielding scaled-up superconducting quantum processors},
journal={npj Quantum Information},
year={2021},
month={Aug},
day={19},
volume={7},
number={1},
pages={129},
abstract={As superconducting quantum circuits scale to larger sizes, the problem of frequency crowding proves a formidable task. Here we present a solution for this problem in fixed-frequency qubit architectures. By systematically adjusting qubit frequencies post-fabrication, we show a nearly tenfold improvement in the precision of setting qubit frequencies. To assess scalability, we identify the types of ``frequency collisions'' that will impair a transmon qubit and cross-resonance gate architecture. Using statistical modeling, we compute the probability of evading all such conditions, as a function of qubit frequency precision. We find that, without post-fabrication tuning, the probability of finding a workable lattice quickly approaches 0. However, with the demonstrated precisions it is possible to find collision-free lattices with favorable yield. These techniques and models are currently employed in available quantum systems and will be indispensable as systems continue to scale to larger sizes.},
issn={2056-6387},
doi={10.1038/s41534-021-00464-5},
url={https://doi.org/10.1038/s41534-021-00464-5}
}

@article{FirstPrinciples,
  title = {First-principles analysis of cross-resonance gate operation},
  author = {Malekakhlagh, Moein and Magesan, Easwar and McKay, David C.},
  journal = {Phys. Rev. A},
  volume = {102},
  issue = {4},
  pages = {042605},
  numpages = {28},
  year = {2020},
  month = {Oct},
  publisher = {American Physical Society},
  doi = {10.1103/PhysRevA.102.042605},
  url = {https://link.aps.org/doi/10.1103/PhysRevA.102.042605}
}

@inproceedings{granata2008trimming,
  title={Trimming of critical current in niobium Josephson devices by laser annealing},
  author={Granata, C and Vettoliere, A and Petti, L and Rippa, M and Ruggiero, B and Mormile, P and Russo, M},
  booktitle={Journal of Physics: Conference Series},
  volume={97},
  number={1},
  pages={012110},
  year={2008},
  organization={IOP Publishing}
}

@article{insitu_bandage,
author = {Osman,A.  and Simon,J.  and Bengtsson,A.  and Kosen,S.  and Krantz,P.  and P. Lozano,D.  and Scigliuzzo,M.  and Delsing,P.  and Bylander,Jonas  and Fadavi Roudsari,A. },
title = {Simplified Josephson-junction fabrication process for reproducibly           high-performance superconducting qubits},
journal = {Applied Physics Letters},
volume = {118},
number = {6},
pages = {064002},
year = {2021},
doi = {10.1063/5.0037093},

url = {https://doi.org/10.1063/5.0037093}

}

@article{reed2010fast,
author = {Reed,M. D.  and Johnson,B. R.  and Houck,A. A.  and DiCarlo,L.  and Chow,J. M.  and Schuster,D. I.  and Frunzio,L.  and Schoelkopf,R. J. },
title = {Fast reset and suppressing spontaneous emission of a superconducting qubit},
journal = {Applied Physics Letters},
volume = {96},
number = {20},
pages = {203110},
year = {2010},
doi = {10.1063/1.3435463},

URL = { 
        https://doi.org/10.1063/1.3435463
    
}

}

@article{DRAG_leak,
  title = {Simple Pulses for Elimination of Leakage in Weakly Nonlinear Qubits},
  author = {Motzoi, F. and Gambetta, J. M. and Rebentrost, P. and Wilhelm, F. K.},
  journal = {Phys. Rev. Lett.},
  volume = {103},
  issue = {11},
  pages = {110501},
  numpages = {4},
  year = {2009},
  month = {Sep},
  publisher = {American Physical Society},
  doi = {10.1103/PhysRevLett.103.110501},
  url = {https://link.aps.org/doi/10.1103/PhysRevLett.103.110501}
}

@article{ChinaZZcl,
  title = {Scalable Method for Eliminating Residual $ZZ$ Interaction between Superconducting Qubits},
  author = {Ni, Zhongchu and Li, Sai and Zhang, Libo and Chu, Ji and Niu, Jingjing and Yan, Tongxing and Deng, Xiuhao and Hu, Ling and Li, Jian and Zhong, Youpeng and Liu, Song and Yan, Fei and Xu, Yuan and Yu, Dapeng},
  journal = {Phys. Rev. Lett.},
  volume = {129},
  issue = {4},
  pages = {040502},
  numpages = {7},
  year = {2022},
  month = {Jul},
  publisher = {American Physical Society},
  doi = {10.1103/PhysRevLett.129.040502},
  url = {https://link.aps.org/doi/10.1103/PhysRevLett.129.040502}
}

@article{NogFastCT,
  title = {Fast parametric two-qubit gates with suppressed residual interaction using the second-order nonlinearity of a cubic transmon},
  author = {Noguchi, Atsushi and Osada, Alto and Masuda, Shumpei and Kono, Shingo and Heya, Kentaro and Wolski, Samuel Piotr and Takahashi, Hiroki and Sugiyama, Takanori and Lachance-Quirion, Dany and Nakamura, Yasunobu},
  journal = {Phys. Rev. A},
  volume = {102},
  issue = {6},
  pages = {062408},
  numpages = {11},
  year = {2020},
  month = {Dec},
  publisher = {American Physical Society},
  doi = {10.1103/PhysRevA.102.062408},
  url = {https://link.aps.org/doi/10.1103/PhysRevA.102.062408}
}

@article{zz_cl_IBM_reso,
  title = {Demonstration of a High-Fidelity cnot Gate for Fixed-Frequency Transmons with Engineered $ZZ$ Suppression},
  author = {Kandala, A. and Wei, K. X. and Srinivasan, S. and Magesan, E. and Carnevale, S. and Keefe, G. A. and Klaus, D. and Dial, O. and McKay, D. C.},
  journal = {Phys. Rev. Lett.},
  volume = {127},
  issue = {13},
  pages = {130501},
  numpages = {6},
  year = {2021},
  month = {Sep},
  publisher = {American Physical Society},
  doi = {10.1103/PhysRevLett.127.130501},
  url = {https://link.aps.org/doi/10.1103/PhysRevLett.127.130501}
}

@article{zz_cl_Princeton,
  title = {Suppression of Qubit Crosstalk in a Tunable Coupling Superconducting Circuit},
  author = {Mundada, Pranav and Zhang, Gengyan and Hazard, Thomas and Houck, Andrew},
  journal = {Phys. Rev. Applied},
  volume = {12},
  issue = {5},
  pages = {054023},
  numpages = {10},
  year = {2019},
  month = {Nov},
  publisher = {American Physical Society},
  doi = {10.1103/PhysRevApplied.12.054023},
  url = {https://link.aps.org/doi/10.1103/PhysRevApplied.12.054023}
}

@article{Brown2016,
author={Brown, Kenneth R.
and Kim, Jungsang
and Monroe, Christopher},
title={Co-designing a scalable quantum computer with trapped atomic ions},
journal={npj Quantum Information},
year={2016},
month={Nov},
day={08},
volume={2},
number={1},
pages={16034},
abstract={The first generation of quantum computers are on the horizon, fabricated from quantum hardware platforms that may soon be able to tackle certain tasks that cannot be performed or modelled with conventional computers. These quantum devices will not likely be universal or fully programmable, but special-purpose processors whose hardware will be tightly co-designed with particular target applications. Trapped atomic ions are a leading platform for first-generation quantum computers, but they are also fundamentally scalable to more powerful general purpose devices in future generations. This is because trapped ion qubits are atomic clock standards that can be made identical to a part in 1015, and their quantum circuit connectivity can be reconfigured through the use of external fields, without modifying the arrangement or architecture of the qubits themselves. In this forward-looking overview, we show how a modular quantum computer with thousands or more qubits can be engineered from ion crystals, and how the linkage between ion trap qubits might be tailored to a variety of applications and quantum-computing protocols.},
issn={2056-6387},
doi={10.1038/npjqi.2016.34},
url={https://doi.org/10.1038/npjqi.2016.34}
}

@ARTICLE{PR_IBM_Analytical,
  author={Murray, Conal E. and Gambetta, Jay M. and McClure, Douglas T. and Steffen, Matthias},
  journal={IEEE Transactions on Microwave Theory and Techniques}, 
  title={Analytical Determination of Participation in Superconducting Coplanar Architectures}, 
  year={2018},
  volume={66},
  number={8},
  pages={3724-3733},
  doi={10.1109/TMTT.2018.2841829}}

@article{magesan2012efficient,
  title = {Efficient Measurement of Quantum Gate Error by Interleaved Randomized Benchmarking},
  author = {Magesan, Easwar and Gambetta, Jay M. and Johnson, B. R. and Ryan, Colm A. and Chow, Jerry M. and Merkel, Seth T. and da Silva, Marcus P. and Keefe, George A. and Rothwell, Mary B. and Ohki, Thomas A. and Ketchen, Mark B. and Steffen, M.},
  journal = {Phys. Rev. Lett.},
  volume = {109},
  issue = {8},
  pages = {080505},
  numpages = {5},
  year = {2012},
  month = {Aug},
  publisher = {American Physical Society},
  doi = {10.1103/PhysRevLett.109.080505},
  url = {https://link.aps.org/doi/10.1103/PhysRevLett.109.080505}
}

@article{mckay2017efficient,
  title = {Efficient $Z$ gates for quantum computing},
  author = {McKay, David C. and Wood, Christopher J. and Sheldon, Sarah and Chow, Jerry M. and Gambetta, Jay M.},
  journal = {Phys. Rev. A},
  volume = {96},
  issue = {2},
  pages = {022330},
  numpages = {8},
  year = {2017},
  month = {Aug},
  publisher = {American Physical Society},
  doi = {10.1103/PhysRevA.96.022330},
  url = {https://link.aps.org/doi/10.1103/PhysRevA.96.022330}
}

@article{AGF_leak,
  title = {Quantification and characterization of leakage errors},
  author = {Wood, Christopher J. and Gambetta, Jay M.},
  journal = {Phys. Rev. A},
  volume = {97},
  issue = {3},
  pages = {032306},
  numpages = {17},
  year = {2018},
  month = {Mar},
  publisher = {American Physical Society},
  doi = {10.1103/PhysRevA.97.032306},
  url = {https://link.aps.org/doi/10.1103/PhysRevA.97.032306}
}

@article{siZZle_2021Mitchell,
  title = {Hardware-Efficient Microwave-Activated Tunable Coupling between Superconducting Qubits},
  author = {Mitchell, Bradley K. and Naik, Ravi K. and Morvan, Alexis and Hashim, Akel and Kreikebaum, John Mark and Marinelli, Brian and Lavrijsen, Wim and Nowrouzi, Kasra and Santiago, David I. and Siddiqi, Irfan},
  journal = {Phys. Rev. Lett.},
  volume = {127},
  issue = {20},
  pages = {200502},
  numpages = {6},
  year = {2021},
  month = {Nov},
  publisher = {American Physical Society},
  doi = {10.1103/PhysRevLett.127.200502},
  url = {https://link.aps.org/doi/10.1103/PhysRevLett.127.200502}
}

@article{CR_procedure,
  title = {Procedure for systematically tuning up cross-talk in the cross-resonance gate},
  author = {Sheldon, Sarah and Magesan, Easwar and Chow, Jerry M. and Gambetta, Jay M.},
  journal = {Phys. Rev. A},
  volume = {93},
  issue = {6},
  pages = {060302(R)},
  numpages = {5},
  year = {2016},
  month = {Jun},
  publisher = {American Physical Society},
  doi = {10.1103/PhysRevA.93.060302},
  url = {https://link.aps.org/doi/10.1103/PhysRevA.93.060302}
}

@article{CR_1,
  title = {Simple All-Microwave Entangling Gate for Fixed-Frequency Superconducting Qubits},
  author = {Chow, Jerry M. and C\'orcoles, A. D. and Gambetta, Jay M. and Rigetti, Chad and Johnson, B. R. and Smolin, John A. and Rozen, J. R. and Keefe, George A. and Rothwell, Mary B. and Ketchen, Mark B. and Steffen, M.},
  journal = {Phys. Rev. Lett.},
  volume = {107},
  issue = {8},
  pages = {080502},
  numpages = {5},
  year = {2011},
  month = {Aug},
  publisher = {American Physical Society},
  doi = {10.1103/PhysRevLett.107.080502},
  url = {https://link.aps.org/doi/10.1103/PhysRevLett.107.080502}
}

@article{opt_f_crowding,
  title = {Optimizing frequency allocation for fixed-frequency superconducting quantum processors},
  author = {Morvan, Alexis and Chen, Larry and Larson, Jeffrey M. and Santiago, David I. and Siddiqi, Irfan},
  journal = {Phys. Rev. Research},
  volume = {4},
  issue = {2},
  pages = {023079},
  numpages = {6},
  year = {2022},
  month = {Apr},
  publisher = {American Physical Society},
  doi = {10.1103/PhysRevResearch.4.023079},
  url = {https://link.aps.org/doi/10.1103/PhysRevResearch.4.023079}
}
@article{FogiCZ_2020krinner,
  title = {Demonstration of an All-Microwave Controlled-Phase Gate between Far-Detuned Qubits},
  author = {Krinner, S. and Kurpiers, P. and Royer, B. and Magnard, P. and Tsitsilin, I. and Besse, J.-C. and Remm, A. and Blais, A. and Wallraff, A.},
  journal = {Phys. Rev. Applied},
  volume = {14},
  issue = {4},
  pages = {044039},
  numpages = {11},
  year = {2020},
  month = {Oct},
  publisher = {American Physical Society},
  doi = {10.1103/PhysRevApplied.14.044039},
  url = {https://link.aps.org/doi/10.1103/PhysRevApplied.14.044039}
}

@article{CR_HF,
  title = {Demonstration of a High-Fidelity cnot Gate for Fixed-Frequency Transmons with Engineered $ZZ$ Suppression},
  author = {Kandala, A. and Wei, K. X. and Srinivasan, S. and Magesan, E. and Carnevale, S. and Keefe, G. A. and Klaus, D. and Dial, O. and McKay, D. C.},
  journal = {Phys. Rev. Lett.},
  volume = {127},
  issue = {13},
  pages = {130501},
  numpages = {6},
  year = {2021},
  month = {Sep},
  publisher = {American Physical Society},
  doi = {10.1103/PhysRevLett.127.130501},
  url = {https://link.aps.org/doi/10.1103/PhysRevLett.127.130501}
}

@article{PhysRevA.101.052308,
  title = {Effective Hamiltonian models of the cross-resonance gate},
  author = {Magesan, Easwar and Gambetta, Jay M.},
  journal = {Phys. Rev. A},
  volume = {101},
  issue = {5},
  pages = {052308},
  numpages = {15},
  year = {2020},
  month = {May},
  publisher = {American Physical Society},
  doi = {10.1103/PhysRevA.101.052308},
  url = {https://link.aps.org/doi/10.1103/PhysRevA.101.052308}
}

@article{Kjaergaard_review,
author = {Kjaergaard, Morten and Schwartz, Mollie E. and Braum\"{u}ller, Jochen and Krantz, Philip and Wang, Joel I.-J. and Gustavsson, Simon and Oliver, William D.},
title = {Superconducting Qubits: Current State of Play},
journal = {Annual Review of Condensed Matter Physics},
volume = {11},
number = {1},
pages = {369-395},
year = {2020},
doi = {10.1146/annurev-conmatphys-031119-050605},

url = { 
    
        https://doi.org/10.1146/annurev-conmatphys-031119-050605

}
,
    abstract = { Superconducting qubits are leading candidates in the race to build a quantum computer capable of realizing computations beyond the reach of modern supercomputers. The superconducting qubit modality has been used to demonstrate prototype algorithms in the noisy intermediate-scale quantum (NISQ) technology era, in which non-error-corrected qubits are used to implement quantum simulations and quantum algorithms. With the recent demonstrations of multiple high-fidelity, two-qubit gates as well as operations on logical qubits in extensible superconducting qubit systems, this modality also holds promise for the longer-term goal of building larger-scale error-corrected quantum computers. In this brief review, we discuss several of the recent experimental advances in qubit hardware, gate implementations, readout capabilities, early NISQ algorithm implementations, and quantum error correction using superconducting qubits. Although continued work on many aspects of this technology is certainly necessary, the pace of both conceptual and technical progress in recent years has been impressive, and here we hope to convey the excitement stemming from this progress. }
}

@article{PhysRevA.67.042308,
  title = {Tackling systematic errors in quantum logic gates with composite rotations},
  author = {Cummins, Holly K. and Llewellyn, Gavin and Jones, Jonathan A.},
  journal = {Phys. Rev. A},
  volume = {67},
  issue = {4},
  pages = {042308},
  numpages = {7},
  year = {2003},
  month = {Apr},
  publisher = {American Physical Society},
  doi = {10.1103/PhysRevA.67.042308},
  url = {https://link.aps.org/doi/10.1103/PhysRevA.67.042308}
}

@article{Cummins_2000,
doi = {10.1088/1367-2630/2/1/006},
url = {https://dx.doi.org/10.1088/1367-2630/2/1/006},
year = {2000},
month = {mar},
publisher = {},
volume = {2},
number = {1},
pages = {6},
author = {H K Cummins and J A Jones},
title = {Use 
of composite rotations to 
correct systematic errors in 
NMR quantum computation},
journal = {New Journal of Physics},
abstract = {We implement an ensemble quantum counting algorithm on three NMR spectrometers with 1 H resonance frequencies of 500, 600 and 750 MHz. At higher frequencies, the results deviate markedly from naive theoretical predictions. These systematic errors can be attributed almost entirely to off-resonance effects, which can be substantially corrected for using fully compensating composite rotation pulse sequences originally developed by Tycko. We also derive an analytic expression for generating such sequences with arbitrary rotation angles.}
}

@article{bSWAP_2012Poletto,
  title = {Entanglement of Two Superconducting Qubits in a Waveguide Cavity via Monochromatic Two-Photon Excitation},
  author = {Poletto, S. and Gambetta, Jay M. and Merkel, Seth T. and Smolin, John A. and Chow, Jerry M. and C\'orcoles, A. D. and Keefe, George A. and Rothwell, Mary B. and Rozen, J. R. and Abraham, D. W. and Rigetti, Chad and Steffen, M.},
  journal = {Phys. Rev. Lett.},
  volume = {109},
  issue = {24},
  pages = {240505},
  numpages = {5},
  year = {2012},
  month = {Dec},
  publisher = {American Physical Society},
  doi = {10.1103/PhysRevLett.109.240505},
  url = {https://link.aps.org/doi/10.1103/PhysRevLett.109.240505}
}

@article{MAP_2013Chow,
doi = {10.1088/1367-2630/15/11/115012},
url = {https://dx.doi.org/10.1088/1367-2630/15/11/115012},
year = {2013},
month = {nov},
publisher = {IOP Publishing},
volume = {15},
number = {11},
pages = {115012},
author = {Jerry M Chow and Jay M Gambetta and Andrew W Cross and Seth T Merkel and Chad Rigetti and M Steffen},
title = {Microwave-activated conditional-phase gate for superconducting qubits},
journal = {New Journal of Physics},
abstract = {We introduce a new entangling gate between two fixed-frequency qubits statically coupled via a microwave resonator bus which combines the following desirable qualities: all-microwave control, appreciable qubit separation for reduction of crosstalk and leakage errors and the ability to function as a two-qubit conditional-phase gate. A fixed, always-on interaction is explicitly designed between higher energy (non-computational) states of two transmon qubits, and then a conditional-phase gate is ‘activated’ on the otherwise unperturbed qubit subspace via a microwave drive. We implement this microwave-activated conditional-phase gate with a fidelity from quantum process tomography of  ∼ 87
}
}

@article{RIP_2016Paik,
  title = {Experimental Demonstration of a Resonator-Induced Phase Gate in a Multiqubit Circuit-QED System},
  author = {Paik, Hanhee and Mezzacapo, A. and Sandberg, Martin and McClure, D. T. and Abdo, B. and C\'orcoles, A. D. and Dial, O. and Bogorin, D. F. and Plourde, B. L. T. and Steffen, M. and Cross, A. W. and Gambetta, J. M. and Chow, Jerry M.},
  journal = {Phys. Rev. Lett.},
  volume = {117},
  issue = {25},
  pages = {250502},
  numpages = {5},
  year = {2016},
  month = {Dec},
  publisher = {American Physical Society},
  doi = {10.1103/PhysRevLett.117.250502},
  url = {https://link.aps.org/doi/10.1103/PhysRevLett.117.250502}
}

@article{CR_2010Rigetti,
  title = {Fully microwave-tunable universal gates in superconducting qubits with linear couplings and fixed transition frequencies},
  author = {Rigetti, Chad and Devoret, Michel},
  journal = {Phys. Rev. B},
  volume = {81},
  issue = {13},
  pages = {134507},
  numpages = {7},
  year = {2010},
  month = {Apr},
  publisher = {American Physical Society},
  doi = {10.1103/PhysRevB.81.134507},
  url = {https://link.aps.org/doi/10.1103/PhysRevB.81.134507}
}

@article{GARBOW1982504,
title = {Bilinear rotation decoupling of homonuclear scalar interactions},
journal = {Chemical Physics Letters},
volume = {93},
number = {5},
pages = {504-509},
year = {1982},
issn = {0009-2614},
doi = {https://doi.org/10.1016/0009-2614(82)83229-6},
url = {https://www.sciencedirect.com/science/article/pii/0009261482832296},
author = {J.R. Garbow and D.P. Weitekamp and A. Pines},
abstract = {A method for obtaining NMR spectra of organic liquids free of J-couplings is described. The coupling between a group of equivalent protons and an adjacent 13C nets as a local decoupling field for the protons. A Pure chemical shift spectrum of ethanol is presented and possible applications to strongly coupled systems are discussed.}
}

@article{Koch2007,
  title = {Charge-insensitive qubit design derived from the Cooper pair box},
  author = {Koch, Jens and Yu, Terri M. and Gambetta, Jay and Houck, A. A. and Schuster, D. I. and Majer, J. and Blais, Alexandre and Devoret, M. H. and Girvin, S. M. and Schoelkopf, R. J.},
  journal = {Phys. Rev. A},
  volume = {76},
  issue = {4},
  pages = {042319},
  numpages = {19},
  year = {2007},
  month = {Oct},
  publisher = {American Physical Society},
  doi = {10.1103/PhysRevA.76.042319},
  url = {https://link.aps.org/doi/10.1103/PhysRevA.76.042319}
}

@article{
doi:10.1126/sciadv.abi6690,
author = {Eric J. Zhang  and Srikanth Srinivasan  and Neereja Sundaresan  and Daniela F. Bogorin  and Yves Martin  and Jared B. Hertzberg  and John Timmerwilke  and Emily J. Pritchett  and Jeng-Bang Yau  and Cindy Wang  and William Landers  and Eric P. Lewandowski  and Adinath Narasgond  and Sami Rosenblatt  and George A. Keefe  and Isaac Lauer  and Mary Beth Rothwell  and Douglas T. McClure  and Oliver E. Dial  and Jason S. Orcutt  and Markus Brink  and Jerry M. Chow },
title = {High-performance superconducting quantum processors via laser annealing of transmon qubits},
journal = {Science Advances},
volume = {8},
number = {19},
pages = {eabi6690},
year = {2022},
doi = {10.1126/sciadv.abi6690},
URL = {https://www.science.org/doi/abs/10.1126/sciadv.abi6690},
abstract = {Scaling the number of qubits while maintaining high-fidelity quantum gates remains a key challenge for quantum computing. Presently, superconducting quantum processors with \&gt;50 qubits are actively available. For these systems, fixed-frequency transmons are attractive because of their long coherence and noise immunity. However, scaling fixed-frequency architectures proves challenging because of precise relative frequency requirements. Here, we use laser annealing to selectively tune transmon qubits into desired frequency patterns. Statistics over hundreds of annealed qubits demonstrate an empirical tuning precision of 18.5 MHz, with no measurable impact on qubit coherence. We quantify gate error statistics on a tuned 65-qubit processor, with median two-qubit gate fidelity of 98.7\%. Baseline tuning statistics yield a frequency-equivalent resistance precision of 4.7 MHz, sufficient for high-yield scaling beyond 103 qubit levels. Moving forward, we anticipate selective laser annealing to play a central role in scaling fixed-frequency architectures. Laser-based thermal annealing is demonstrated for high-yield scaling of high-performance superconducting quantum processors.}}

@article{doi:10.1063/5.0102092,
author = {Kim,Hyunseong  and Jünger,Christian  and Morvan,Alexis  and Barnard,Edward S.  and Livingston,William P.  and Altoé,M. Virginia P.  and Kim,Yosep  and Song,Chengyu  and Chen,Larry  and Kreikebaum,John Mark  and Ogletree,D. Frank  and Santiago,David I.  and Siddiqi,Irfan },
title = {Effects of laser-annealing on fixed-frequency superconducting qubits},
journal = {Applied Physics Letters},
volume = {121},
number = {14},
pages = {142601},
year = {2022},
doi = {10.1063/5.0102092},

URL = { 
        https://doi.org/10.1063/5.0102092
    
}

}

@article{JOHANSSON20121760,
title = {QuTiP: An open-source Python framework for the dynamics of open quantum systems},
journal = {Computer Physics Communications},
volume = {183},
number = {8},
pages = {1760-1772},
year = {2012},
issn = {0010-4655},
doi = {https://doi.org/10.1016/j.cpc.2012.02.021},
url = {https://www.sciencedirect.com/science/article/pii/S0010465512000835},
author = {J.R. Johansson and P.D. Nation and Franco Nori},
keywords = {Open quantum systems, Lindblad master equation, Quantum Monte Carlo, Python},
abstract = {We present an object-oriented open-source framework for solving the dynamics of open quantum systems written in Python. Arbitrary Hamiltonians, including time-dependent systems, may be built up from operators and states defined by a quantum object class, and then passed on to a choice of master equation or Monte Carlo solvers. We give an overview of the basic structure for the framework before detailing the numerical simulation of open system dynamics. Several examples are given to illustrate the build up to a complete calculation. Finally, we measure the performance of our library against that of current implementations. The framework described here is particularly well suited to the fields of quantum optics, superconducting circuit devices, nanomechanics, and trapped ions, while also being ideal for use in classroom instruction.
Program summary
Program title: QuTiP: The Quantum Toolbox in Python Catalogue identifier: AEMB_v1_0 Program summary URL: http://cpc.cs.qub.ac.uk/summaries/AEMB_v1_0.html Program obtainable from: CPC Program Library, Queenʼs University, Belfast, N. Ireland Licensing provisions: GNU General Public License, version 3 No. of lines in distributed program, including test data, etc.: 16 482 No. of bytes in distributed program, including test data, etc.: 213 438 Distribution format: tar.gz Programming language: Python Computer: i386, x86-64 Operating system: Linux, Mac OSX, Windows RAM: 2+ Gigabytes Classification: 7 External routines: NumPy (http://numpy.scipy.org/), SciPy (http://www.scipy.org/), Matplotlib (http://matplotlib.sourceforge.net/) Nature of problem: Dynamics of open quantum systems. Solution method: Numerical solutions to Lindblad master equation or Monte Carlo wave function method. Restrictions: Problems must meet the criteria for using the master equation in Lindblad form. Running time: A few seconds up to several tens of minutes, depending on size of underlying Hilbert space.}
}

@article{JOHANSSON20131234,
title = {QuTiP 2: A Python framework for the dynamics of open quantum systems},
journal = {Computer Physics Communications},
volume = {184},
number = {4},
pages = {1234-1240},
year = {2013},
issn = {0010-4655},
doi = {https://doi.org/10.1016/j.cpc.2012.11.019},
url = {https://www.sciencedirect.com/science/article/pii/S0010465512003955},
author = {J.R. Johansson and P.D. Nation and Franco Nori},
keywords = {Open quantum systems, Lindblad, Bloch–Redfield, Floquet–Markov, Master equation, Quantum Monte Carlo, Python},
abstract = {We present version 2 of QuTiP, the Quantum Toolbox in Python. Compared to the preceding version [J.R. Johansson, P.D. Nation, F. Nori, Comput. Phys. Commun. 183 (2012) 1760.], we have introduced numerous new features, enhanced performance, and made changes in the Application Programming Interface (API) for improved functionality and consistency within the package, as well as increased compatibility with existing conventions used in other scientific software packages for Python. The most significant new features include efficient solvers for arbitrary time-dependent Hamiltonians and collapse operators, support for the Floquet formalism, and new solvers for Bloch–Redfield and Floquet–Markov master equations. Here we introduce these new features, demonstrate their use, and give a summary of the important backward-incompatible API changes introduced in this version.
Program Summary
Program title: QuTiP: The Quantum Toolbox in Python Catalog identifier: AEMB_v2_0 Program summary URL:http://cpc.cs.qub.ac.uk/summaries/AEMB_v2_0.html Program obtainable from: CPC Program Library, Queen’s University, Belfast, N. Ireland Licensing provisions: GNU General Public License, version 3 No. of lines in distributed program, including test data, etc.: 33625 No. of bytes in distributed program, including test data, etc.: 410064 Distribution format: tar.gz Programming language: Python. Computer: i386, x86-64. Operating system: Linux, Mac OSX. RAM: 2+ Gigabytes Classification: 7. External routines: NumPy, SciPy, Matplotlib, Cython Catalog identifier of previous version: AEMB_v1_0 Journal reference of previous version: Comput. Phys. Comm. 183 (2012) 1760 Does the new version supercede the previous version?: Yes Nature of problem: Dynamics of open quantum systems Solution method: Numerical solutions to Lindblad, Floquet–Markov, and Bloch–Redfield master equations, as well as the Monte Carlo wave function method. Reasons for new version: Compared to the preceding version we have introduced numerous new features, enhanced performance, and made changes in the Application Programming Interface (API) for improved functionality and consistency within the package, as well as increased compatibility with existing conventions used in other scientific software packages for Python. The most significant new features include efficient solvers for arbitrary time-dependent Hamiltonians and collapse operators, support for the Floquet formalism, and new solvers for Bloch–Redfield and Floquet–Markov master equations. Restrictions: Problems must meet the criteria for using the master equation in Lindblad, Floquet–Markov, or Bloch–Redfield form. Running time: A few seconds up to several tens of hours, depending on size of the underlying Hilbert space.}
}

@article{PRXQuantum.2.040336,
  title = {Cross-Cross Resonance Gate},
  author = {Heya, Kentaro and Kanazawa, Naoki},
  journal = {PRX Quantum},
  volume = {2},
  issue = {4},
  pages = {040336},
  numpages = {15},
  year = {2021},
  month = {Nov},
  publisher = {American Physical Society},
  doi = {10.1103/PRXQuantum.2.040336},
  url = {https://link.aps.org/doi/10.1103/PRXQuantum.2.040336}
}

@article{SuppressZZ_TQ_CSFQ,
  title = {Suppression of Unwanted $ZZ$ Interactions in a Hybrid Two-Qubit System},
  author = {Ku, Jaseung and Xu, Xuexin and Brink, Markus and McKay, David C. and Hertzberg, Jared B. and Ansari, Mohammad H. and Plourde, B. L. T.},
  journal = {Phys. Rev. Lett.},
  volume = {125},
  issue = {20},
  pages = {200504},
  numpages = {6},
  year = {2020},
  month = {Nov},
  publisher = {American Physical Society},
  doi = {10.1103/PhysRevLett.125.200504},
  url = {https://link.aps.org/doi/10.1103/PhysRevLett.125.200504}
}

@article{PhysRevA.91.043846,
  title = {Microwave-induced amplitude- and phase-tunable qubit-resonator coupling in circuit quantum electrodynamics},
  author = {Zeytino\ifmmode \breve{g}\else \u{g}\fi{}lu, S. and Pechal, M. and Berger, S. and Abdumalikov, A. A. and Wallraff, A. and Filipp, S.},
  journal = {Phys. Rev. A},
  volume = {91},
  issue = {4},
  pages = {043846},
  numpages = {9},
  year = {2015},
  month = {Apr},
  publisher = {American Physical Society},
  doi = {10.1103/PhysRevA.91.043846},
  url = {https://link.aps.org/doi/10.1103/PhysRevA.91.043846}
}

@article{PhysRevLett.94.123602,
  title = {ac Stark Shift and Dephasing of a Superconducting Qubit Strongly Coupled to a Cavity Field},
  author = {Schuster, D. I. and Wallraff, A. and Blais, A. and Frunzio, L. and Huang, R.-S. and Majer, J. and Girvin, S. M. and Schoelkopf, R. J.},
  journal = {Phys. Rev. Lett.},
  volume = {94},
  issue = {12},
  pages = {123602},
  numpages = {4},
  year = {2005},
  month = {Mar},
  publisher = {American Physical Society},
  doi = {10.1103/PhysRevLett.94.123602},
  url = {https://link.aps.org/doi/10.1103/PhysRevLett.94.123602}
}

@article{PhysRevA.97.062334,
  title = {Local sensing with the multilevel ac Stark effect},
  author = {Schneider, Andre and Braum\"uller, Jochen and Guo, Lingzhen and Stehle, Patrizia and Rotzinger, Hannes and Marthaler, Michael and Ustinov, Alexey V. and Weides, Martin},
  journal = {Phys. Rev. A},
  volume = {97},
  issue = {6},
  pages = {062334},
  numpages = {8},
  year = {2018},
  month = {Jun},
  publisher = {American Physical Society},
  doi = {10.1103/PhysRevA.97.062334},
  url = {https://link.aps.org/doi/10.1103/PhysRevA.97.062334}
}

@book{winkler2003spin,
      author        = "Winkler, Roland",
      title         = "{Spin-Orbit Coupling Effects in Two-Dimensional Electron and Hole Systems}",
      publisher     = "Springer",
      address       = "Berlin",
      series        = "Tracts in Modern Physics",
      year          = "2003",
      url           = "https://link.springer.com/book/10.1007/b13586#book-header",
      doi           = "10.1007/b13586",
}

@article{PRXQuantum.2.040324,
  title = {Experimental Deep Reinforcement Learning for Error-Robust Gate-Set Design on a Superconducting Quantum Computer},
  author = {Baum, Yuval and Amico, Mirko and Howell, Sean and Hush, Michael and Liuzzi, Maggie and Mundada, Pranav and Merkh, Thomas and Carvalho, Andre R. R. and Biercuk, Michael J.},
  journal = {PRX Quantum},
  volume = {2},
  issue = {4},
  pages = {040324},
  numpages = {12},
  year = {2021},
  month = {Nov},
  publisher = {American Physical Society},
  doi = {10.1103/PRXQuantum.2.040324},
  url = {https://link.aps.org/doi/10.1103/PRXQuantum.2.040324}
}

@article{KHANEJA2005296,
title = {Optimal control of coupled spin dynamics: design of NMR pulse sequences by gradient ascent algorithms},
journal = {Journal of Magnetic Resonance},
volume = {172},
number = {2},
pages = {296-305},
year = {2005},
issn = {1090-7807},
doi = {https://doi.org/10.1016/j.jmr.2004.11.004},
url = {https://www.sciencedirect.com/science/article/pii/S1090780704003696},
author = {Navin Khaneja and Timo Reiss and Cindie Kehlet and Thomas Schulte-Herbrüggen and Steffen J. Glaser},
keywords = {Pulse design, Sequence optimization, Time-optimal coherence transfer, Relaxation-optimized experiments, Time-optimal realization of unitary operators, Quantum gates, GRAPE algorithm, Optimal control theory},
abstract = {In this paper, we introduce optimal control algorithm for the design of pulse sequences in NMR spectroscopy. This methodology is used for designing pulse sequences that maximize the coherence transfer between coupled spins in a given specified time, minimize the relaxation effects in a given coherence transfer step or minimize the time required to produce a given unitary propagator, as desired. The application of these pulse engineering methods to design pulse sequences that are robust to experimentally important parameter variations, such as chemical shift dispersion or radiofrequency (rf) variations due to imperfections such as rf inhomogeneity is also explained.}
}

@article{doi:10.1063/1.5089550,
author = {Krantz,P.  and Kjaergaard,M.  and Yan,F.  and Orlando,T. P.  and Gustavsson,S.  and Oliver,W. D. },
title = {A quantum engineer's guide to superconducting qubits},
journal = {Applied Physics Reviews},
volume = {6},
number = {2},
pages = {021318},
year = {2019},
doi = {10.1063/1.5089550},

URL = { 
        https://doi.org/10.1063/1.5089550
}

}

﻿@Article{Chen2021,
author={Chen, Zijun
and Satzinger, Kevin J.
and Atalaya, Juan
and Korotkov, Alexander N.
and Dunsworth, Andrew
and Sank, Daniel
and Quintana, Chris
and McEwen, Matt
and Barends, Rami
and Klimov, Paul V.
and Hong, Sabrina
and Jones, Cody
and Petukhov, Andre
and Kafri, Dvir
and Demura, Sean
and Burkett, Brian
and Gidney, Craig
and Fowler, Austin G.
and Paler, Alexandru
and Putterman, Harald
and Aleiner, Igor
and Arute, Frank
and Arya, Kunal
and Babbush, Ryan
and Bardin, Joseph C.
and Bengtsson, Andreas
and Bourassa, Alexandre
and Broughton, Michael
and Buckley, Bob B.
and Buell, David A.
and Bushnell, Nicholas
and Chiaro, Benjamin
and Collins, Roberto
and Courtney, William
and Derk, Alan R.
and Eppens, Daniel
and Erickson, Catherine
and Farhi, Edward
and Foxen, Brooks
and Giustina, Marissa
and Greene, Ami
and Gross, Jonathan A.
and Harrigan, Matthew P.
and Harrington, Sean D.
and Hilton, Jeremy
and Ho, Alan
and Huang, Trent
and Huggins, William J.
and Ioffe, L. B.
and Isakov, Sergei V.
and Jeffrey, Evan
and Jiang, Zhang
and Kechedzhi, Kostyantyn
and Kim, Seon
and Kitaev, Alexei
and Kostritsa, Fedor
and Landhuis, David
and Laptev, Pavel
and Lucero, Erik
and Martin, Orion
and McClean, Jarrod R.
and McCourt, Trevor
and Mi, Xiao
and Miao, Kevin C.
and Mohseni, Masoud
and Montazeri, Shirin
and Mruczkiewicz, Wojciech
and Mutus, Josh
and Naaman, Ofer
and Neeley, Matthew
and Neill, Charles
and Newman, Michael
and Niu, Murphy Yuezhen
and O'Brien, Thomas E.
and Opremcak, Alex
and Ostby, Eric
and Pat{\'o}, B{\'a}lint
and Redd, Nicholas
and Roushan, Pedram
and Rubin, Nicholas C.
and Shvarts, Vladimir
and Strain, Doug
and Szalay, Marco
and Trevithick, Matthew D.
and Villalonga, Benjamin
and White, Theodore
and Yao, Z. Jamie
and Yeh, Ping
and Yoo, Juhwan
and Zalcman, Adam
and Neven, Hartmut
and Boixo, Sergio
and Smelyanskiy, Vadim
and Chen, Yu
and Megrant, Anthony
and Kelly, Julian
and AI, Google Quantum},
title={Exponential suppression of bit or phase errors with cyclic error correction},
journal={Nature},
year={2021},
month={Jul},
day={01},
volume={595},
number={7867},
pages={383-387},
abstract={Realizing the potential of quantum computing requires sufficiently low logical error rates1. Many applications call for error rates as low as 10−15 (refs. 2--9), but state-of-the-art quantum platforms typically have physical error rates near 10−3 (refs. 10--14). Quantum error correction15--17 promises to bridge this divide by distributing quantum logical information across many physical qubits in such a way that errors can be detected and corrected. Errors on the encoded logical qubit state can be exponentially suppressed as the number of physical qubits grows, provided that the physical error rates are below a certain threshold and stable over the course of a computation. Here we implement one-dimensional repetition codes embedded in a two-dimensional grid of superconducting qubits that demonstrate exponential suppression of bit-flip or phase-flip errors, reducing logical error per round more than 100-fold when increasing the number of qubits from 5 to 21. Crucially, this error suppression is stable over 50 rounds of error correction. We also introduce a method for analysing error correlations with high precision, allowing us to characterize error locality while performing quantum error correction. Finally, we perform error detection with a small logical qubit using the 2D surface code on the same device18,19 and show that the results from both one- and two-dimensional codes agree with numerical simulations that use a simple depolarizing error model. These experimental demonstrations provide a foundation for building a scalable fault-tolerant quantum computer with superconducting qubits.},
issn={1476-4687},
doi={10.1038/s41586-021-03588-y},
url={https://doi.org/10.1038/s41586-021-03588-y}
}

@article{Erik_GP,
author = {Sjöqvist, Erik},
title = {Geometric phases in quantum information},
journal = {International Journal of Quantum Chemistry},
volume = {115},
number = {19},
pages = {1311-1326},
keywords = {geometric phase, quantum computation, mixed quantum states, quantum entanglement},
doi = {https://doi.org/10.1002/qua.24941},
url = {https://onlinelibrary.wiley.com/doi/abs/10.1002/qua.24941},
abstract = {The rise of quantum information science has opened up a new venue for applications of the geometric phase (GP), as well as triggered new insights into its physical, mathematical, and conceptual nature. Here, we review this development by focusing on three main themes: the use of GPs to perform robust quantum computation, the development of GP concepts for mixed quantum states, and the discovery of a new type of topological phases for entangled quantum systems. We delineate the theoretical development as well as describe recent experiments related to GPs in the context of quantum information. © 2015 Wiley Periodicals, Inc.},
year = {2015}
}

@Article{Barends2014,
author={Barends, R.
and Kelly, J.
and Megrant, A.
and Veitia, A.
and Sank, D.
and Jeffrey, E.
and White, T. C.
and Mutus, J.
and Fowler, A. G.
and Campbell, B.
and Chen, Y.
and Chen, Z.
and Chiaro, B.
and Dunsworth, A.
and Neill, C.
and O'Malley, P.
and Roushan, P.
and Vainsencher, A.
and Wenner, J.
and Korotkov, A. N.
and Cleland, A. N.
and Martinis, John M.},
title={Superconducting quantum circuits at the surface code threshold for fault tolerance},
journal={Nature},
year={2014},
month={Apr},
day={01},
volume={508},
number={7497},
pages={500-503},
abstract={A universal set of logic gates in a superconducting quantum circuit is shown to have gate fidelities at the threshold for fault-tolerant quantum computing by the surface code approach, in which the quantum bits are distributed in an array of planar topology and have only nearest-neighbour couplings.},
issn={1476-4687},
doi={10.1038/nature13171},
url={https://doi.org/10.1038/nature13171}
}

@misc{Supplemental,
note = {See Supplemental Material for experimental setup, details about theoretical and numerical calculations, and auxiliary experimental results, which includes Refs. \cite{winkler2003spin, Floquet_Leskes, AGF_leak, PhysRevLett.94.123602, PhysRevA.97.062334, PhysRevA.91.043846, FogiCZ_2020krinner, PhysRevA.101.052308}.},
}

\title{
Supplemental Material for ``All-microwave manipulation of superconducting qubits with a fixed-frequency transmon coupler"
}
\maketitle
\appendix
\onecolumngrid
\beginsupplement
\section{Sample and the experimental setup}
\begin{figure*}[h]
    \centering
    \includegraphics[width=17.8cm]{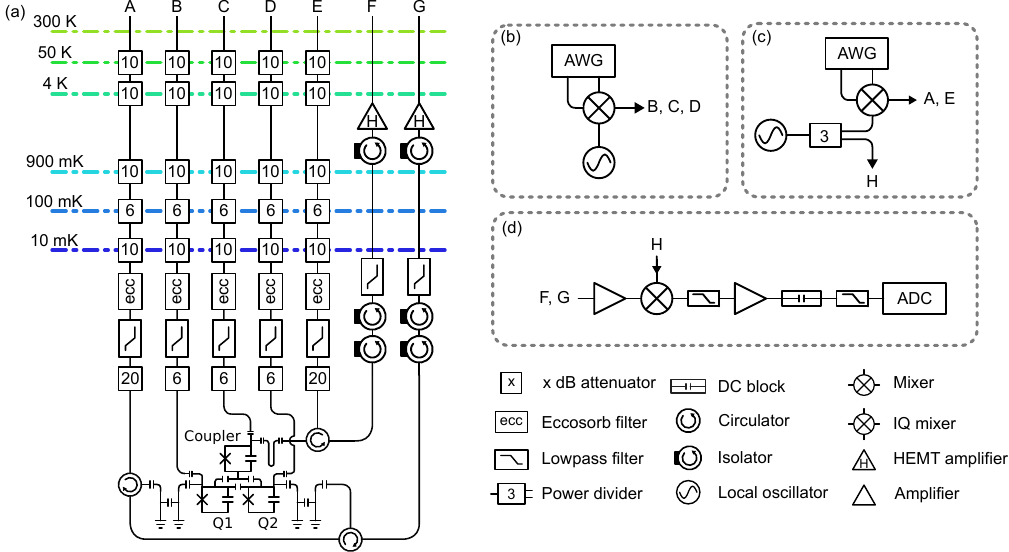}
    \caption{Experimental setup. (a) Connections from the sample chip to ports A-G at room temperature. (b) Pulse-generating systems for qubit control. (c)~Pulse-generation systems for qubit readout. (d) Readout system.}
    \label{fig:setup_paper}
\end{figure*}

The transmon qubits and the resonators are fabricated on a high-resistivity Si substrate. They are made from a sputtered and lithographically-patterned TiN film and Al/AlO$_x$/Al Josephson junctions evaporated and lifted off with the in-situ bridge-free bandage technique. As shown in Fig.~\ref{fig:setup_paper}(a), each input line of the dilution refrigerator has about 56-dB attenuation at 8\,GHz including the cable loss. Each input line also has an eccosorb filter, an 8-GHz lowpass filter, and an extra 6-dB~(20-dB) attenuator for the qubit\,(resonator) drive line. The sample is mounted inside a three-layer magnetic shield and cooled down to $\sim$10\,mK. Microwave pulses are generated by the single sideband modulation (SSB)~[Figs.\,\ref{fig:setup_paper}(b) and (c)]. The reflection pulses of the readout resonators are amplified with a low-noise HEMT amplifier at the 4-K stage and demodulated to IQ signals for the data processing~[Fig.\,\ref{fig:setup_paper}(d)]. The readout resonator frequencies are $\omega_{r}^1/2\pi\simeq7.436\,$GHz, $\omega_{r}^2/2\pi\simeq7.375\,$GHz and $\omega_{r}^c/2\pi\simeq7.551\,$GHz, respectively. The dispersive shifts of the readout resonators are less than 1\,MHz, and the qubit energy relaxation through the readout line is negligible.

\section{Derivation of equations}
As described in the main text, the system and drive Hamiltonians we consider are
\begin{align}
    \hat{H} &= \hat{H}_0 + \hat{H}_c \label{eq:S:H}\\
    \hat{H}_0/\hbar &= \sum_i \left(\omega_i\Ap_i\Am_i + \frac{\alpha_i}{2}\Ap_i\Ap_i\Am_i\Am_i\right), \\ 
    \hat{H}_c/\hbar &= \sum_{i\neq c} g_{ic}(\Ap_i \Am_c + \Am_i\Ap_c), \\
    \hat{H}_d/\hbar &= \Omega_d \cos{\omega_d t} \left(\Ap_c + \Am_c \right), \label{eq:S:Hd}
\end{align}
where $i \in \{1,2,c\}$. By following the procedure of Schrieffer-Wolff transformation \cite{winkler2003spin}, we obtain an anti-Hermitian operator $\hat{S}=\hat{S}_1+\hat{S}_2$ from the solutions of the following equations,
\begin{align}
    [\hat{H}_0,\hat{S}_1] &+ \hat{O}_1 = 0 , \label{eq:S:SWT1} \\
    [\hat{H}_0,\hat{S}_2] &+ \hat{O}_2 = 0. \label{eq:S:SWT2}
\end{align}
Here, $\hat{O}_1 = \hat{H}_c$ is considered as an off-diagonal perturbation term. We also define the diagonal and off-diagonal terms of $\frac{1}{2} \left[ \hat{O}_1, \hat{S}_1 \right] $ as $\hat{D}_2$ and $\hat{O}_2$, respectively. To derive the explicit forms of $\hat{S}_1$ and $\hat{S}_2$, we assume that each transmon is a four-level system and algebraically solve Eqs.\,\eqref{eq:S:SWT1} and \eqref{eq:S:SWT2} with a Python program. Under this setup, we first derive the CAS transition frequencies in the weak drive amplitude limit. Upon the transformation, the anti-Hermitian operators $\hat{S}_1$ and $\hat{S}_2$ cancel the off-diagonal terms in $\hat{H}$, and we obtain a diagonalized Hamiltonian valid up to the second order of $g_{ic}$,
\begin{align}
    \hat{H}' &= \hat{H}_0 + \hat{D}_2. \label{eq:S:effective_hamiltonian}
\end{align}
With this equation, we derive the analytical expressions of the CAS transition frequencies in the weak drive amplitude limit,
\begin{align}
    \omega_b' &= \bra{101}\hat{H}'\ket{101}/\hbar - \bra{010}\hat{H}'\ket{010}/\hbar, \notag\\
              &= \omega_{c} + \Delta_{12} + \frac{2 g_{1c}^{2}(\alpha_{1} + \alpha_{c})}{(\Delta_{1c}-\alpha_{c})(\Delta_{1c}+\alpha_{1})} - \frac{2 g_{2c}^{2}}{\Delta_{2c}}, \label{eq:S:freq_blue_eff}\\
    \omega_r' &= \bra{011}\hat{H}'\ket{011}/\hbar - \bra{100}\hat{H}'\ket{100}/\hbar, \notag\\
              &= \omega_{c} - \Delta_{12} + \frac{2 g_{2c}^{2}(\alpha_{2} + \alpha_{c})}{(\Delta_{2c}-\alpha_{c})(\Delta_{2c}+\alpha_{2})} - \frac{2 g_{1c}^{2}}{\Delta_{1c}}. \label{eq:S:freq_red_eff}
\end{align}
Next, we derive the effective CAS oscillation frequencies. We move to the reference frame rotating at $\omega_d$ and transform the drive Hamiltonian Eq.\,\eqref{eq:S:Hd} into
\begin{equation}
    \hat{H}_d^r/\hbar \approx \frac{1}{2} \Omega \left(\Ap_c + \Am_c \right),
    \label{eq:S:Hd_r} 
\end{equation}
where we use the rotating-wave approximation. Note that, the form of $\hat{S}$ is the same in the rotating frame. Using the obtained $\hat{S}$ and the Baker-Campbell-Hausdorff formula, we expand the drive Hamiltonian Eq.\,\eqref{eq:S:Hd_r} as
\begin{align}
    e^{\hat{S}}\hat{H}_d^re^{-\hat{S}} &= \hat{H}_d^r + \left[ \hat{S}, \hat{H}_d^r \right] + \frac{1}{2!} \left[ \hat{S}, \left[ \hat{S}, \hat{H}_d^r \right] \right] + \cdots, \notag\\
    &\approx \hat{H}_d^r + \left[ \hat{S}_1 + \hat{S}_2, \hat{H}_d^r \right] + \frac{1}{2!} \left[ \hat{S}_1, \left[ \hat{S}_1, \hat{H}_d^r \right] \right] \equiv \hat{H}_d^{'r}. \label{eq:S:Hd_r_eff}
\end{align}
In the last line of the formula, we keep only the terms up to the second order of $g_{ic}$ by assuming $|g_{ic}/\Delta_{ic}| \ll 1$. Finally, we reach the expressions of the effective CAS oscillation frequencies presented in the main text:
\begin{align}
    \Omega_b &\approx 2\bra{010}\hat{H}_d^{'r}\ket{101}/\hbar \notag \\
    &= \frac{2g_{1c} g_{2c} \alpha_c \Omega_d}{\Delta_{12}(\delta_c-\delta_1+\alpha_c)(\delta_c-\delta_2)},\notag\\
    &= \frac{2g_{1c} g_{2c} \alpha_c \Omega_d}{\Delta_{12}(\omega_c-\omega_1+\alpha_c)(\omega_c-\omega_2)},\label{eq:S:gb}\\
    \Omega_r &\approx 2\bra{100}\hat{H}_d^{'r}\ket{011}/\hbar \notag \\
    &= \frac{-2g_{1c} g_{2c} \alpha_c \Omega_d}{\Delta_{12}(\delta_c-\delta_2+\alpha_c)(\delta_c-\delta_1)}, \notag\\
    &= \frac{-2g_{1c} g_{2c} \alpha_c \Omega_d}{\Delta_{12}(\omega_c-\omega_2+\alpha_c)(\omega_c-\omega_1)}, \label{eq:S:gr}
\end{align}
where $\delta_i=\omega_i-\omega_d,$ $(i\in\{1,2,c\})$.
Moreover, we derive an analytical expression of the ac-field-dependent ZZ coupling induced by the ac Stark shift. For concreteness, we consider the case where the drive frequency $\omega_d$ is off-resonant but close to $\omega_b$. As a first step, we expand the drive Hamiltonian using $\hat{S}$ and then move to the reference frame rotating at $\omega_b$, which is determined by the drive power $\Omega_d$. Applying the rotating-wave approximation and dropping fast oscillating terms, we get the following time-dependent effective drive Hamiltonian
\begin{align}
    \hat{H}_d^r(t)/\hbar \approx \frac{\Omega_b}{2} \left( \dyad{101}{010} e^{-i\delta t} + \dyad{010}{101} e^{i\delta t} \right),
    \label{eq:S:Hdr_time}
\end{align}
where $\delta = \omega_d-\omega_b$. For further analysis, we assume a form of system Hamiltonian
\begin{align}
    \hat{H}_{\mathrm{sys}} = \hat{H}^{(0)} + \hat{H}(t),
\end{align}
where $\hat{H}^{(0)}$ is the time-independent part and $\hat{H}(t)$ is the time-periodic part. When $\hat{H}(t)$ has the characteristic frequency $\omega$, it can be expanded in a Fourier series of the form
\begin{align}
    \hat{H}(t) = \sum_{n \neq 0} \hat{H}_n e^{in\omega t}.
\end{align}
We now apply the van Vleck transformation \cite{Floquet_Leskes} and obtain the time-averaged Hamiltonian to first order
\begin{align}
    \hat{H}_{\mathrm{sys}}' \approx \hat{H}^{(0)} - \frac{1}{2} \sum_{n \neq 0} \frac{\left[ \hat{H}_{-n}, \hat{H}_n\right]}{n \hbar \omega}.
    \label{eq:S:Have_vv}
\end{align}
Comparing Eqs.\,\eqref{eq:S:Hdr_time} and the last terms of \eqref{eq:S:Have_vv}, we obtain an expression of the ac-field-tunable part of the ZZ coupling,
\begin{align}
    \xi_{\mathrm{ac}} = -\frac{\Omega_b^2}{8\delta},
\end{align}
where we assume that the coupler is in the ground state. With this term, the entire ZZ interaction under the off-resonant microwave drive can be expressed as 
\begin{align}
    \xi_{\mathrm{ZZ}}(\omega_d, \Omega_d) = \xi_0 - \frac{\Omega_b^2}{8(\omega_d - \omega_b)}, \label{eq:S:tunable_ZZ}
\end{align}
where $\xi_0=\frac{2g_\mathrm{eff}^2(\alpha_1+\alpha_2)}{(\Delta_{12}+\alpha_1)(\alpha_2-\Delta_{12})}$ is the static residual ZZ coupling that is valid up to the second order of the effective transverse coupling, $g_\mathrm{eff}=\frac{g_{1c}g_{2c}}{2}\left( \frac{1}{\Delta_{1c}} + \frac{1}{\Delta_{2c}} \right) + g_{12}$, between the data qubits.

\section{Numerical simulation method}
As mentioned in the main text, we consider up to the third excited state of each transmon for numerical calculations. To evaluate the CAS oscillation frequency, we numerically diagonalize the Hamiltonian represented in the reference frame rotating at $\omega_d$, 
\begin{align}
    \hat{H}_r &= \hat{H}_0^r + \hat{H}_c^r + \hat{H}_d^r, \label{eq:S:Hr_num}\\
    \hat{H}_0^r/\hbar &= \sum_i \left(\delta_i\Ap_i\Am_i + \frac{\alpha_i}{2}\Ap_i\Ap_i\Am_i\Am_i\right), \\ 
    \hat{H}_c^r/\hbar &= \sum_{i\neq c} g_{ic}(\Ap_i \Am_c + \Am_i\Ap_c) + g_{12}(\Ap_1 \Am_2 + \Am_1\Ap_2), \\
    \hat{H}_d^r/\hbar &\approx \frac{1}{2} \Omega_d \left(\Ap_c + \Am_c \right). \label{eq:S:Hd_at_rot}
\end{align}
Here we take into account the direct coupling $g_{12}$ between the data transmons. For each drive amplitude $\Omega_d$, we sweep the drive frequency $\omega_d$ and obtain the resonant CAS oscillation frequencies $\Omega_b$ ($\Omega_r$) as the splitting at the anticrossing between the states $\ket{010} \mathrm{and} \ket{101}$ $\left(\ket{100} \mathrm{and} \ket{011}\right)$.

Next, we estimate the coherence limit of the average fidelity of the CAS-based CZ gate. We use Eq.\,\eqref{eq:S:Hr_num} as the starting point and numerically simulate the JAZZ sequence in Fig.~3(a) in the main text. For the measurement angle 0 in the JAZZ sequence, the population of the state $\ket{110}$ ideally becomes unity at the end of the controlled phase is $\pi$. We thus numerically maximize the $\ket{110}$ population by iteratively solving the time-dependent Schr\"{o}dinger equation taking account of the flat-top Gaussian pulse shape and obtaining an optimal parameter set of the drive frequency and amplitude. Note that we assume perfect accuracy of the state preparation, measurement, $\pi$-pulse, and $\pi/2$-pulse. Using the result, we solve the master equation taking into account the coherence times shown in Table\,\ref{tab:cohe} and reconstruct a noisy quantum channel $\mathcal{E}_{\mathrm{CZ}}'$, which is locally equivalent to a CZ gate, for the entire system. We thus apply local phase rotations to the data qubits with perfect accuracy and obtain the noisy CAS-based CZ gate channel $\mathcal{E}_{\mathrm{CZ}}$. Finally, we express the average gate fidelity of a quantum channel $\mathcal{E}$ using the following equation \cite{AGF_leak}
\begin{align}
    \Bar{F}(\mathcal{E}) = \frac{\mathrm{Tr} \left[(P_1 \otimes P_1)\mathcal{S}_\mathcal{E} \right] + \mathrm{Tr} \left[P_1 \mathcal{E}\left( P_1 \right) \right]}{d(d+1)}, \label{eq:S:Fave}
\end{align}
where $P_1$ is a projector onto the $d$-dimensional computational subspace and $\mathcal{S}_\mathcal{E}$ is the superoperator form of the quantum channel $\mathcal{E}$. Using Eq.\,\eqref{eq:S:Fave}, we evaluate the average gate fidelity of $\mathcal{E}_{\mathrm{CZ}}$ considering a composition between two channels $\tilde{\mathcal{E}}=\mathcal{U}_{\mathrm{CZ}} \circ \mathcal{E}_{\mathrm{CZ}}$, where $\mathcal{U}_{\mathrm{CZ}}$ is the ideal CAS-based CZ gate channel. The value obtained is $\bar{F} (\tilde{\cal{\varepsilon}}) \approx 0.978$. 
\begin{table}[ht]
    \caption{Coherence times of the qubits.}
    \centering
    \begin{tabular}{c@{\hspace{1em}}c@{\hspace{1em}}c@{\hspace{1em}}c}
            & $T_1\,(\mathrm{\mu s})$ & $T_2^*\,(\mathrm{\mu s})$ & $T_2^e\,(\mathrm{\mu s})$ \\ \hline \hline
    Data transmon, Q$_1$           & $95 \pm 10$ & $76 \pm 10$ & $88 \pm 3$ \\
    Data transmon, Q$_2$           & $108 \pm 6$ & $81 \pm 8$ & $166 \pm 9$ \\
    Coupler transmon, Q$_c$ & $15 \pm 1$ & $15 \pm 2$ & $18 \pm 2$ \\ \hline 
    \end{tabular}
    \label{tab:cohe}
\end{table}

\section{Experiment for the ac-field tunable ZZ interaction and estimation of the direct coupling}
\begin{figure}[ht]
    \centering
    \includegraphics[width=8.6cm]{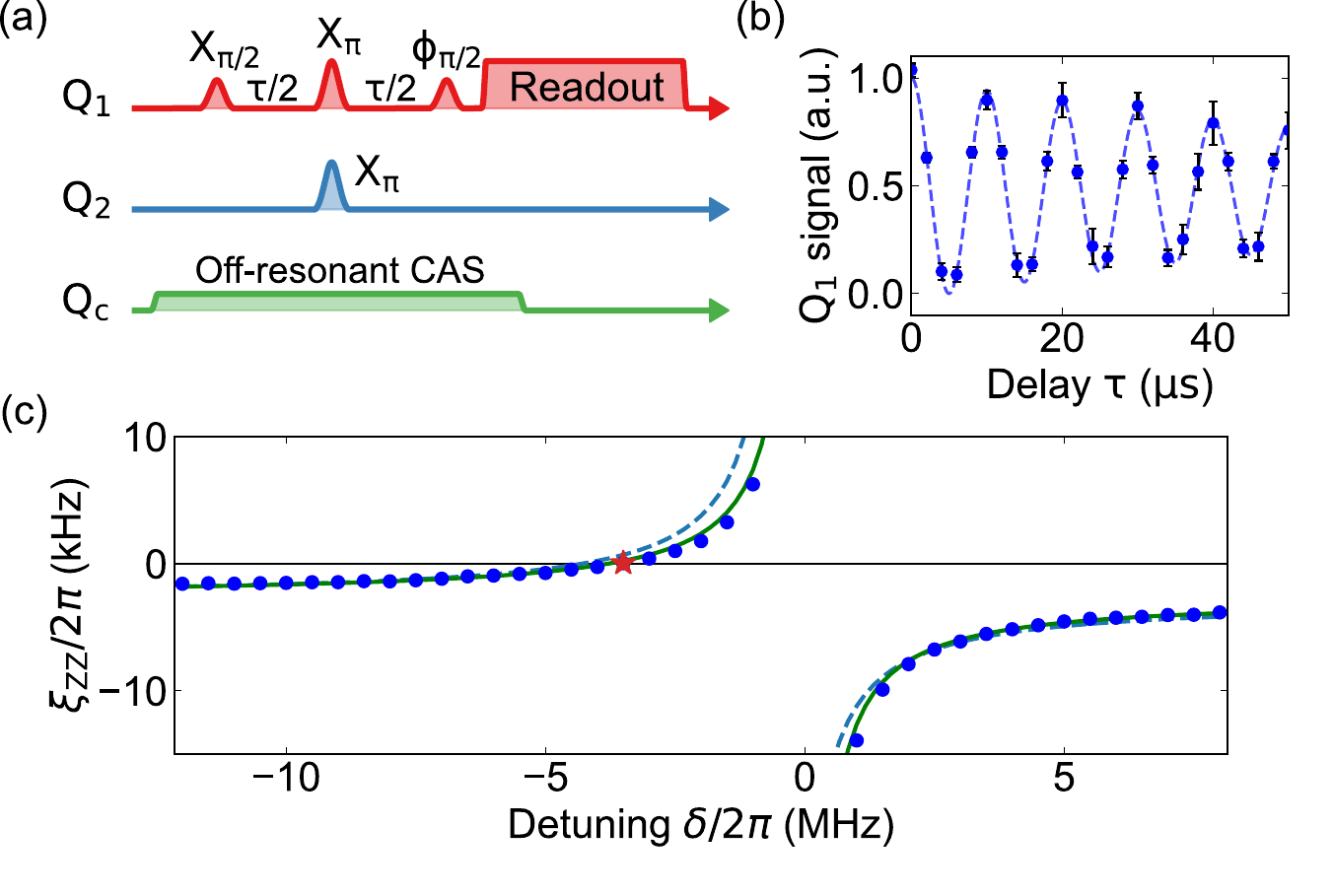}
    \caption{(a) Pulse sequence for measuring the tunable ZZ interaction strength using the JAZZ protocol. (b) Example of the experimental data obtained with the JAZZ protocol. The data was taken at the red star shown in (c). The dashed line is the fitting curve to an exponentially decaying sinusoidal function. (c) Dependence of the ZZ interaction strength on the drive detuning from the blue CAS transition with a fixed drive power $\Omega_d/2\pi$ = 7.3 MHz. The static ZZ interaction is canceled at the condition indicated by the red star. The green solid line shows the fitting result using Eq.\,\eqref{eq:S:tunable_ZZ}, and the blue dashed line is the numerical fit using the direct coupling $g_{12}$ as the only free parameter.}
    \label{fig:S:calib_zz}
\end{figure}

We estimate the direct transverse coupling strength and the residual ZZ interaction strength using the JAZZ experiment described in the main text. Figure \ref{fig:S:calib_zz}(a) shows a pulse sequence, where we constantly apply a relatively weak coupler drive $\Omega_d/2\pi \approx 7.3$ MHz, while sweeping the delay time $\tau$ between the echo pulses and the coupler drive detuning from the blue CAS transition, $\delta = \omega_d - \omega_b$. Furthermore, to know the sign as well as the magnitude of the ZZ interaction, the measurement angle $\phi$ is swept together with $\tau$ according to the relation $\phi/\tau = 2\pi \times 100 \,\mathrm{kHz}$ as shown in Fig.\,\ref{fig:S:calib_zz}(b) as an example. Figure\,\ref{fig:S:calib_zz}(c) shows the measured ZZ interaction strength $\xi_\mathrm{ZZ}$ depending on the coupler drive detuning. By fitting this modulation with numerically calculated values of the residual ZZ interaction diagonalizing Eq.\,\eqref{eq:S:Hr_num}, we obtain the direct transverse coupling strength of $g_{12}/2\pi\simeq1.9\,$MHz. The bare residual ZZ interaction strength is also estimated to be $\xi_0/2\pi\simeq-1.5\,$kHz, where all other parameters we use are presented in the main text and $g_{12}$ is the only free parameter.

\section{ac Stark shift of the CAS transitions}
\begin{figure}[ht]
    \centering
    \includegraphics[width=8.6cm]{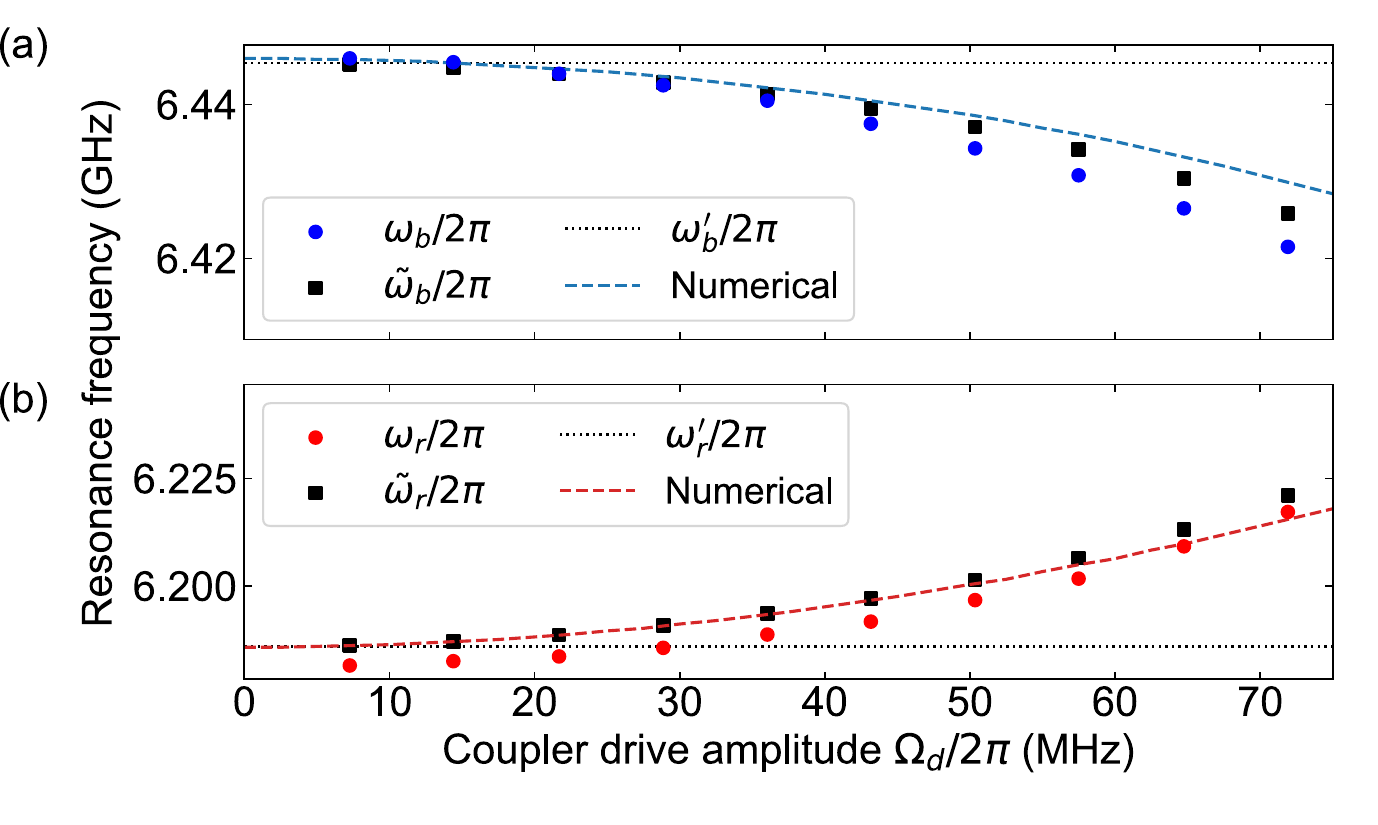}
    \caption{Resonance frequencies of the (a) blue and (b) red CAS transitions. The filled circles are the experimental results obtained from the fitting of the chevron patterns for each drive amplitude. The filled squares are analytically calculated ac-Srark-shifted CAS transition frequencies~[Eqs.\,\eqref{eq:S:wb_ac} and~\eqref{eq:S:wr_ac}] using the same parameters as in the experiment. The dotted lines are the analytically evaluated CAS transition frequencies in the limit of the weak drive [Eqs.\,\eqref{eq:S:freq_blue_eff} and~\eqref{eq:S:freq_red_eff}]. The dashed lines are obtained numerically by diagonalizing Eqs.\,\eqref{eq:S:Hr_num}--\eqref{eq:S:Hd_at_rot}.}
    \label{fig:S:acStark}
\end{figure}

In Figs.\,\ref{fig:S:acStark}(a) and (b), we show the experimental results of the ac-Stark-shifted blue and red CAS transition frequencies as a function of the coupler drive amplitude. The CAS transition frequencies are determined by fitting the chevron pattern at each point. We model the frequency shift with the ac Stark shift of the coupler transmon $\Delta_c^\mathrm{ac}=\frac{\alpha_c\Omega_d^2}{2\delta_c(\delta_c + \alpha_c)}$ \cite{PhysRevLett.94.123602, PhysRevA.97.062334}. Using Eqs.\,\eqref{eq:S:freq_blue_eff} and \eqref{eq:S:freq_red_eff}, we define the analytically evaluated ac-Stark-shifted CAS transition frequencies as
\begin{align}
    \Tilde{\omega}_b &= \omega_b' + \Delta_c^\mathrm{ac}, \label{eq:S:wb_ac}\\
    \Tilde{\omega}_r &= \omega_r' + \Delta_c^\mathrm{ac}, \label{eq:S:wr_ac}
\end{align}
where we ignore the ac Stark shifts of the data transmons, which are negligible compared to $\Delta_c^\mathrm{ac}$. In Figs.\,\ref{fig:S:acStark}(a) and (b), we see semiquantitative agreement in the weak drive limit. The deviations between the numerical and experimental results at larger drive amplitudes could be explained by the higher-order nonlinear terms dropped in the Duffing-oscillator model \cite{FogiCZ_2020krinner, PhysRevA.91.043846}.

\section{Rabi oscillations in the blue CAS subspace}
We measure the associated oscillations of the population of each qubit involved in the blue CAS transition. The pulse sequence used is identical to the one in Fig.\,2(a) in the main text except for the qubits to be read out. As predicted by the theoretical model, we observe signals corresponding to the Rabi oscillations between the states $\ket{010}$ and $\ket{101}$.
\begin{figure}
    \centering
    \includegraphics[width=8.6cm]{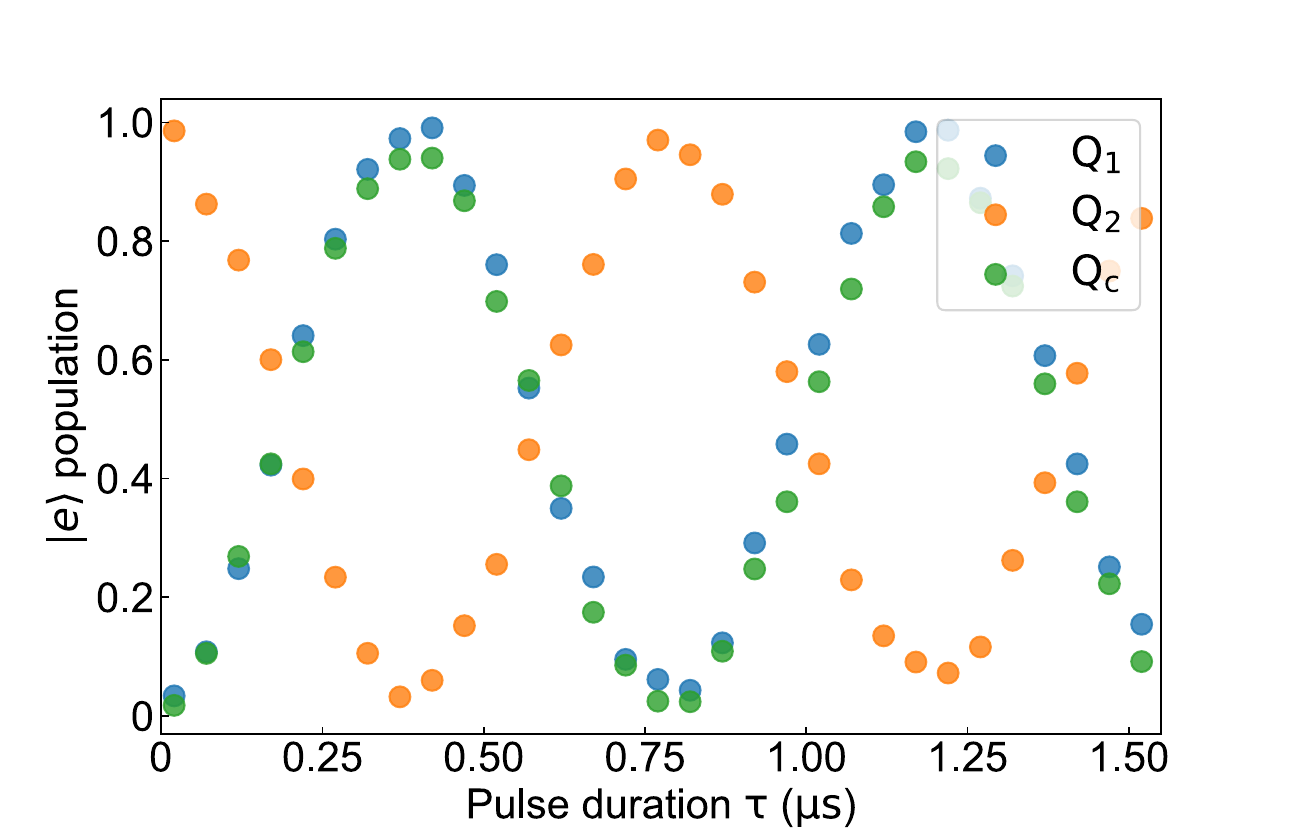}
    \caption{Associated oscillations of the excited-state population of each transmon involved in the blue CAS transition. The vertical axis is normalized using the response signals of the ground and first-excited states of each transmon, corresponding to the excited state population of each qubit. The horizontal axis is the length of the drive pulse to the coupler transmon. The drive amplitude is $\Omega_d/2\pi\simeq75\,$MHz. Note that this data was obtained at a different cooldown from the one for the experiments in the main text.}
    \label{fig:S:bcas_rabi}
\end{figure}

\section{Comparison of the drive efficiency and residual ZZ interaction strength with the CR gate}
\begin{figure*}[ht]
    \centering
    \includegraphics[width=17.8cm]{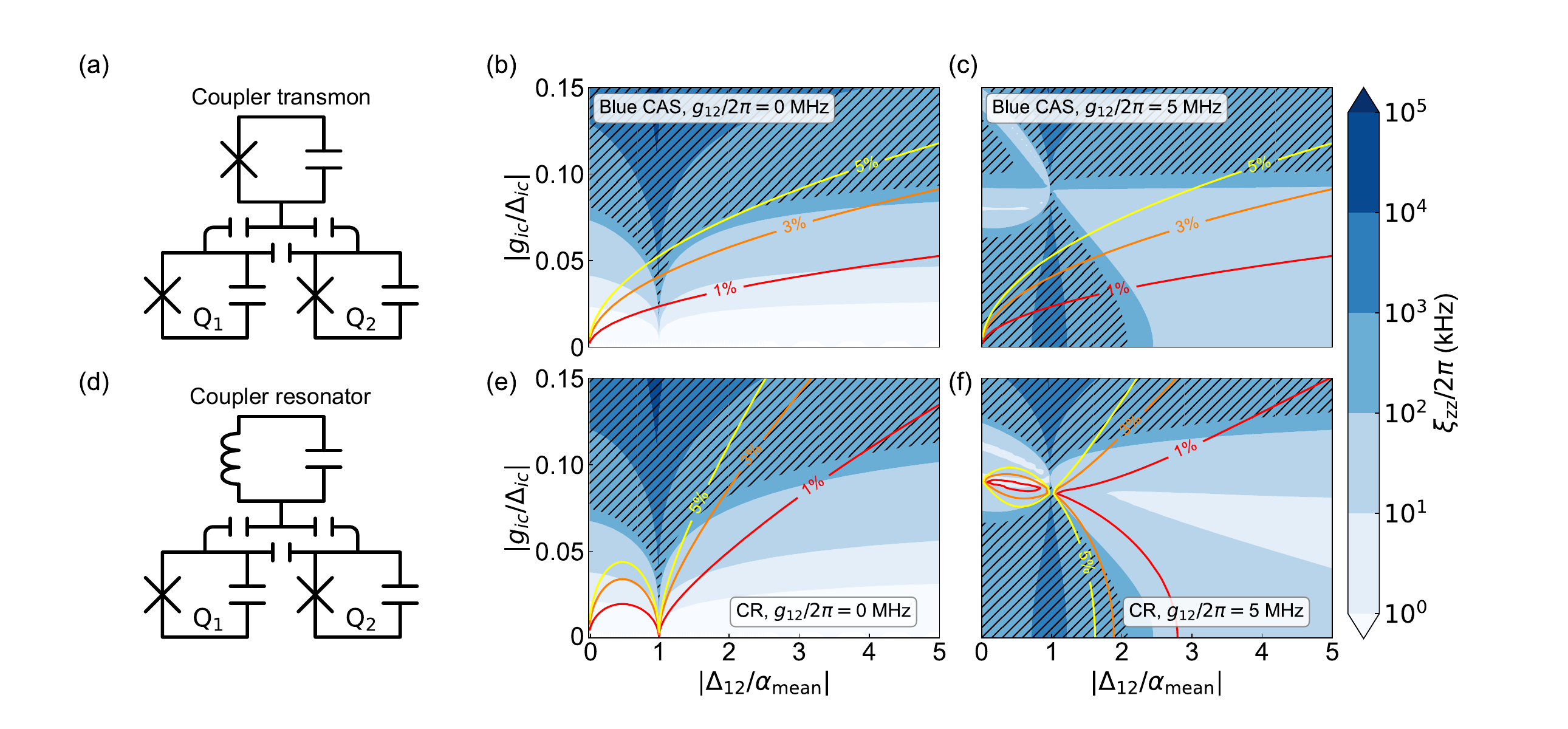}
    \caption{(a) Circuit diagram to implement the CAS-based gates described in the main text. The residual ZZ interaction strength $\xi_\mathrm{ZZ}$~(filled contour plot) and the drive efficiency $\eta_b$ of the blue CAS transition~(contour line plot) are shown in (b)~without and (c)~with~[same as in Fig.~4(b)] the direct transverse coupling $g_{12}$. (d) Typical circuit diagram to implement the CR gate, where we consider a linear coupler (off-resonant LC resonator) as opposed to the transmon coupler in~(a). The residual ZZ interaction strength $\xi_\mathrm{ZZ}$ and drive efficiency $\eta_\mathrm{CR}$ of the CR gate, as a function of $|\Delta_{12}/\alpha_\mathrm{mean}|$ and $|g_{ic}/\Delta_{ic}|$ are shown in (e)~without and (f)~with the direct transverse coupling $g_{12}$. Here, $\xi_\mathrm{ZZ}$ is calculated through numerical diagonalization of Eq.\,\eqref{eq:S:Hr_num}. The drive efficiency is defined as $\eta_\mathrm{CR} = 2\times\frac{2g_\mathrm{eff}\alpha_1}{\Delta_{12}(\Delta_{12}+\alpha_1)}$ from Eq.\,(4.26) in Ref.~\cite{PhysRevA.101.052308}. The additional multiplying factor of 2 explicitly indicates the fact that a $\pi/2$-rotation of the CR gate is locally equivalent to the CNOT gate. For the calculations, we use the same parameters as in the case with the CAS transitions except for the anharmonicities of the linear coupler, $\alpha_c/2\pi = 0$~GHz, and the data transmons, $\alpha_1/2\pi=\alpha_2/2\pi=-0.3$~GHz. The latter value is typical and indeed more favorable for CR gates. }
    \label{fig:S:comparison}
\end{figure*}

Lastly, we compare the expected properties of the blue CAS-based CZ gate with those of the CR gate, which is most commonly used in architectures with fixed-frequency transmons. The results are shown in Fig.~\ref{fig:S:comparison}. In both cases, we see the decrease of the residual ZZ interaction by introducing $g_{12}$ in the regions with large enough $g_{ic}/\Delta_{ic}$ for a high drive efficiency. However, the CR gate only achieves sufficient drive efficiency $\eta_\mathrm{CR}$ in the regime where $\xi_\mathrm{ZZ}$ rapidly increases with $g_\mathrm{eff}$. In contrast, the blue CAS drive efficiency is independent of $g_{12}$~[See Eqs.\,\eqref{eq:S:gb} and~\,\eqref{eq:S:gr}], allowing for the wide range of detuning and coupling strength with large $\eta_\mathrm{b}$ and small $\xi_\mathrm{ZZ}$.

\end{document}